\documentclass{aa}

\usepackage[varg]{txfonts}
\usepackage{graphicx}
\usepackage{amsmath}
\usepackage{amssymb}
\usepackage{microtype}
\usepackage{booktabs}
\usepackage{multirow}
\usepackage{subcaption}
\usepackage{tablefootnote}
\usepackage{threeparttable}
\usepackage{siunitx}

\usepackage{hyperref}
\hypersetup{colorlinks=true, linkcolor=blue, citecolor=blue}

\title{Linking the dust and chemical evolution: Taurus and Perseus}
\subtitle{New collisional rates for HCN, HNC, and their C, N, and H isotopologues}

\author{D. Navarro-Almaida \inst{1}
        \and 
        C. T. Bop \inst{2}
        \and
        F. Lique \inst{2}
        \and 
        G. Esplugues \inst{3}
        \and
        M. Rodr\'iguez-Baras \inst{3}
        \and
        C. Kramer \inst{4}
        \and
        C. E. Romero \inst{5}
        \and
        A. Fuente \inst{3}
        \and
        P. Caselli \inst{6}
        \and
        P. Rivi\'ere-Marichalar \inst{3}
        \and
        J. M. Kirk \inst{7}
        \and
        A. Chac\'on-Tanarro \inst{3}
        \and
        E. Roueff \inst{8}
        \and
        T. Mroczkowski\inst{9}
        \and
        T. Bhandarkar\inst{10}
        \and
        M. Devlin\inst{10}
        \and
        S. Dicker\inst{10}
        \and
        I. Lowe\inst{10}
        \and
        B. Mason\inst{11}
        \and
        C. L. Sarazin\inst{12}
        \and
        J. Sievers\inst{13}
        }

\institute{D\'epartement d'Astrophysique (DAp), Commissariat \`a l'\'Energie Atomique et aux \'Energies Alternatives (CEA), Orme des Merisiers, B\^at. 709, 91191 Gif sur Yvette, Paris-Saclay, France\\ \email{david.navarroalmaida@cea.fr}
        \and
        Univ. Rennes, CNRS, IPR (Institut de Physique de Rennes) – UMR 6251, 35000 Rennes, France
        \and
        Observatorio Astron\'omico Nacional (OAN), Alfonso XII, 3,  28014, Madrid, Spain
        \and
        Institut de Radioastronomie Millim\'etrique (IRAM), 300 Rue de la Piscine, 38406 Saint Martin d'H\`eres, France
        \and
        Center for Astrophysics, Harvard and Smithsonian, 60 Garden Street, Cambridge, MA 02143, USA
        \and
        Centre for Astrochemical Studies, Max Planck Institute for Extraterrestrial Physics, Giessenbachstrasse 1, 85748 Garching, Germany
        \and
        Jeremiah Horrocks Institute, University of Central Lancashire, Preston, PR1 2HE, UK
        \and
        Sorbonne Universit\'es, Observatoire de Paris, Universit\'e PSL, CNRS, LERMA, 92190 Meudon, France
        \and
        European Southern Observatory (ESO), Karl-Schwarzschild-Strasse 2, D-85748 Garching, Germany
        \and
        Department of Physics and Astronomy, University of Pennsylvania, 209 South 33rd Street, Philadelphia, PA, 19104, USA 
        \and
        NRAO, 520 Edgemont Rd, Charlottesville, VA, 22903, USA
        \and
        Department of Astronomy, University of Virginia, 530 McCormick Road, Charlottesville, VA, 22904-4325, USA
        \and
        Department of Physics, McGill University, 3600 University Street Montreal, QC, H3A 2T8, Canada
        }

\date{}

\begin{document}

\abstract {HCN, HNC, and their isotopologues are ubiquitous molecules that can serve as chemical thermometers and evolutionary tracers to characterize star-forming regions. Despite their importance in carrying information that is vital to studies of the chemistry and evolution of star-forming regions, the collision rates of some of these molecules have not been available for rigorous studies in the past.}
{Our goal is to perform an up-to-date gas and dust chemical characterization of two different star-forming regions, TMC 1-C and NGC 1333-C7, using new collisional rates of HCN, HNC, and their isotopologues. We investigated the possible effects of the environment and stellar feedback in their chemistry and their evolution.}
{We used updated collisional rates of HCN, HNC, and their isotopologues in our analysis of the chemistry of TMC 1-C (Taurus) and NGC 1333-C7 (Perseus). With millimeter observations, we derived their column densities, the C and N isotopic fractions, the isomeric ratios, and the deuterium fractionation. The continuum data at 3 mm and 850 $\mu$m allowed us to compute the emissivity spectral index and look for grain growth as an evolutionary tracer.}
{The H$^{13}$CN/HN$^{13}$C ratio is anticorrelated with the deuterium fraction of HCN, thus it can readily serve as a proxy for the temperature. The spectral index $(\beta\sim 1.34-2.09)$ shows a tentative anticorrelation with the H$^{13}$CN/HN$^{13}$C ratio, suggesting grain growth in the evolved, hotter, and less deuterated sources. Unlike TMC 1-C, the south-to-north gradient in dust temperature and spectral index observed in NGC 1333-C7 suggests feedback from the main NGC 1333 cloud.}
{With this up-to-date characterization of two star-forming regions, we found that the chemistry and the physical properties are tightly related. The dust temperature, deuterium fraction, and the spectral index are complementary evolutionary tracers. The large-scale environmental factors may dominate the chemistry and evolution in clustered star-forming regions.}


\maketitle

\section{Introduction}

The two isomers of the H-C-N system, hydrogen cyanide (HCN) and hydrogen isocyanide (HNC), are ubiquitous in the interstellar medium (ISM). They have been detected in diffuse clouds \citep{Liszt2001}, dark clouds \citep{Hirota1998}, starless cores \citep{Tennekes2006, Hily2010}, and low- and high-mass star forming regions \citep{Schilke1992, Godard2010}. Moreover, HCN is widely used in extragalactic research as tracer of the dense gas that forms stars. Together with HNC and HCO$^+$, it also serves as a chemical diagnostic for differentiating between active galactic nuclei (AGNs) and starbursts (see, e.g., \citealp{Aalto2007, Aalto2012}). The [HCN]/[HNC] ratio has also been proposed as a tracer of the gas kinetic temperature in massive star-forming regions \citep{Graninger2014, Hacar2020} and proto-planetary disks \citep{Long2021}. Therefore, a full comprehension of the HCN and HNC chemistry is of critical importance in the study of the physics and chemistry of interstellar gas. 

Although chemical models appear to reasonably account for the abundances of these compounds to a reasonable extent, notorious exceptions show that the HCN/HNC chemistry is not yet fully understood. Although HNC is less stable than HCN by 55 kJ/mol, with an isomerization barrier for passage from HNC to HCN calculated to be equal to 124 kJ/mol, steady-state gas phase models predict an HNC abundance comparable to that of HCN at $\sim$10 K \citep{Hirota1998, Tennekes2006, Sarrasin2010}. In cold clouds, the two isomers form primarily via the dissociative recombination of HCNH$^+$ and then the recombination branching ratio (similar for the two isomers) regulates their abundance ratio, provided that they are efficiently protonated via reactions with ions, such as H$_3^+$ and HCO$^+$ \citep{Churchwell1984}. At high temperatures, new paths to form HCN via neutral-neutral reactions produce [HCN]/[HNC] ratios larger than 1. This simple scenario was able to explain in general terms the wealth of observational data gathered thus far: the [HCN]/[HNC] ratio is observed close to 1 in cold cores \citep{Daniel2013, Lefloch2021}, slightly higher than 1 (between 1 and 3) in giant molecular clouds (GMCs), and increasing to values much greater than 1 in hot cores \citep{Jin2015}, Young Stellar Objects \citep[YSOs,][]{Jorgensen2004} and luminous galaxies \citep[ULIRGs,][]{Baan2010}. However, some exceptions challenge our understanding of the chemistry of nitriles. In particular, low values of the [HCN]/[HNC] ratio ($\sim$0.3-0.5) have been measured in starless cores \citep{Tennekes2006, Hily2010} and infrared dark clouds \citep{Liu2013}. These exceptions are difficult to reproduce on the basis of current state-of-the-art chemical models.

A great theoretical effort has been undertaken to expand the understanding of nitrogen chemistry. \citet{Loison2014} reviewed the reactions involving the formation and destruction of HCN and HNC in dark clouds, updating their gas-grain time-dependent chemical model with the latest laboratory measurements of reaction rates and branching ratios. They found that the [HCN]/[HNC] ratio is strongly dependent on time and on the amount of atomic carbon available in the gas phase, and could reach values $\sim$1 even at low temperatures. Laboratory experiments by \citet{Wu2012} showed that N$_3$, C$_n$N (n = 1$-$3), CN$_2$, (CN)$_2$, HCN$_2$, HC$_2$N, C(NH)$_2$, HN$_3$, HNC, HCN, HCCNH$^+$, and NCCN$^+$ are formed when a mix of solid N$_2$ and CH$_4$ is irradiated with UV photons. The release of these compounds to gas phase by thermal evaporation, photo-desorption, or sputtering would produce an increase in the abundance of nitriles (CN, HCN, and HNC). Subsequent experiments show that nitriles can also be produced by irradiation of ices formed with N$_2$ and other C-bearing species such as acetylene \citep{Wu2014,Chen2015}.

Understanding isotopic abundance ratios is a major goal in modern astrophysics. On one hand, rarer isotopologues are used to determine physical conditions and as a proxy of the most abundant isotopologues because they usually do not suffer from opacity problems. In addition, isotopic ratios are tools to investigate the link between the Solar System objects and galactic interstellar environments as discussed by \citet{Aleon2010}. In particular, \citet{Roueff2015} presented a time-dependent gas phase model including isotopic fractionation of C, N, and deuterium, and showed that the [H$^{13}$CN]/[HCN] and [HNC]/[HN$^{13}$C] ratios could differ from the standard ratio of $\sim$68 \citep{Milam2005, Colzi2020}, which is the one generally assumed by the observers. In detail, the fractionation of CO reduces the amount of $^{13}$C atoms in gas phase, leading to values of the [HCN]/[H$^{13}$CN] and [HNC]/[HN$^{13}$C] ratios that approach $110-120$. 

We mainly focus our study on interstellar environments where, due to the low-temperature conditions, most molecules are frozen out onto dust grains presenting depleted abundances \citep{Caselli1999, Tafalla2006, Pagani2007}. Deuterated compounds are thought to be good tracers of the cold gas in the interior of dense starless cores. At the low temperatures prevailing in these environments, the exothermicity of the reaction,

\begin{equation}
\label{eq:deuteration}
H_3^+ + HD \rightarrow  H_2D^+ +H_2,
\end{equation}
in the forward direction (of $\sim$232 K, see \citealp{Gerlich2002}) promotes the formation of H$_2$D$^+$ and inhibits the reverse reaction. The deuteration proceeds further via formation of D$_2$H$^+$ and D$_3^+$ in reactions with HD and D$_2$ . These deuterated ions react with a variety of neutral molecules, such as CO and N$_2$, leading to high abundances of other deuterated species (see, e.g., \citealp{Ceccarelli2014, Roueff2015}). Thus, deuterated molecules become important diagnostic tools to determine the evolutionary stage of a dense core \citep{Caselli2002, Emprechtinger2009,  Chen2011, Fontani2015, Colzi2018, Navarro2021, Esplugues2022}. Complex chemical models including deuterated species and spin chemistry have been developed to account for these observations \citep[see, e.g.,][]{Roueff2015, Majumdar2017, Sipila2019}. Still, dating using deuterated molecules is limited by large uncertainties in the isotopic abundance ratios. Usually, the collisional coefficients are ignored and the collisional rates of the rarer isotopologues are set to be equal to those of the most abundant species, thus introducing a bias into the resulting values. 


The evolution of prestellar and protostellar objects in star-forming regions may also be inferred from the dust emission. The expected evolution of dust properties from the diffuse interstellar medium to dense cores is still an open issue. In dense regions, grains are covered by icy mantles that make grains sticky, favoring grain coagulation and producing a different dust size distribution, thereby changing the dust emissivity \citep{Ormel2009, Ormel2011}. The knowledge of grain properties (e.g., emissivity and grain size distribution) is therefore important to better understand the star formation process. Moreover, grain coagulation would affect the charge balance in the gas phase, thus modifying the coupling of the gas with the magnetic field. In addition, grain surface chemistry plays an important role in the formation of molecules that are key to chemical networks (e.g., H$_2$, H$_2$O) and to the freeze-out of molecules (e.g., CO) in the cold interiors of starless cores. Summarizing, the precise knowledge of dust properties is of paramount importance to the understanding of the chemical gas evolution and the dynamics of starless, prestellar, and protostellar cores.

In this paper, we use spectroscopic observations and dust continuum emission to explore the chemical and physical evolution of the prestellar and protostellar envelopes during the first stages of the star formation process. To explore the chemical evolution, we investigated the isotopic and isomeric ratios between the C, N  and  H isotopologues of HCN and HNC. To obtain the most precise determinations, we estimated the specific collisional coefficients for each isotopologue (see Sect.~\ref{sec:colCoeff}), and, for the first time, we used these coefficients to obtain the isotopic and isomeric ratios (see Sect.~\ref{sec:mol}). Then, to explore the dust evolution, we carried out the mapping of the continuum emission at 3mm using the MUSTANG-2 bolometer of the Green Bank Telescope. These observations were combined with {\em Herschel} data to obtain the mm dust spectral index. Finally, we compared the chemical gas composition with the derived mm spectral index to link the evolution of these two phases of the interstellar medium.

\section{Collisional coefficients}
\label{sec:colCoeff}

Rate coefficients of HCN and HNC induced by collision with para-H$_2$ (\textit{p}-H$_2$) were computed by \cite{hernandez2017rotational} for temperatures up to 500 K. These authors used high accurate ab initio potential energy surfaces (PESs) of the HCN-H$_2$ \citep{denis2013new} and HNC-H$_2$ \citep{dumouchel2011rotational} van der Waals complexes. In the case of HCN-H$_2$, the accuracy of the PES was validated through comparison with spectroscopic measurements.

\subsection{Derivation of collisional rates}

To the best of our knowledge, rate coefficients induced by collision with H$_2$ of the isotopologues of HCN and HNC, namely, DCN, D$^{13}$CN, H$^{13}$CN, HC$^{15}$N, DNC, DN$^{13}$C, HN$^{13}$C, and H$^{15}$NC, are missing from the literature. We note that the data for DCN reported by \cite{denis2015isotopic} were obtained using helium as collision partner. Several studies \citep{dumouchel2017hyperfine,flower2015excitation} have shown that isotopic substitution can lead to substantial changes in the study of rotational energy transfer induced by collision. Therefore, it is worth determining the actual collisonal rate coefficients of the isotopologues of HCN and HNC for a better interpretation of their observational spectra.

For isotopologues (e.g., HCN-H$_2$ and DCN-H$_2$) in their ground electronic state, the interaction potentials calculated using the Born-Oppenheimer approximation are the same. They depend only on the interatomic distances. Hence, we used the HCN-H$_2$ and HNC-H$_2$ PESs to compute inelastic cross-sections of the isotopologues of HCN and HNC due to collision with H$_2$, respectively. In practice, we incorporated the shift of the center of mass, for each isotopologue, to describe the PESs as function of Jacobi coordinates.

Using the \texttt{MOLSCAT} computer code \citep{hutson1994molscat}, we calculated state-to-state inelastic cross sections of DCN, D$^{13}$CN, H$^{13}$CN, HC$^{15}$N, DNC, DN$^{13}$C, HN$^{13}$C, and H$^{15}$NC induced by collision with \textit{p}-H$_2$ ($j_2$\footnote{The quantum numbers $j_1$ and $j_2$ denote the rotational levels of the HCN isotopologues and H$_2$, respectively.} = 0) for total energies up to 250 cm$^{-1}$. For reasons of convergence, we used wide rotational bases including $j_2=2$ for H$_2$ and $j_1$ up to 18. The other parameters are defined so that the convergence threshold is less than 1\%. By thermally averaging these cross sections over the Maxwell-Boltzmann collision energy distribution, we derived rate coefficients for the eight low-lying energy levels of each isotopologue ($j_1=0-7$) up to a kinetic temperature of 30 K.

\begin{figure*}
    \centering
    \includegraphics[width = 0.98\textwidth]{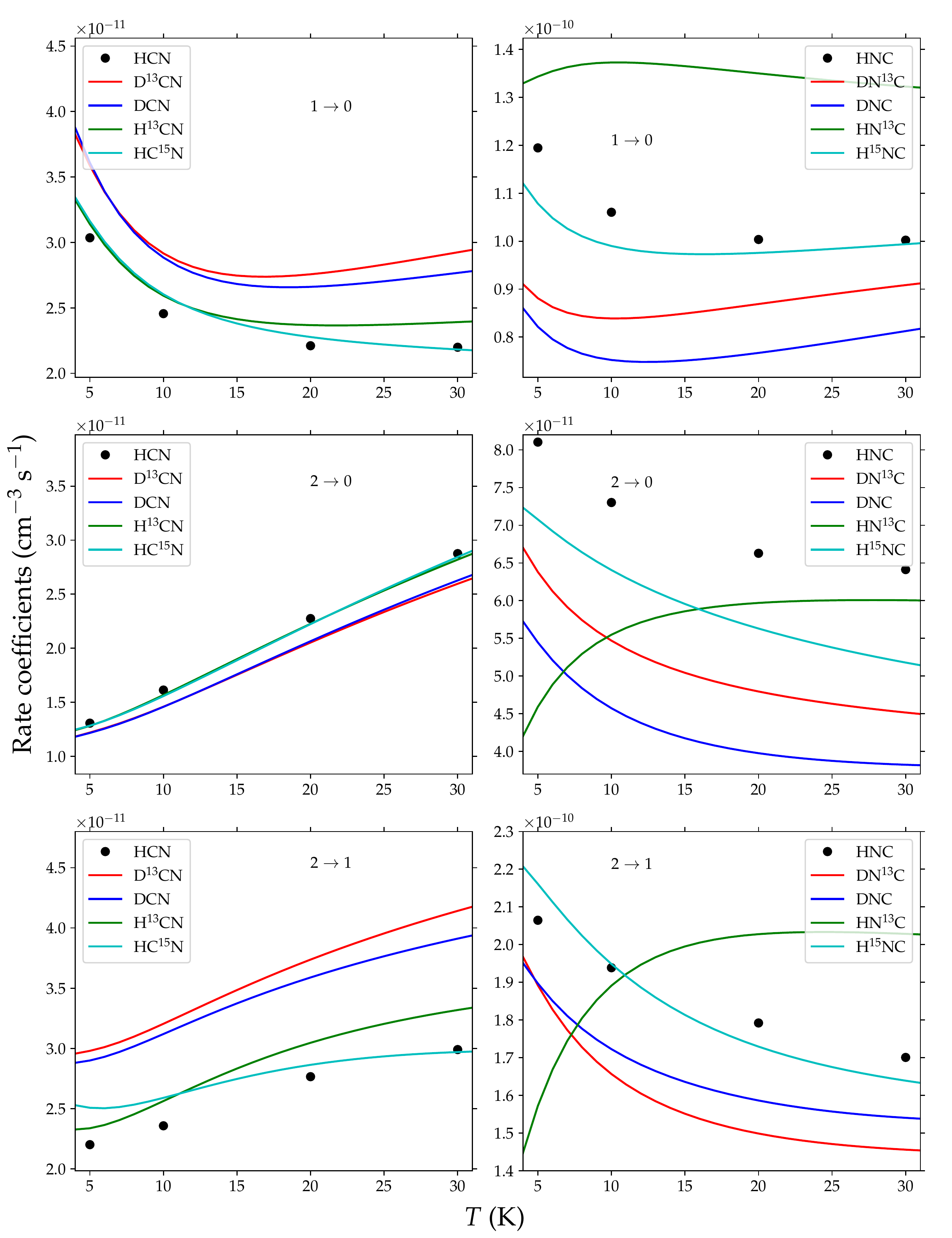}
    \caption{Temperature dependence of the rate coefficients of the isotopologues of HCN (left panels) and HNC (right panels). The data of the parent species (HCN and HNC) are from \citet{hernandez2017rotational}.}
    \label{fig:rates}
\end{figure*}

Figure~\ref{fig:rates} shows the variation of the rate coefficients of the isotopologues of HCN and HNC as a function of the temperature for the $1\to0$, $2\to0$, and $2\to1$ transitions. The effect of isotopic substitution increases with increasing temperature in the case of HCN. Concerning HNC, we obtained large differences at a low temperature, but no systematic pattern is observed. For example, at 10 K (the typical temperature of cold molecular clouds), the change in the HCN and HNC rate coefficients due to isotopic substitution can be as high as $\sim50\%$. Globally, we can say that HNC is more sensitive than HCN to isotopic substitution and the impact depends on the substituted isotope, the temperature, and the transition. This can be explained by recalling the strong anisotropy and deep global minimum of the HNC-H$_2$ PES with respect to that of HCN-H$_2$ (see \cite{denis2013new} and \cite{dumouchel2011rotational}).

\subsection{Radiative transfer computations}

\begin{figure*}
    \centering
        \begin{subfigure}[b]{0.49\textwidth}\includegraphics[width=\textwidth,keepaspectratio]{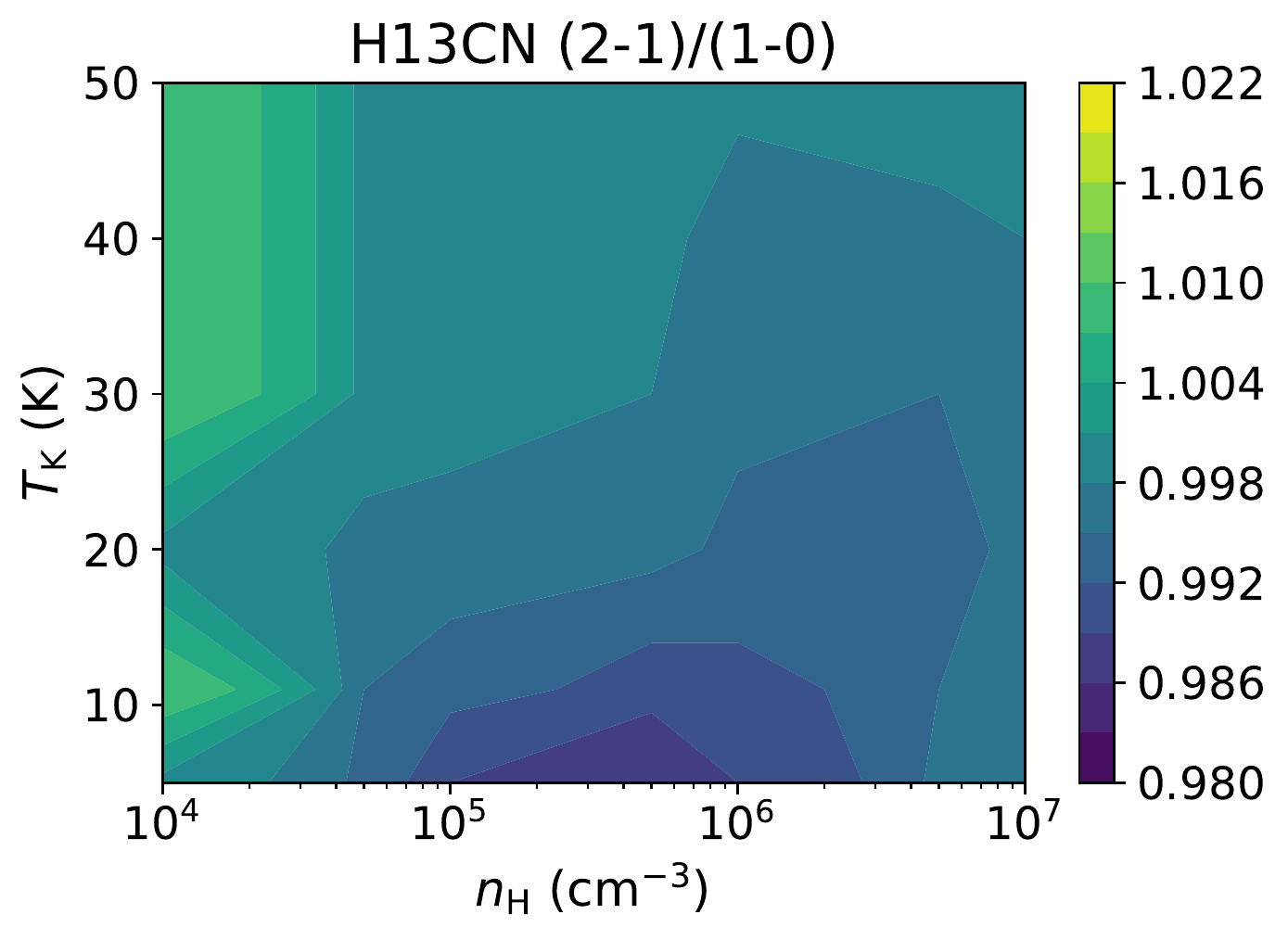}
		\end{subfigure}
        \begin{subfigure}[b]{0.49\textwidth}\includegraphics[width=\textwidth,keepaspectratio]{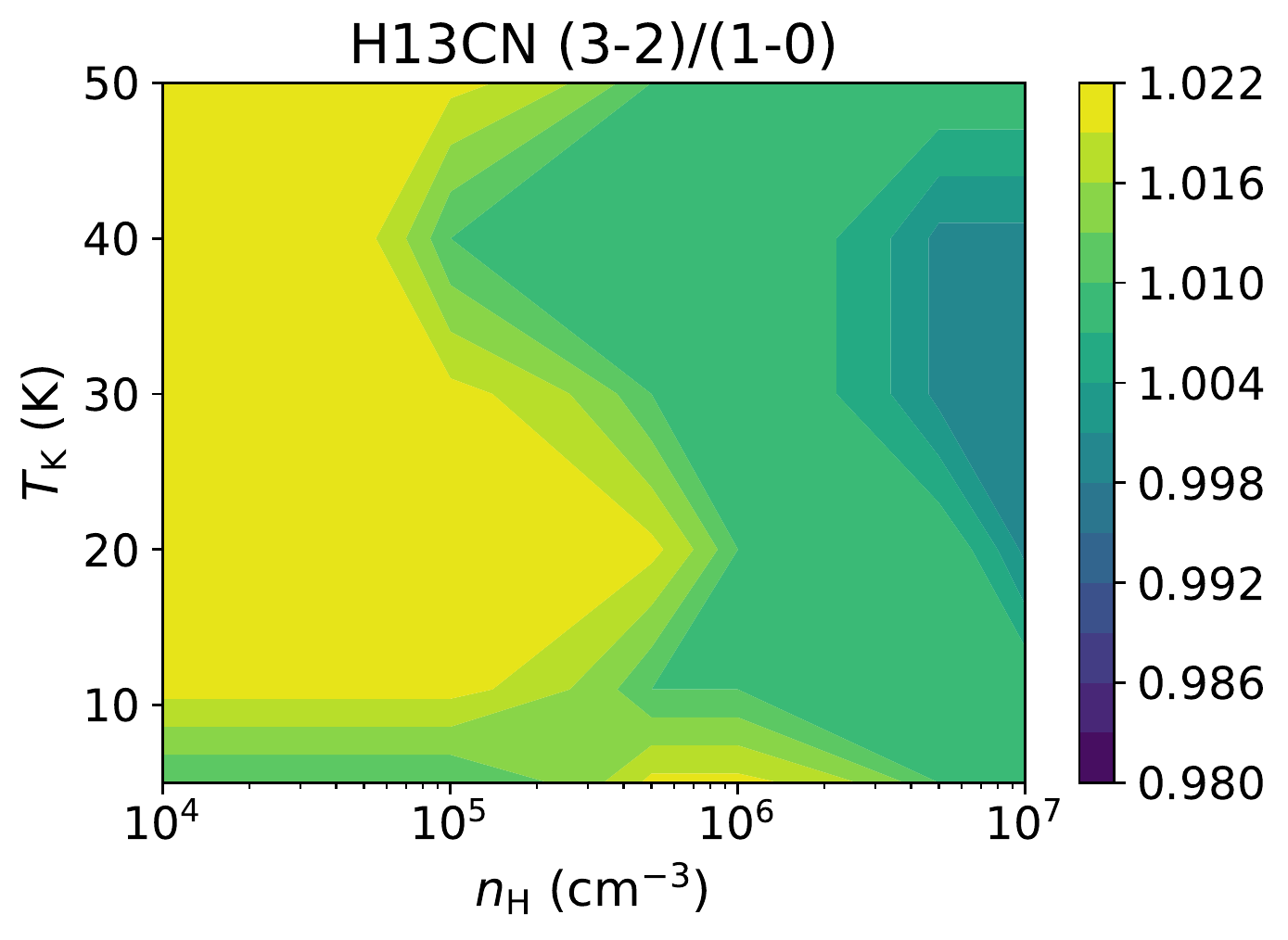}
        \end{subfigure}
        \caption{Contours display the ratio between quotients of radiation temperatures $T_{r}(2-1)/T_{r}(1-0)$ (left), and $T_{r}(3-2)/T_{r}(1-0)$ (right) calculated with the previous and the new collisional coefficients for H$^{13}$CN.}
	    \label{figure:H13CN}
\end{figure*}

\begin{figure*}
    \centering
        \begin{subfigure}[b]{0.49\textwidth}\includegraphics[width=\textwidth,keepaspectratio]{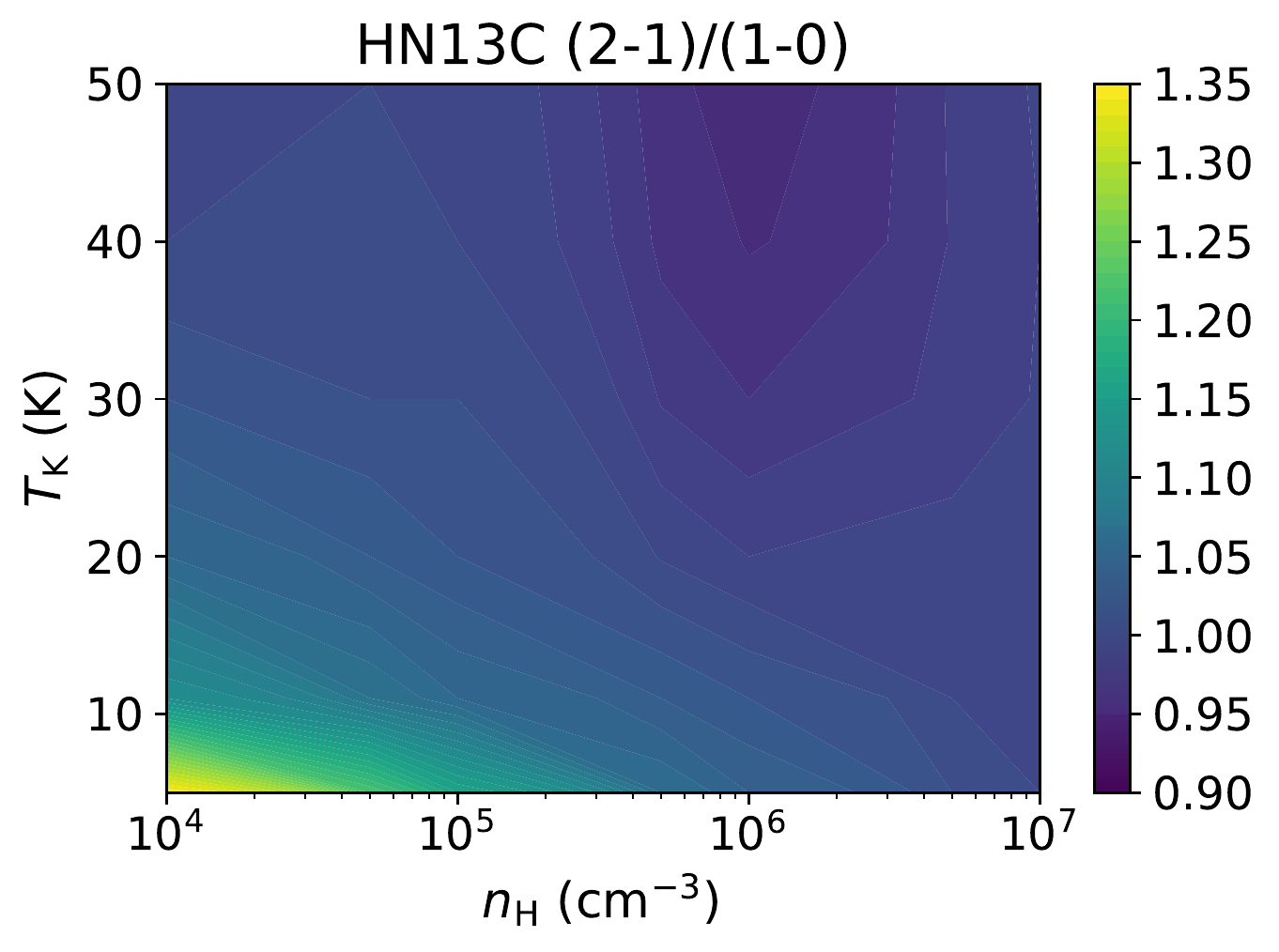}
		\end{subfigure}
        \begin{subfigure}[b]{0.49\textwidth}\includegraphics[width=\textwidth,keepaspectratio]{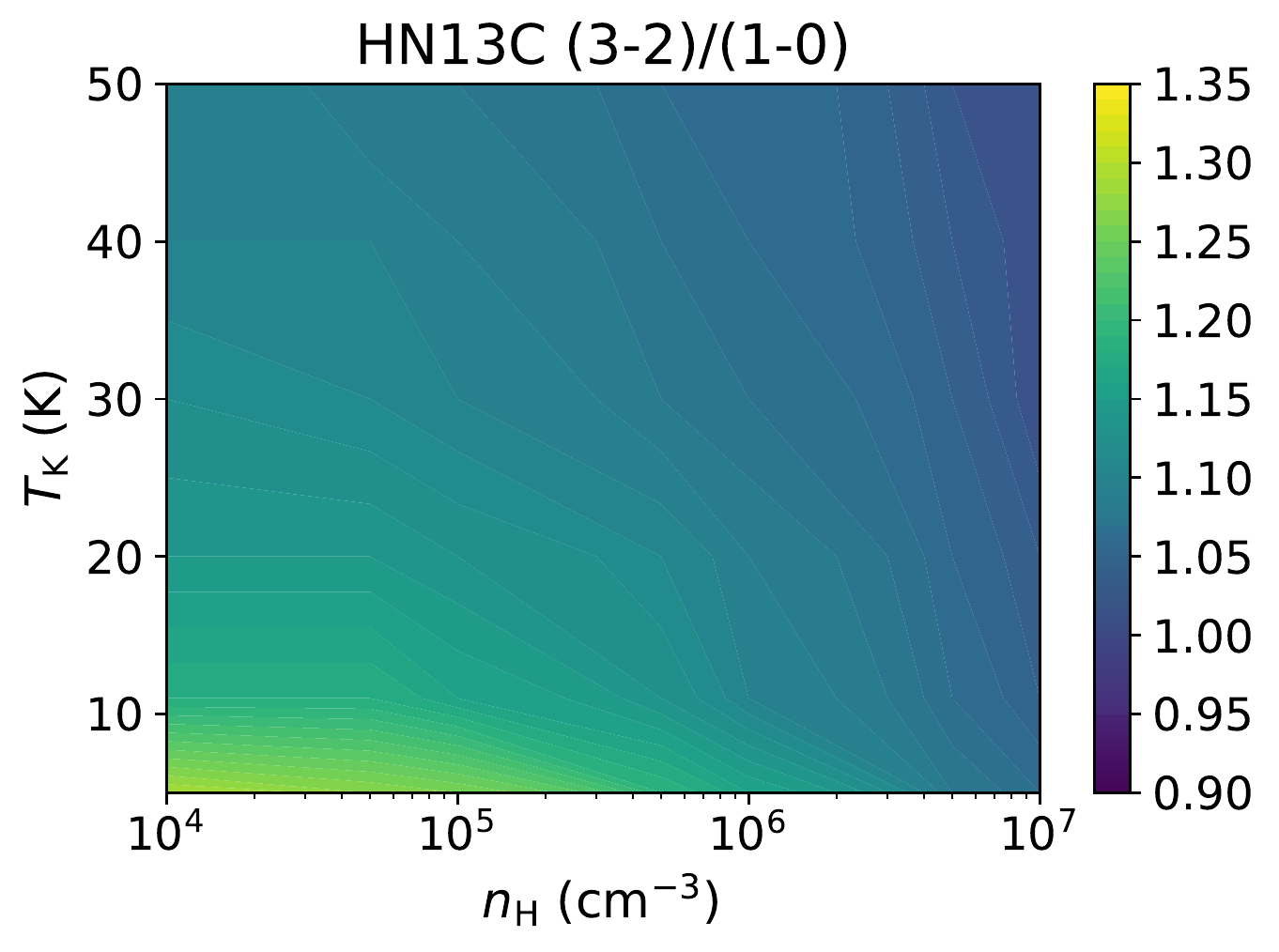}
        \end{subfigure}
	    \caption{Contours display the ratio between quotients of radiation temperatures $T_{r}(2-1)/T_{r}(1-0)$ (left), and $T_{r}(3-2)/T_{r}(1-0)$ (right) calculated with the previous and the new collisional coefficients for HN$^{13}$C.}
	    \label{figure:HN13C}
\end{figure*}

\begin{figure*}
    \centering
        \begin{subfigure}[b]{0.49\textwidth}\includegraphics[width=\textwidth,keepaspectratio]{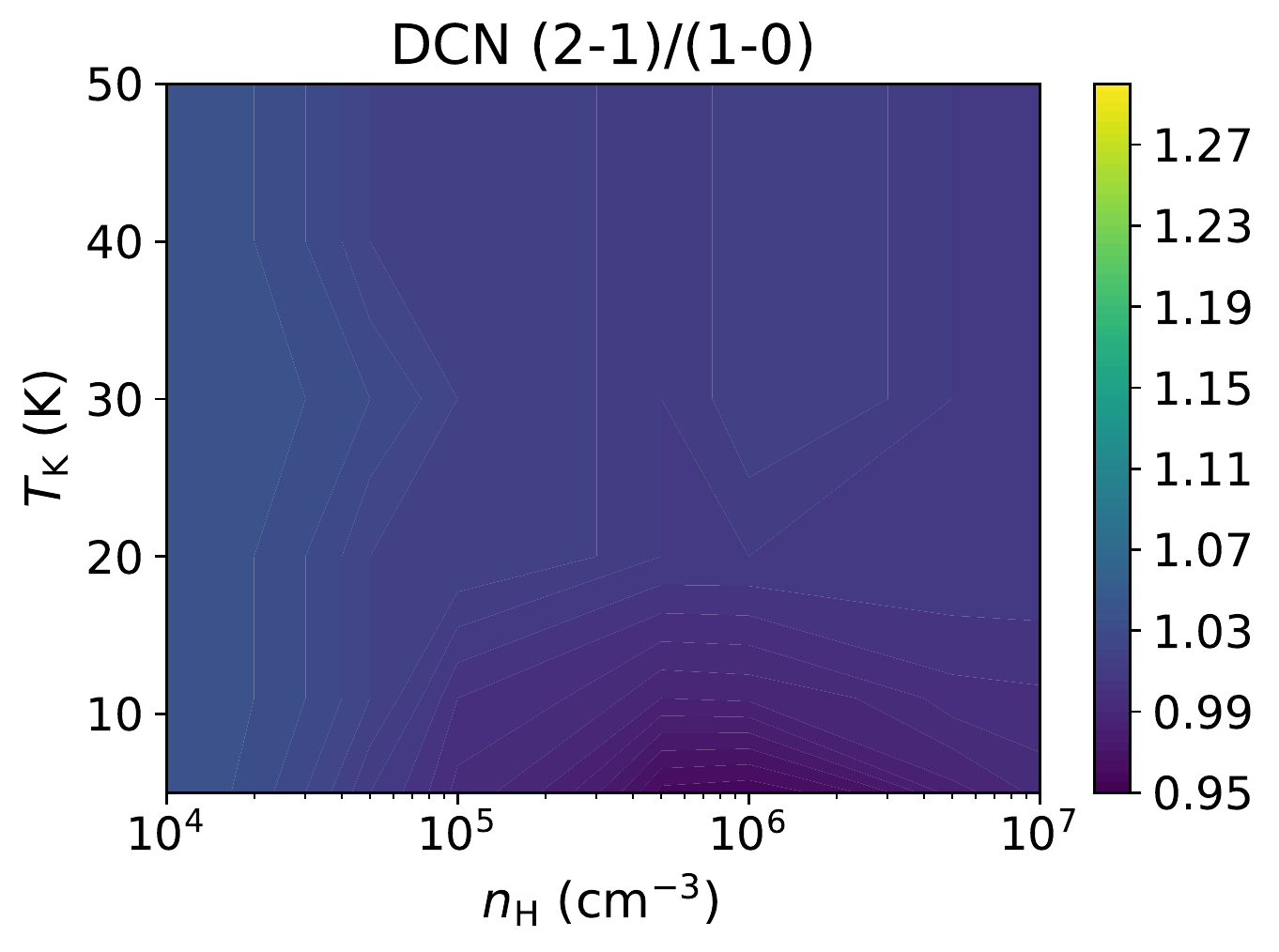}
		\end{subfigure}
        \begin{subfigure}[b]{0.49\textwidth}\includegraphics[width=\textwidth,keepaspectratio]{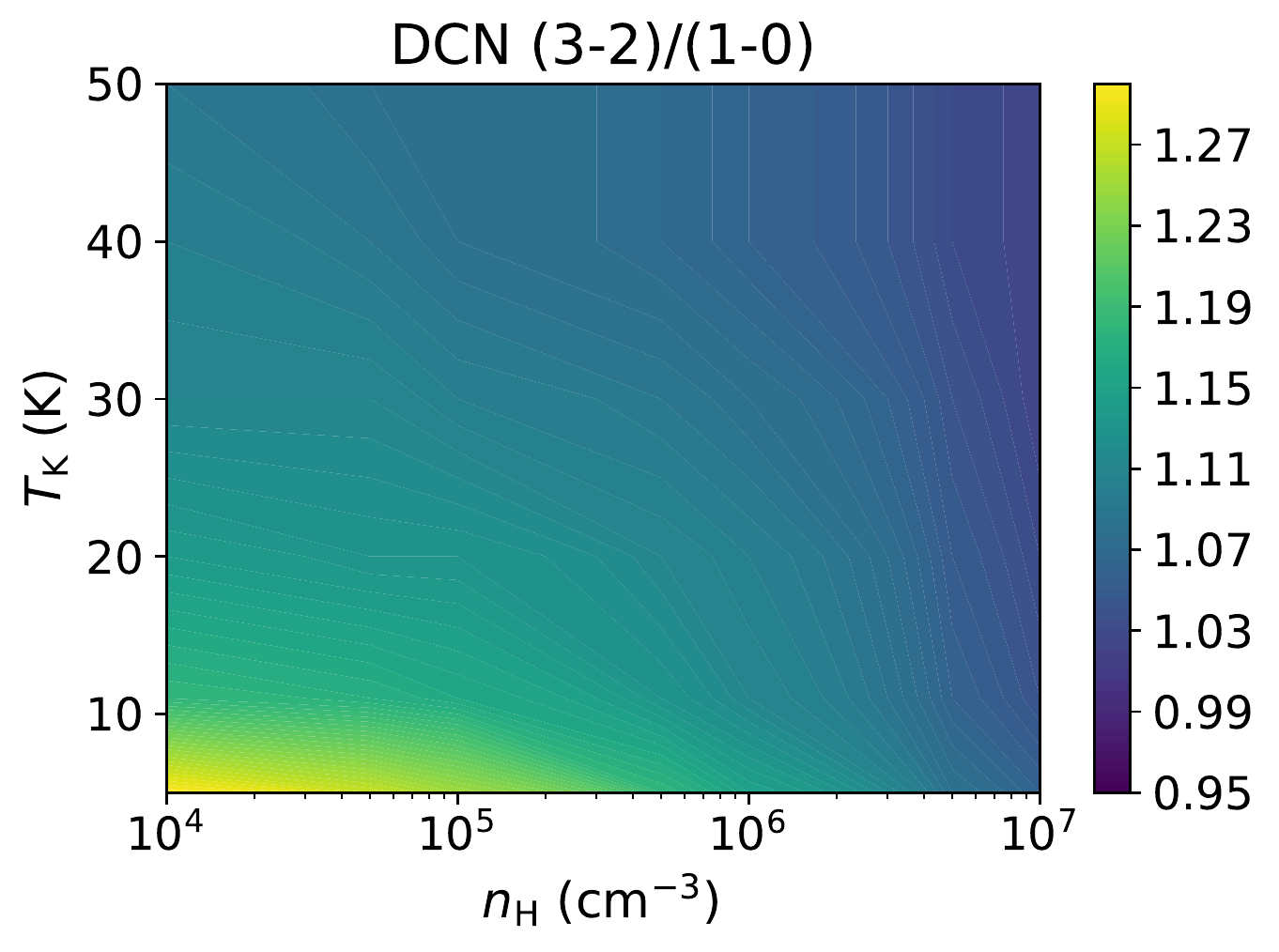}
        \end{subfigure}
	    \caption{Contours display the ratio between quotients of radiation temperatures $T_{r}(2-1)/T_{r}(1-0)$ (left), and $T_{r}(3-2)/T_{r}(1-0)$ (right) calculated with the previous and the new collisional coefficients for DCN.}
	    \label{figure:DCN}
\end{figure*}

\begin{figure*}
    \centering
        \begin{subfigure}[b]{0.49\textwidth}\includegraphics[width=\textwidth,keepaspectratio]{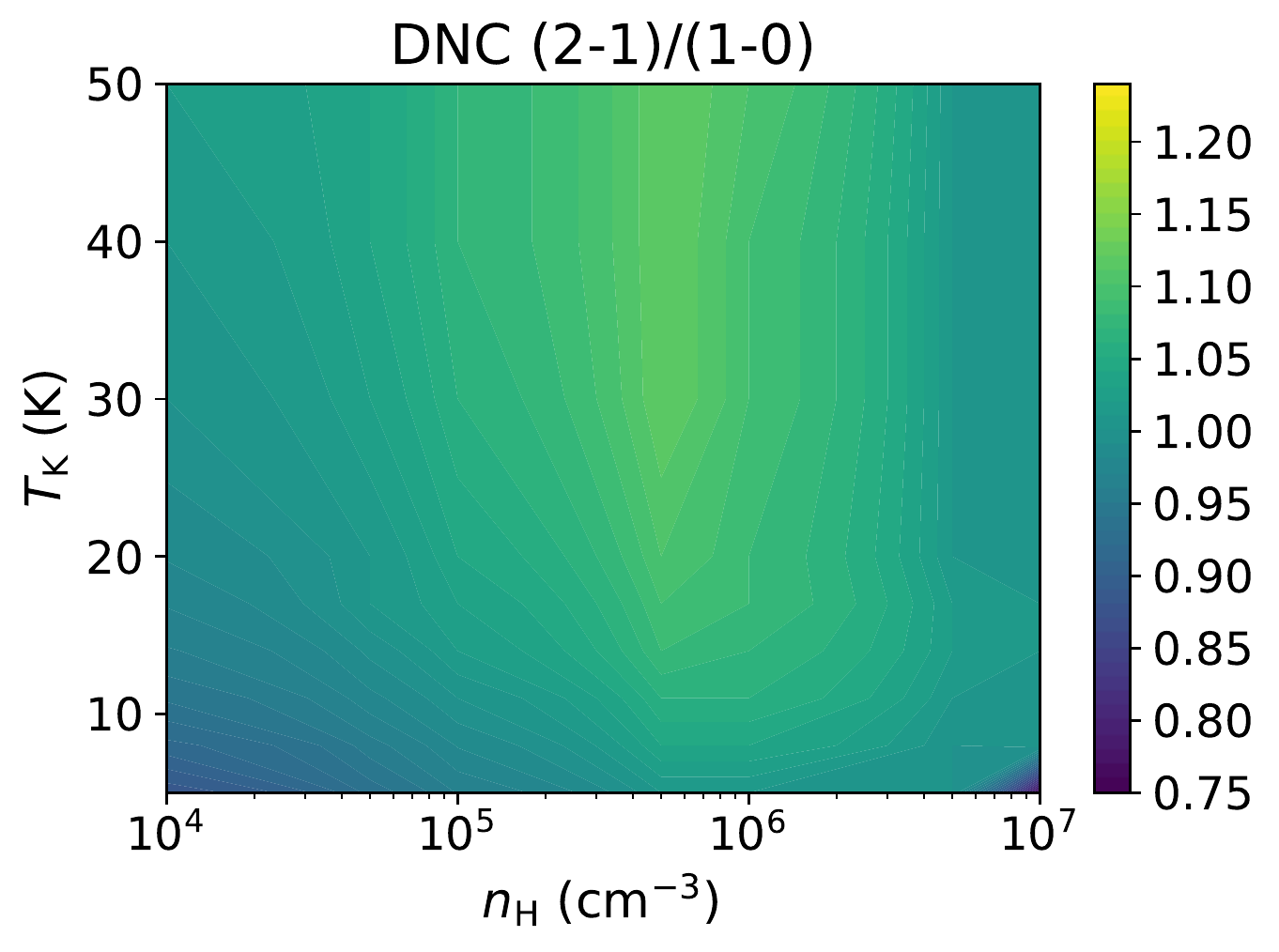}
		\end{subfigure}
        \begin{subfigure}[b]{0.49\textwidth}\includegraphics[width=\textwidth,keepaspectratio]{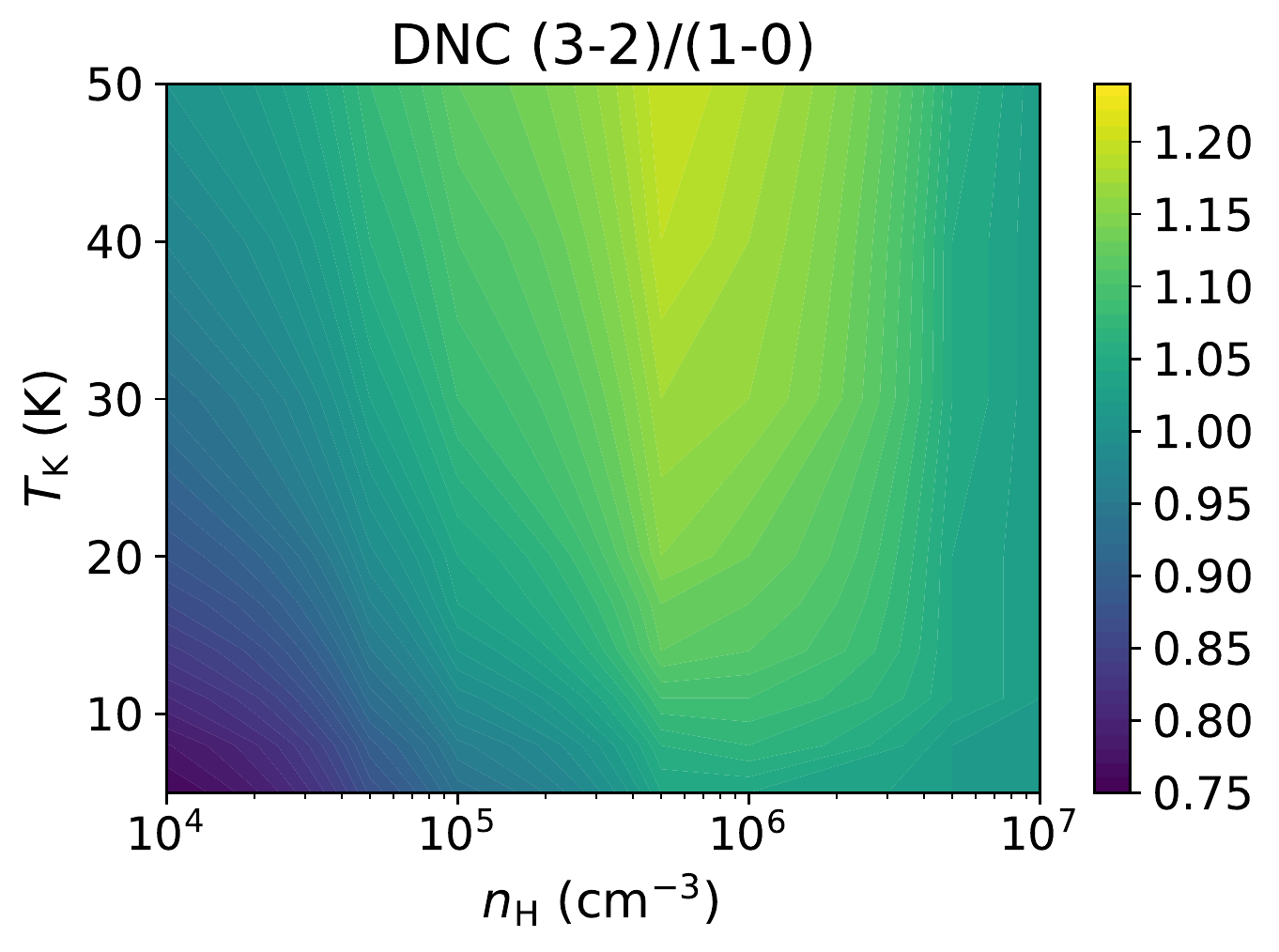}
        \end{subfigure}
	    \caption{Contours display the ratio between quotients of radiation temperatures $T_{r}(2-1)/T_{r}(1-0)$ (left), and $T_{r}(3-2)/T_{r}(1-0)$ (right) calculated with the previous and the new collisional coefficients for DNC.}
	    \label{figure:DNC}
\end{figure*}

\begin{figure*}
    \centering
        \begin{subfigure}[b]{0.49\textwidth}\includegraphics[width=\textwidth,keepaspectratio]{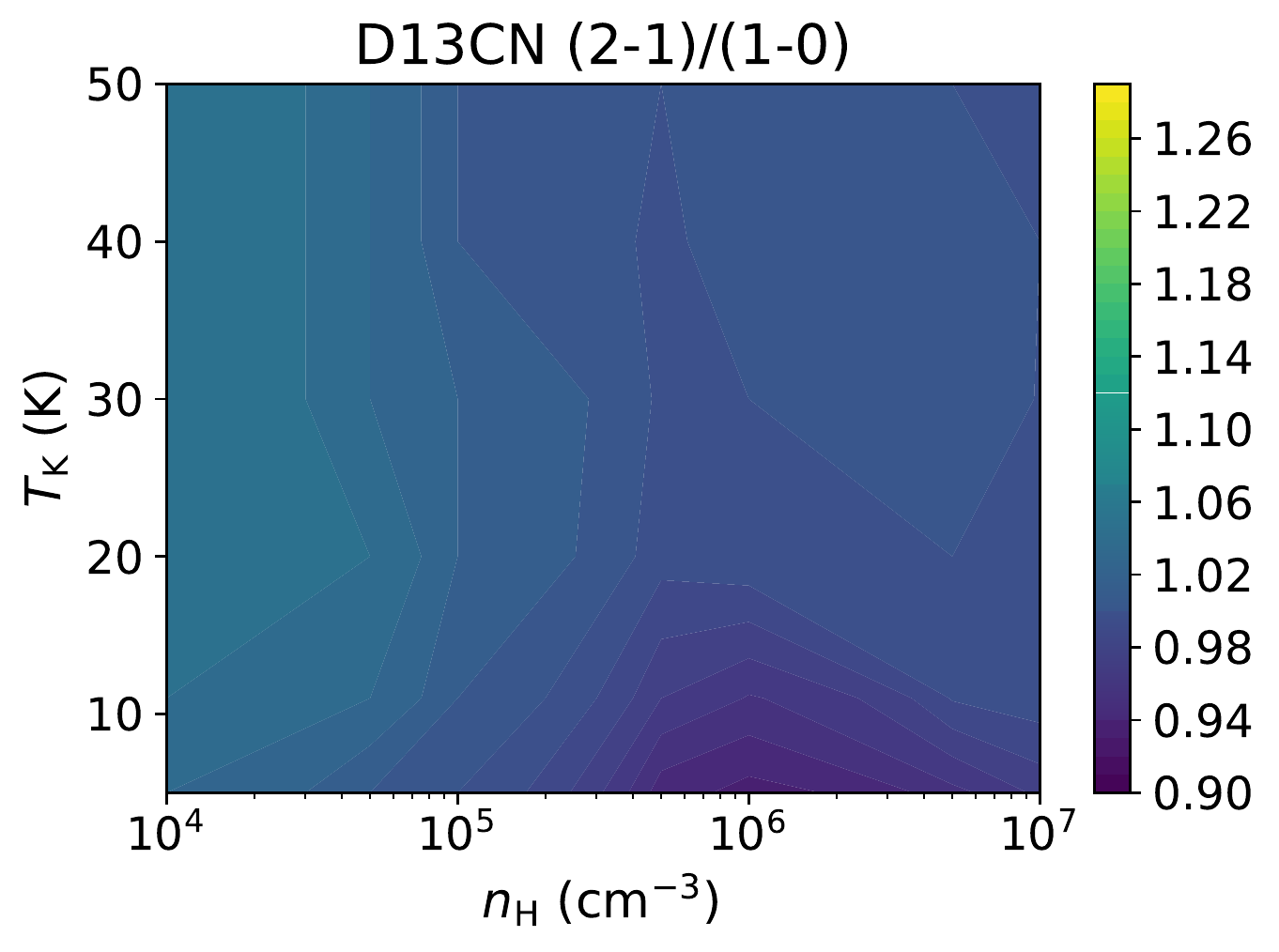}
		\end{subfigure}
        \begin{subfigure}[b]{0.49\textwidth}\includegraphics[width=\textwidth,keepaspectratio]{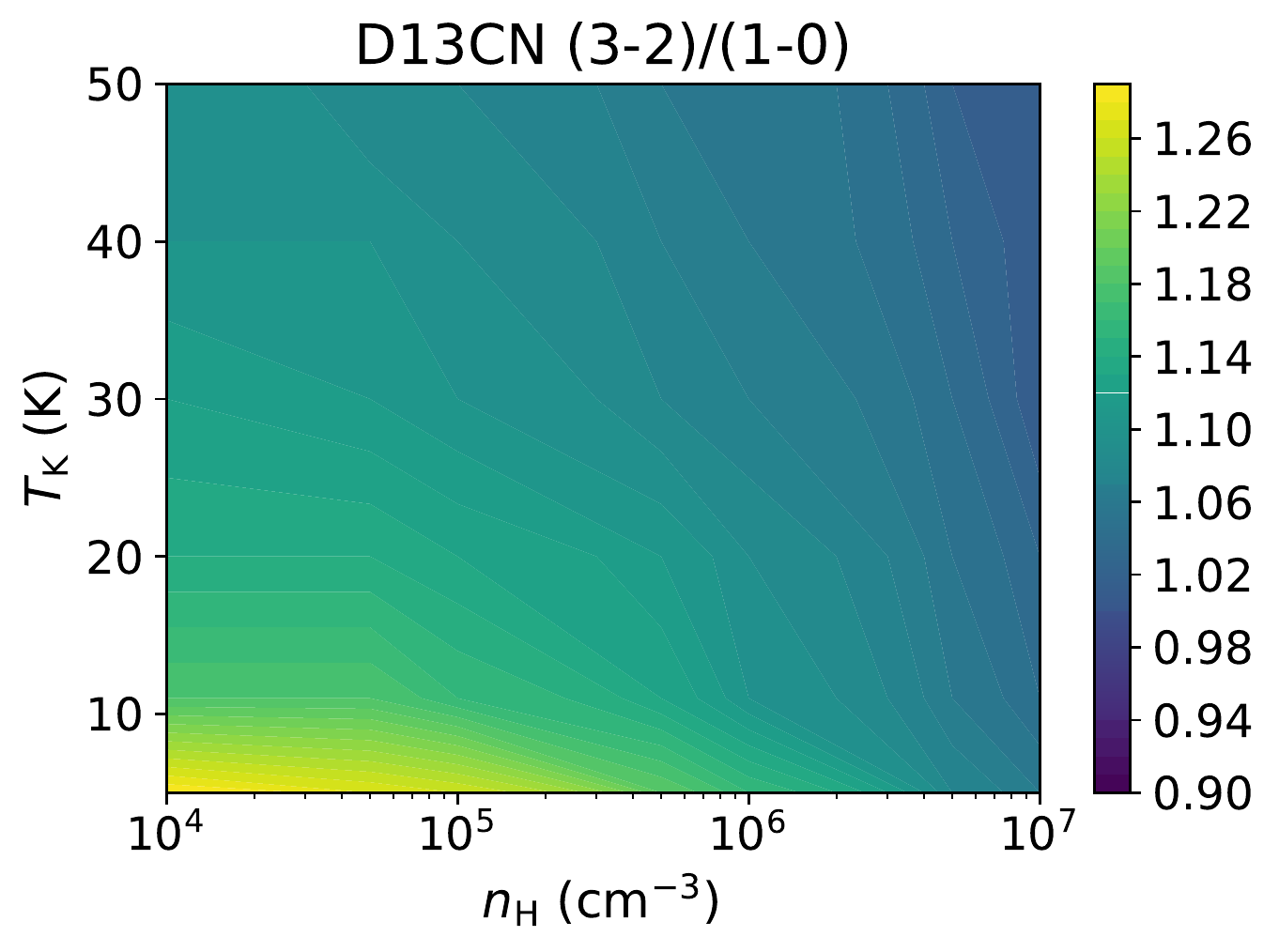}
        \end{subfigure}
	    \caption{Contours display the ratio between quotients of radiation temperatures $T_{r}(2-1)/T_{r}(1-0)$ (left), and $T_{r}(3-2)/T_{r}(1-0)$ (right) calculated with the previous and the new collisional coefficients for D$^{13}$CN.}
	    \label{figure:D13CN}
\end{figure*}

\begin{figure*}
    \centering
        \begin{subfigure}[b]{0.49\textwidth}\includegraphics[width=\textwidth,keepaspectratio]{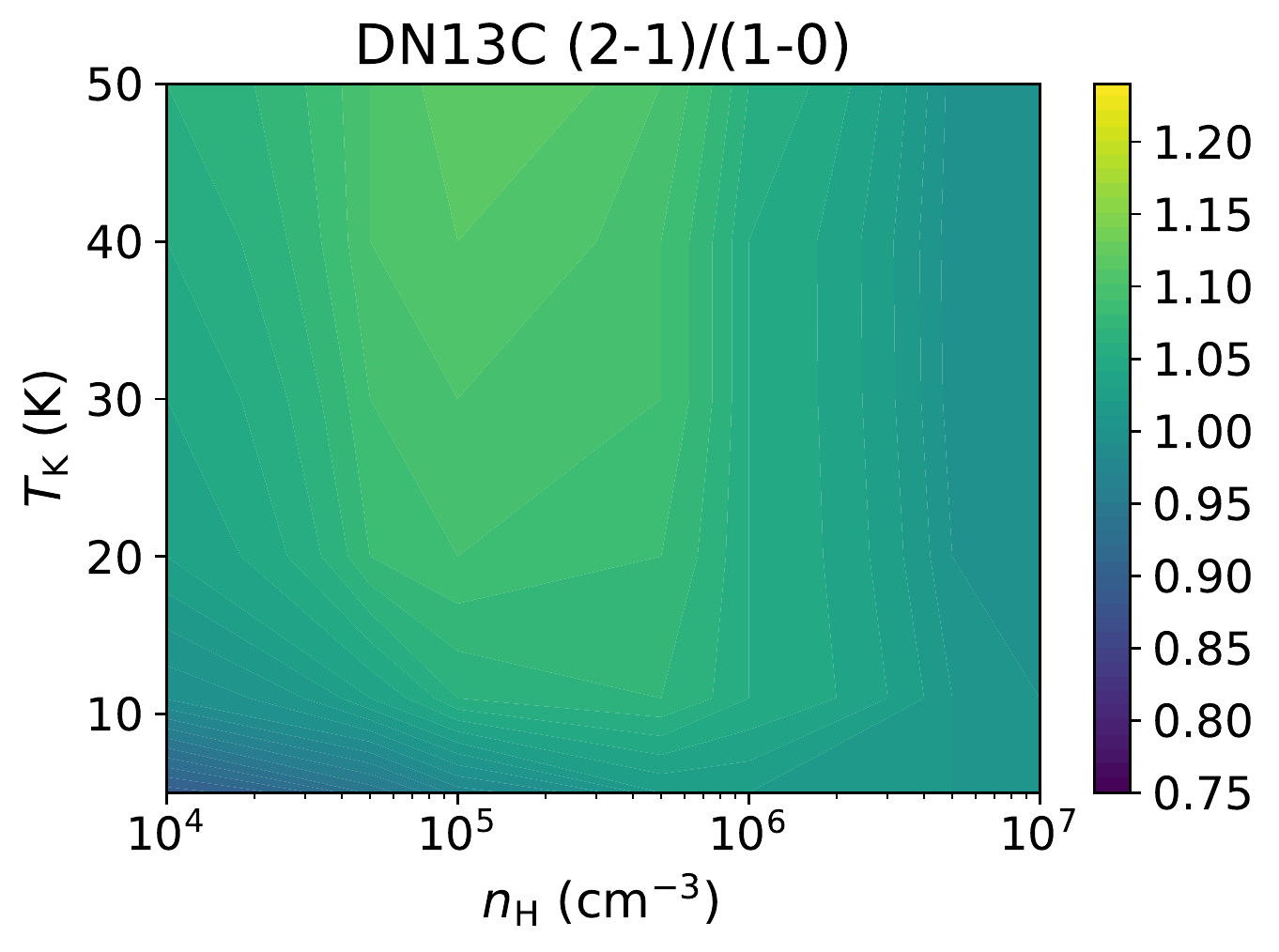}
		\end{subfigure}
        \begin{subfigure}[b]{0.49\textwidth}\includegraphics[width=\textwidth,keepaspectratio]{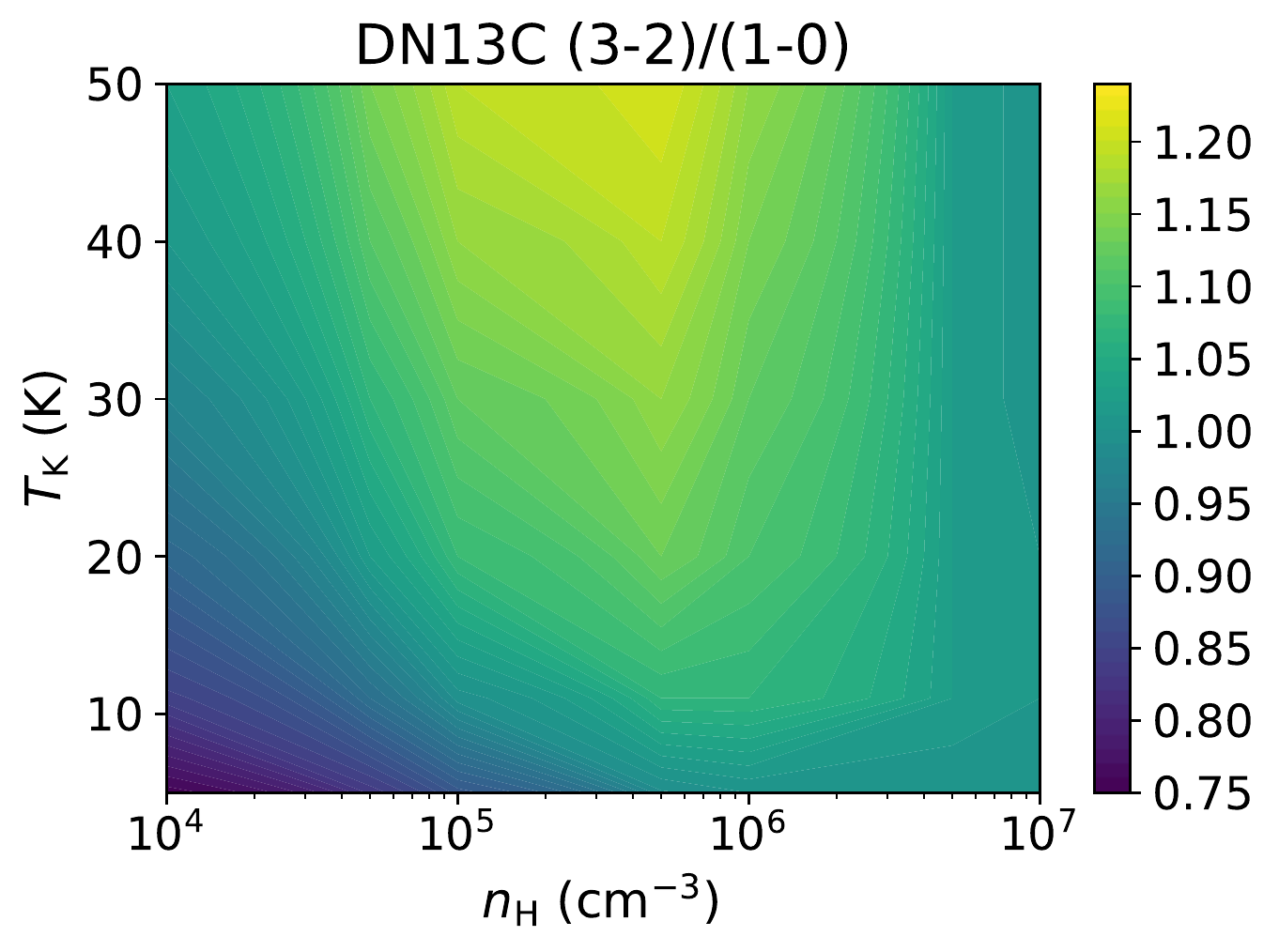}
        \end{subfigure}
	    \caption{Contours display the ratio between quotients of radiation temperatures $T_{r}(2-1)/T_{r}(1-0)$ (left), and $T_{r}(3-2)/T_{r}(1-0)$ (right) calculated with the previous and the new collisional coefficients for DN$^{13}$C.}
	    \label{figure:DN13C}
\end{figure*}


In this section, we carried out RADEX \citep{VanDerTak2007} calculations for typical physical conditions in the ISM to compare results using the previous and our new collisional coefficients for H$^{13}$CN, HN$^{13}$C, DNC, DCN, D$^{13}$CN, and DN$^{13}$C. In particular, we analyzed the ratio between quotients of radiation temperatures, $T_{r}$, for different transitions with the previously available collisional rates and the new ones we present here. These ratios were computed in a grid of models with kinetic temperatures $T_{K}$ in the $5-50$ K range, hydrogen number densities $n_{H}$ in the $10^{4}-10^{7}$ cm$^{-3}$ range, and molecular hydrogen assumed to be in its para-H$_{2}$ form. For kinetic temperatures higher than 30 K, the maximum value of $T_{K}$ that these collisional rates were calculated for, RADEX extrapolates the collisional rates assuming constant rates equal to those corresponding to this maximum value of kinetic temperature \citep{VanDerTak2007}. Figures \ref{figure:H13CN}-\ref{figure:DN13C} show the results for the different chemical species considered and for the ratios between the transitions $(3-2)$ and $(2-1)$ with $(1-0)$, namely, ratios $(2-1)/(1-0)$ and $(3-2)/(1-0)$. Table \ref{tab:radex_physical_parameters_assumed} lists the line-widths and column densities assumed for each case. 

\begin{table}
	\centering
	\caption{Assumed line widths and column densities in the grid.}
	\label{tab:radex_physical_parameters_assumed}
	\resizebox{0.35\textwidth}{!}{
	\begin{tabular}{lll}
		\toprule
		Species     & FWHM (km/s) & $N$(X) (cm$^{-2}$)  \\
		\midrule
		DCN         & 0.5         & $5\times 10^{12}$ \\
		DNC         & 0.5         & $5\times 10^{12}$ \\
		H$^{13}$CN  & 0.5         & $2\times 10^{13}$ \\ 
		HN$^{13}$C  & 0.5         & $2\times 10^{13}$ \\
		D$^{13}$CN  & 0.5         & $1\times 10^{11}$ \\
		DN$^{13}$C  & 0.5         & $1\times 10^{11}$ \\
    		\bottomrule
	\end{tabular}
	}
	\flushleft
\end{table}

The differences in the radiation temperature between the previous and the new collisional coefficients are small (up to $2\%$) for H$^{13}$CN (Figure \ref{figure:H13CN}). The update of collisional coefficients presents a bigger impact in HN$^{13}$C (see Figure \ref{figure:HN13C}), with a difference of up to $\sim 30\%$. This bigger impact is found for the $(2-1)/(1-0)$ ratio at low kinetic temperatures ($T_{K}\lesssim10$ K) and for densities $<5\times10^5$ cm$^{-3}$. This is relevant in a cold core scenario, where these new coefficients would therefore introduce an increase of $\sim 30\%$ in the molecular hydrogen number density.  

For the deuterated versions of hydrogen (iso)cyanide (Figs. \ref{figure:DCN} and \ref{figure:DNC}), the differences between using previous and new collisional coefficients are comparable to that for HN$^{13}$C, with up to $\sim 20\%$ for DNC and $\sim 30\%$ for DCN. It is particularly interesting to observe that in the case of DNC, these differences are more accentuated as $T_{K}$ increases, that is, more for the ratio $(3-2)/(1-0)$ than for $(2-1)/(1-0)$. For the case of starless cores, however, only the DCN $(3-2)/(1-0)$ ratio shows variations. Similar results were also found for the $^{13}$C isotopologues of DCN and DNC (Figs. \ref{figure:D13CN} and \ref{figure:DN13C}), where we also observed that the values of $T_{K}$ for DN$^{13}$C are more affected than those for D$^{13}$CN. Also the impact of updating the collisional coefficients is larger for the $(3-2)/(1-0)$ ratio than for the $(2-1)/(1-0)$ one.

\section{Case study: TMC 1-C and NGC 1333}

\begin{figure*}
	        \centering
		    \begin{subfigure}[b]{0.49\textwidth}\includegraphics[width=\textwidth,keepaspectratio]{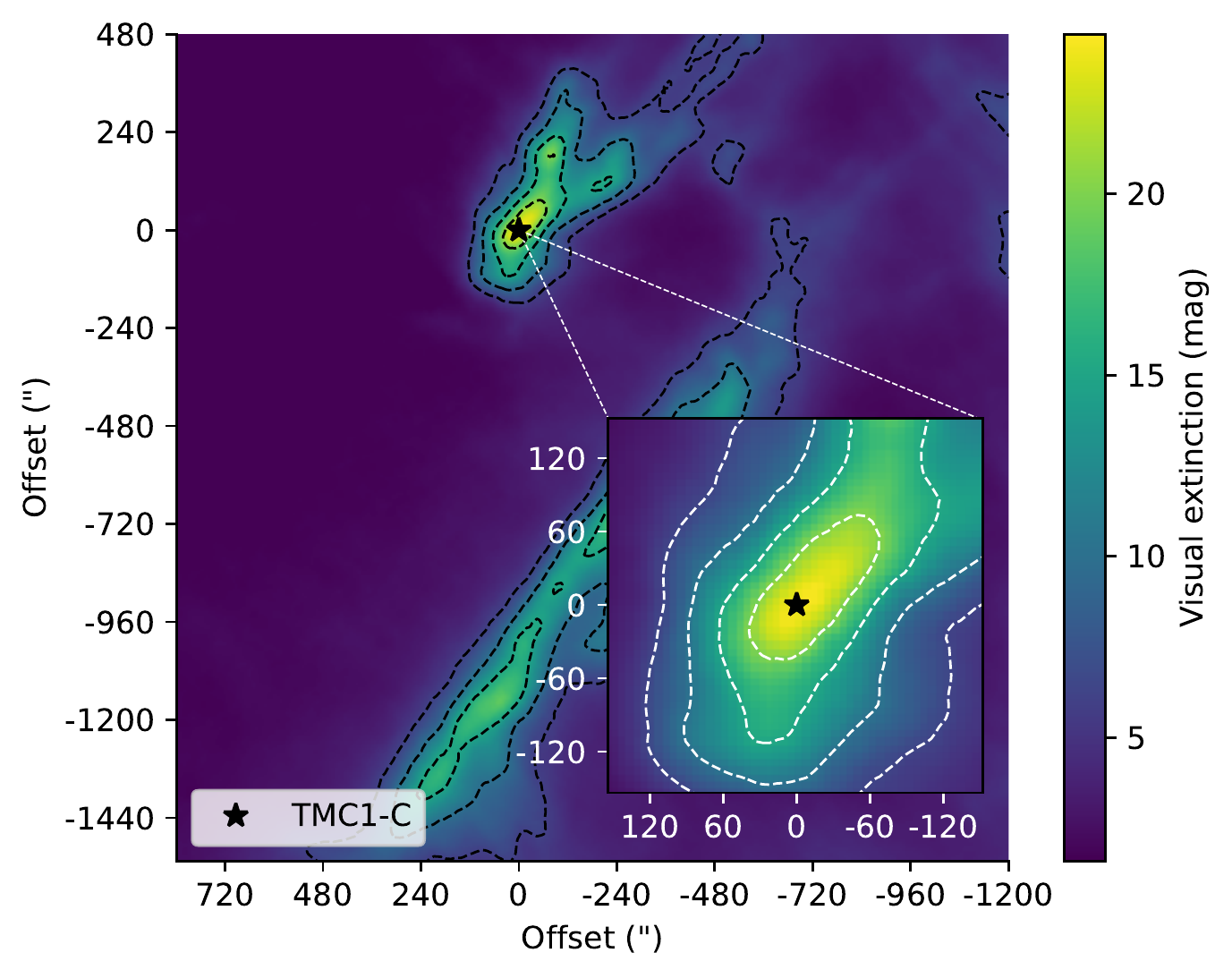}
		    \end{subfigure}
		    \begin{subfigure}[b]{0.49\textwidth}\includegraphics[width=\textwidth,keepaspectratio]{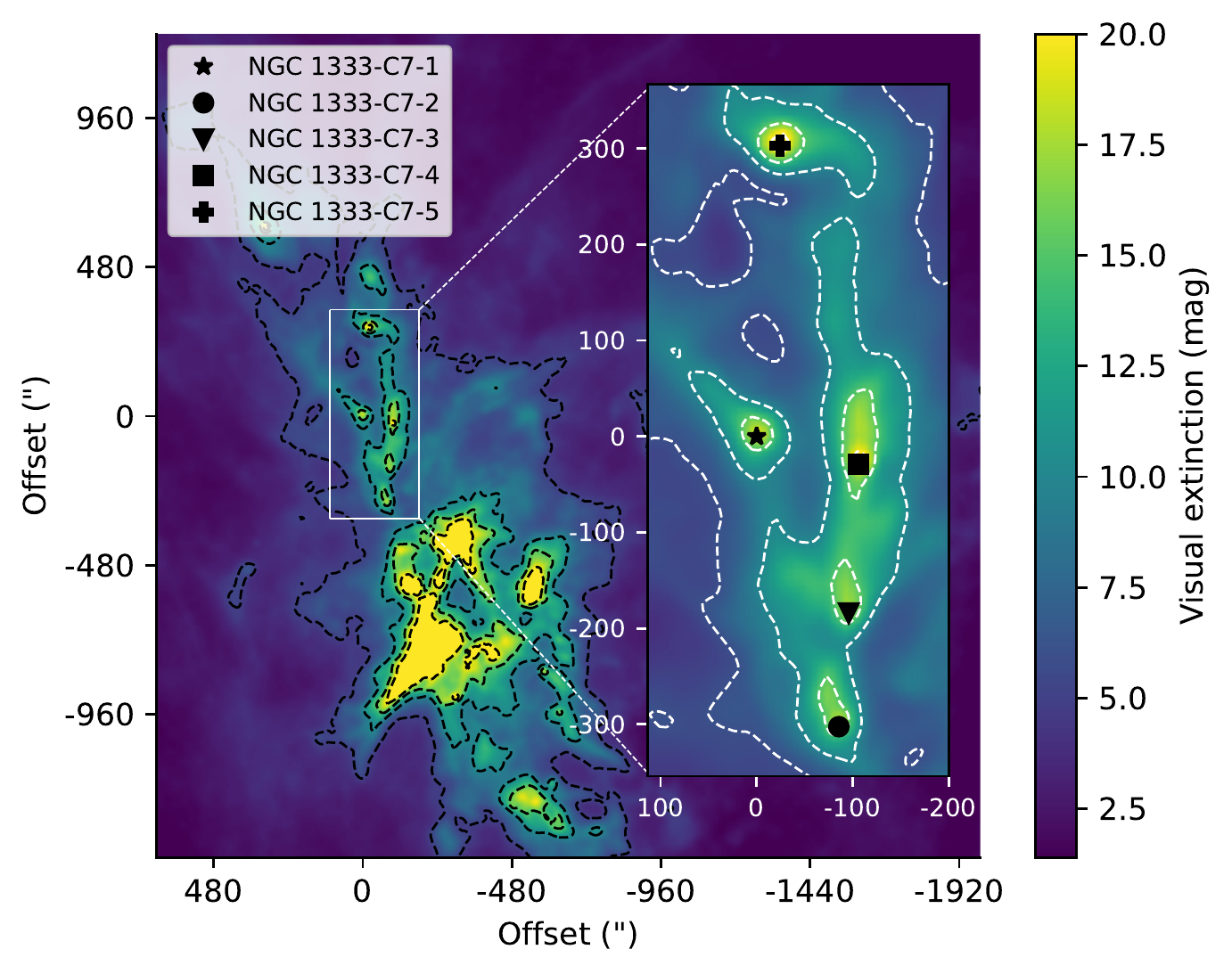}
		    \end{subfigure}
		    \caption{Visual extinction maps of the regions under study. \emph{Left panel:} visual extinction map of TMC 1-C from the {\em Herschel} data (Kirk et al. in prep.). Contours represent levels of $[6,10,15,20,25]$ mag. The origin (0,0) of this map corresponds to the position TMC 1-C (see Table \ref{tab:targets}). \emph{Right panel:} visual extinction map of the northern sector of the NGC 1333 protocluster. The origin (0,0) of this map corresponds to the position NGC1333-C7-1 (see Table \ref{tab:targets}). Contours represent levels of $[5,10,15,20]$ mag.}
		    \label{fig:mapsTMC1NGC1333}
        \end{figure*}

   To assess the impact of these new collisional coefficients in the astrochemical modeling of prestellar and starless cores, we used the emission spectra of the $J=1\rightarrow 0$ rotational transitions of HCN, HNC, and their $^{13}$C, $^{15}$N, and D isotopologues (see Table \ref{tab:summarylines}) to explore the chemical evolution in a sample of dense cores located in TMC 1 and NGC 1333 (see Figure \ref{fig:mapsTMC1NGC1333}). With the updated coefficients, we carried out an up-to-date chemical characterization of the regions under consideration, deriving the chemical abundances of the molecules and obtaining several quantities of interest such as the C and N isotopic ratios, the deuteration fraction, and the isomeric ratios HCN/HNC. The determination of the HCN/HNC ratio is linked to the gas temperature \citep{Hacar2020} and therefore linked to the deuterium fraction. In the following chemical analysis, we look for correlations of these quantities to elucidate the physical conditions and the dynamical state of the region.
    
    We selected two nearby star-forming regions with different star formation activity. TMC 1 is an archetype of isolated, low-mass star forming region, while NGC 1333, located in the Perseus Complex, is the closest intermediate-mass star-forming region, whose activity is often organized in clusters. Given their different star-forming regimes (isolated vs. clustered), we explored the possible effects that stellar feedback may have in the chemistry of a star-forming region.
    

    \begin{table}
	        \centering
	        \caption{Observed transitions, frequencies, beam sizes, and telescope efficiencies}
	        \label{tab:summarylines}
	        \resizebox{0.49\textwidth}{!}{
	            \begin{tabular}{rrcc}
		            \toprule
	                {Line} &  {Freq. (MHz)$^{(\rm a)}$} & ${{\theta}}_{\rm HPBW}$(")$^{(\rm b)}$ & ${\eta}_{\rm mb}^{(\rm b)}$ \\ \midrule
		            D$^{13}$CN $1\rightarrow0$ & 71175.07 & 24 & 0.36\\ 
				    D$^{15}$N$^{13}$C $1\rightarrow0$ & 72310.79 & 24 & 0.36\\
				    DCN $1\rightarrow0$ & 72414.69 & 24 & 0.36\\
				    DN$^{13}$C $1\rightarrow0$ & 73367.75 & 24 & 0.35\\
				    D$^{15}$NC $1\rightarrow0$ & 75286.77 & 24 & 0.33\\
				    DNC $1\rightarrow0$ & 76305.70 & 23 & 0.32\\
				    H$^{13}$C$^{15}$N $1\rightarrow0$ & 83727.58 & 21 & 0.27\\
				    H$^{15}$N$^{13}$C $1\rightarrow0$ & 85258.92 & 21 & 0.26\\
				    HC$^{15}$N $1\rightarrow0$ & 86054.97 & 21 & 0.25\\
				    H$^{13}$CN $1\rightarrow0$ & 86339.92 & 21 & 0.25\\
				    HN$^{13}$C $1\rightarrow0$ & 87090.83 & 21 & 0.24\\
				    HCN $1\rightarrow0$ & 88631.60 & 20 & 0.24\\
				    H$^{15}$NC $1\rightarrow0$ & 88865.69 & 20 & 0.24\\
				    HNC $1\rightarrow0^{(\rm c)}$ & 90663.57 & 27 & 0.85\\
    		        \bottomrule
	            \end{tabular}
	        }
            \flushleft
			    {\small
			\ \ \ \ $^{{(\rm a)}}$ The frequencies shown here do not consider hyperfine splitting. The frequencies are taken from the CDMS catalog \citep{Muller2001, Muller2005, Endres2016} except those of the DN$^{13}$C and D$^{15}$NC lines, taken from the SLAIM catalog, available in \href{https://splatalogue.online}{Splatalogue} \citep{Remijan2007}.\\
			\ \ \ \ $^{{(\rm b)}}$ Yebes 40m and IRAM 30m telescope efficiencies taken from the \href{https://rt40m.oan.es/rt40m_en.php}{Yebes 40m technical information webpage} and \href{https://publicwiki.iram.es/Iram30mEfficiencies}{IRAM 30m efficiencies wiki}, respectively.\\
			\ \ \ \ $^{{(\rm c)}}$ Data from the IRAM 30m Large Program GEMS (PI. Asunci\'on Fuente). 
			}
        \end{table}
        
        
        \begin{table}
	        \centering
	        \caption{Target positions, spectral classes, and radii.}\label{tab:targets}
            \resizebox{0.48\textwidth}{!}{
	        \begin{tabular}{lccccc}
		        \toprule
		        Source &  RA (2000) &  Dec (2000) & Class & Refs. & Radius (")\\ \midrule
				TMC~1-C & 04:41:37.58 & 26:00:31.10 & SCore$^{*}$ & 1,2,3 & 147.3$^{(a)}$  \\
				NGC 1333-C7-1 & 03:29:25.57 & 31:28:14.83 & Class 0 & 4,5 & 15.6$^{(b)}$ \\ 
				NGC 1333-C7-2 & 03:29:19.05 & 31:23:14.45 & Class I & 6,7,8 & 30.6$^{(b)}$ \\ 
				NGC 1333-C7-3 & 03:29:18.29 & 31:25:08.34 & SCore$^{*}$ & 6 & 34.3$^{(b)}$ \\
				NGC 1333-C7-4 & 03:29:17.38 & 31:27:46.09 & Class 0 & 6 & 25.8$^{(b)}$ \\
				NGC 1333-C7-5 & 03:29:23.95 & 31:33:18.15 & Class 0 & 6 & 18.8$^{(b)}$ \\
    		    \bottomrule
	        \end{tabular}
         }
         \flushleft
	        $^{*}$SCore stands for starless core. NGC1333-C7-3 class is uncertain as it was classified as starless core by default since it did not fit the criteria of any other class. References: ${(a)}$ \citet{Goodman1998}; ${(b)}$ \citet{Kirk2007}; 1) \citet{Schnee2007}; 2) \citet{Schnee2010}; 3)  \citet{Navarro2021}; 4)  \citet{Tobin2016}; 5)  \citet{Maureira2020};
	        6) \citet{Hatchell2007a}; 7) \citet{Sadavoy2014}; 8) \citet{Rebull2015}. 
        \end{table}
      
        \subsection{TMC 1-C}
            
            The Taurus Molecular Cloud (TMC), at a distance of $141.8\pm 0.2$ pc \citep{Yan2019}, is one of the closest molecular cloud complexes and is considered an archetypal of low-mass, isolated, star-forming region. The most massive molecular cloud in Taurus is the Heiles cloud 2 (HCL 2), which hosts the well-known region TMC 1. TMC 1 is a cold and dense cloud at the center of TMC that has been the target of several cloud evolution, chemical, and star formation studies (see e.g., \citet{Ungerechts1987, Mizuno1995, Goldsmith2008}), having been extensively mapped in CO \citep{Cernicharo1987, Onishi1996, Narayanan2008} and its visual extinction \citep{Cambresy1999, Padoan2002, Schmalzl2010}. 
            
            This cloud is in an environment less exposed to external radiation fields compared to, for example, photodissociation regions (PDRs) \citep{Xu2016}, leading to a high richness in molecular complexity \citep{Agundez2013, Gratier2016, Feher2016} with a plethora of hydrocarbon chains and cycles being currently detected \citep[see, e.g., QUIJOTE Large Program,][]{Cernicharo2021b, Cernicharo2021a, Cabezas2022, Cernicharo2022, Fuentetaja2022}. It features a differentiated chemistry between its northern and southern areas that may be linked to different evolutionary stages \citep{Sipila2018, Navarro2021}. Indeed, the TMC 1-CP position in the southern part of this filament displays a high abundance of cyanopolyynes, in contrast to the TMC 1-NH3 position showing a strong emission peak in ammonia.
            
            The position TMC 1-C, located at the northern part of TMC 1 (see Figure~\ref{fig:mapsTMC1NGC1333}), is a $\sim 1$ M$_\odot$ prestellar core embedded in a relative dense and cold environment n(H$_{2})\simeq 10^4$ cm$^-3$, T$_{\rm kin}\sim 10$ K \citep{Schnee2007, Fuente2019, Navarro2020} that shows a more evolved chemistry than the southern areas \citep{Navarro2021}, a high CO depletion, and mass accretion \citep{Schnee2007, Schnee2010}. This core, given the wealth of data available in the literature, is therefore ideal for performing evolutionary studies of starless cores and testing the new collisional coefficients of different species we present in this work.

        \subsection{NGC 1333}
            NGC 1333 manifests itself in the visible portion of the electromagnetic spectrum as a bright reflection nebula. It is currently the most active star-forming region in the Perseus molecular cloud, at a distance of $293\pm22$ pc from the Sun \citep{OrtizLeon2018}. As in the case of TMC 1, the NGC 1333 cluster has become one of the best characterized low-to-intermediate mass star-forming regions in the solar neighbourhood due to its proximity \citep{Walawender2008} and it is surrounded by other well-known star-forming regions, namely, B1, L1455, and L1448. It has been extensively surveyed along the electromagnetic spectrum, covering different regimes: Spitzer surveys \citep{Jorgensen2005}, mid/far-infrared \citep{Harvey1984, Jennings1987, Sadavoy2014}, near/mid-infrared \citep{Aspin1994, Aspin2003, Lada1996, Greissl2007}, and millimeter/submillimeter \citep{Lefloch1998, Hatchell2005, Hatchell2007a, Hatchell2007b, Kirk2006, Tobin2016}. It harbors plenty of dense cores and young stellar objects \citep{Hatchell2005, Hatchell2007a, Jorgensen2008}. Physical properties of this protocluster like the total column density N(H$_{2}$) and dust temperature, $T_{\rm d}$, are available in the maps presented by \citet{Zari2016}. Unlike TMC 1, a significant fraction of these protostars are found in protoclusters, unveiling a different star formation regime. The ubiquitous presence of these objects in the central regions (see Figure~\ref {fig:mapsTMC1NGC1333}) leads to higher gas and dust temperatures as well as a shock-driven chemistry \citep{Jorgensen2004, Wakelam2005}. The different environments they develop in, compared to TMC 1, offers the opportunity to study how these differences impact their chemical evolution.
            
            We selected five dense cores located in the northern and more quiescent part of NGC~1333 (see Figure~\ref {fig:mapsTMC1NGC1333} and Table~\ref{tab:targets}) to investigate the potential effects that stellar feedback from the other members of the sample and the main NGC 1333 cluster may have on the chemistry of young stellar objects. The core NGC1333-C7-1 was first identified in James Clerk Maxwell Telescope (JCMT), Submillimeter Common User Bolometry Array (SCUBA), and Caltech Submillimeter Observatory (CSO) surveys \citep{Hatchell2005, Enoch2006}. While it was first thought to be a starless core (SCore) by \citet{Hatchell2007b} and \citet{Enoch2008}, it is now known to host a Class 0 protostar \citep[see Per-Bolo-58][]{Tobin2016, Maureira2020}. NGC1333-C7-2 has been classified as a Class I object \citep[Source \#63][]{Hatchell2007a, Rebull2015} and its chemistry has been investigated in, for instance, \citet{Imai2018} (position NGC 1333-12). In particular, this survey shows a low HNC deuterium fraction in this source as a consequence of its high bolometric temperature. NGC1333-C7-3 class is uncertain. It is considered to be a starless core since it did not meet any criteria in the classification done by \citet{Hatchell2007a} (source \#57). NGC1333-C7-4 and NGC1333-C7-5 \citep[Sources \#61 and \#58, respectively, ][]{Hatchell2007a} are classified as Class 0 objects. Together with TMC~1-C,  our sample is thus composed of: two starless cores, three Class 0 YSOs, and one Class I protostar.

    \section{Observations}
        \subsection{Yebes 40m}
        
        A high-sensitivity spectral survey between 71.150 GHz -  89.590 GHz was carried out towards TMC 1-C and NGC 1333, using the Yebes 40m radiotelescope \citep{Tercero2021}. This telescope is equipped with a 72-90.5 GHz (W band) receiver that allows for observations in a 18 GHz bandwidth. The backends consist of two sets of eight Fast Fourier Transform Spectrometers (FFTS) covering the whole bandwidth, each providing a spectrum of up to 65536 channels with a channel spacing of 38 kHz. The observing procedure for TMC 1-C was position-switching, with the off-position being RA(J2000) = 04$^{\rm h}$42$^{\rm m}$32$^{\rm s}$.16 and Dec(J2000):25$^{\circ}$59$'$42.0$''$. In a different observation campaign, NGC 1333 was observed using frequency-switching with a throw of 6 MHz. The intensity scale is the main-beam temperature scale $T_{\rm mb}$, such that $T_{\rm A} = \eta_{\rm mb}\ T_{\rm mb}$, where the main-beam efficiencies $\eta_{\rm mb}$ are shown in Table \ref{tab:summarylines}, along with the half power beam width (HPBW) in arcseconds.
    
        \subsection{IRAM 30m and GEMS data}
        
        The HNC $1\rightarrow0$ line is not included in the Yebes 40m telescope W-band setup. We complemented our observations with the HNC $1\rightarrow0$ data from the 30m IRAM GEMS Large Program in TMC 1 and NGC 1333-C7-1 \citep{Fuente2019, Navarro2021, RodriguezBaras2021}. These observations were carried out using the IRAM 30m telescope using the Eight Mixer Receivers (EMIR) and the FFTS with a spectral resolution of 49 kHz. The observing mode was frequency-switching, with a frequency throw of 6 MHz. The intensity scale is the main-beam temperature scale, with the main-beam efficiency at the corresponding frequency shown in Table \ref{tab:summarylines}.
        
        \subsection{MUSTANG-2}
        Our continuum data at 90 GHz (3 mm) comes from the MUSTANG-2 instrument on the 100-meter Robert C. Byrd Green Bank Telescope (GBT). This instrument features a 215 element array of feedhorn-coupled transition edge sensor (TES) bolometers, with a bandwidth from 75 to 105 GHz. The data was obtained in two observation campaigns over November and December, 2019 (November 2,7; December 6,7,22). NGC 1333 was imaged during the 2020 campaign (November 7,15). MUSTANG-2 observations towards TMC~1-C were done with on-the-fly mapping using a Lissajous daisy scan pattern \citep[see, e.g., ][]{Dicker2014, Romero2020}, which provides an instantaneous field of view (FOV) of 4.25 arcminutes and a mapping speed of $\sim 98\ \mu\rm K h^{1/2}$. In the case of NGC 1333, we used a slightly new observing strategy of offsetting scans at different pointing centers which increases the size of the map to $\sim 5'$ in diameter with basically the same noise. The theoretical beam size is $8.5"$. However, the illumination pattern plus surface inaccuracies result in slightly larger beam sizes of $\sim 9"-10"$, which we characterized with frequent measurements of bright secondary calibrator sources in each observing run. At the start of each run, observations of a bright compact source are used to solve for primary aperture wavefront phase errors using the ``out of focus'' (OOF) holography technique \citep{Nikolic2007}. The solutions were applied to the active GBT surface to achieve the best sensitivity. The data were processed with the MIDAS pipeline \citep{Romero2020}.

    \section{Molecular column densities}
    \label{sec:mol}
        
        To characterize the chemical properties of the selected sources (Table \ref{tab:targets}), we analyzed the line emission of the different transitions listed in Table \ref{tab:summarylines}. From the line emission, we derived the column densities of the species using the new collisional coefficients and the {\scshape Radex} \citep{VanDerTak2007} radiative transfer code. In all positions belonging to NGC 1333, the gas temperature is assumed to be $\sim 15$ K, similar to the dust temperature derived from the {\em Herschel} space telescope maps in \citet{Zari2016}. In TMC 1-C, we adopted $T_{\rm kin}=8$ K following the calculations of \citet{Navarro2020, Navarro2021}. Unfortunately, we have only observed one rotational line of each isotopologue, which has not allowed us to use molecular excitation calculations to estimate molecular hydrogen densities and molecular column densities in a self-consistent way. In order to estimate the possible impact that the uncertainties in the gas density could introduce in our study, we repeated our calculations with the assumption of a couple of hydrogen nuclei number density values, namely, $n_{\rm H} = 4\times 10^{5}$ cm$^{-3}$ and $n_{\rm H} = 10^{6}$ cm$^{-3}$, that are representative of the densest gas in prestellar and protostellar cores. In the case of the NGC1333-C7-1 we assumed the density obtained by \citet{RodriguezBaras2021}. The differences between the values obtained with these two densities are minor when the lines are optically thin. By default, we only address the values obtained with $n_{\rm H} = 4\times 10^{5}$ cm$^{-3}$, unless we find significant differences.
        
        \subsection{Resolved hyperfine structures: DCN, \texorpdfstring{H$^{13}$CN}, and HCN}\label{sec:colDensHfs}

        
        The observed line profiles are shown in Figure \ref{fig:spectraTMC1C} and Figures \ref{fig:spectraNGC1333_C7_1}-\ref{fig:spectraNGC1333_C7_5}, and their properties are summarized in Table \ref{tab:lineResultsTMC1C} and Tables \ref{tab:lineResultsNGC1333_C7_1}-\ref{tab:lineResultsNGC1333_C7_5}. These properties were obtained fitting Gaussian profiles to the lines using the \emph{minimize} routine of the CLASS-GILDAS software. DCN $1\rightarrow0$, H$^{13}$CN $1\rightarrow0$, and HCN $1\rightarrow0$ hyperfine structures exhibit similarities in terms of their number of components, velocity structure, and the line-width of the weakest hyperfine component. In TMC 1-C, the HCN $1\rightarrow0$ line presents self-absorption features that are absent in NGC 1333. In NGC 1333, it is worth noting the presence of wings in, for instance, the HCN $1\rightarrow0$ line in NGC1333-C7-2 or NGC1333-C7-3, which might be linked to protostellar activity. In fact, the line profile of HCN $1\rightarrow0$ in NGC1333-C7-3 suggests that this source might be in a later evolutionary stage than a starless core. The new collisional coefficients presented here do not account for the hyperfine splitting of the molecules. However, to derive the column densities of the species that show hyperfine splitting, we took advantage of the relative intensity between the hyperfine components and the scaling of the optical depth with the intensity to estimate the line opacities.
        
        More precisely, the column density was obtained with {\scshape Radex} in such a way that the resulting integrated intensity from the radiative transfer reproduces the observations of the weakest hyperfine component. Then, the total column density was obtained by scaling the obtained column density value using the relative intensities of the hyperfine components taken from the CDMS catalog. The weakest hyperfine component of the different species across our sample share similar line-widths (see Table \ref{tab:lineResultsTMC1C} and Tables \ref{tab:lineResultsNGC1333_C7_1}-\ref{tab:lineResultsNGC1333_C7_5}), ensuring consistency across our calculations of column density ratios. This method is a good approximation as long as the opacity of the weakest hyperfine component is moderate. In order to check this assumption, taking a single value for the line width and the excitation temperature for all the hyperfine components, we estimated the opacity of the lines using the HFS method of GILDAS-CLASS. The opacities of the hyperfine components are listed in Table \ref{tab:lineResultsTMC1C} and Tables \ref{tab:lineResultsNGC1333_C7_1}-\ref{tab:lineResultsNGC1333_C7_5}; also, except the HCN $1\rightarrow 0$ line, the opacity of the weakest hyperfine component is $\tau < 1$. Even in the case of HCN, the opacity of the weakest hyperfine line is $<$1 in all our targets except TMC 1-C and NGC 1333-C7-1. In these two sources, our HCN $1\rightarrow 0$ column density might be underestimated and we focus on the more reliable estimate of the abundance ratios involving H$^{13}$CN in our discussion.
        
        \subsection{Unresolved hyperfine structures}
        \label{sec:colDensNoHfs}
        
        The observed line profiles with unresolved hyperfine structure are also shown in the corresponding panels of Figure \ref{fig:spectraTMC1C} and Figures \ref{fig:spectraNGC1333_C7_1}-\ref{fig:spectraNGC1333_C7_5}, while their properties are summarized in Table \ref{tab:lineResultsTMC1C} and Tables \ref{tab:lineResultsNGC1333_C7_1}-\ref{tab:lineResultsNGC1333_C7_5}. Unresolved hyperfine structures across our sample display varying line widths according to the amplitude of their unresolved hyperfine splitting. On one hand, HC$^{15}$N and H$^{15}$NC show line widths compatible with that of the weakest hyperfine component of the transitions described in Section \ref{sec:colDensHfs}. On the other hand, HN$^{13}$C $1\rightarrow0$, DNC $1\rightarrow0$ or DN$^{13}$C $1\rightarrow0$ show larger line widths (see, e.g., Figure \ref{fig:spectraTMC1C}) due to the larger, but unresolved, hyperfine splitting. Although these unresolved hyperfine structures appear as only one component centered around the same velocity, it is worth noting the presence of self-absorption features in the HNC $1\rightarrow0$ lines observed toward TMC 1-C that give their two-component shape in Figure \ref{fig:spectraTMC1C}. This feature has prevented us from using them to compute reliable column densities and ratios. In the case of HNC emission measured towards NGC1333-C7-1, the high opacity of this line account for its wider line width. If the hyperfine structure is unresolved, we can only compare the measured integrated intensity of the lines with that resulting from the radiative transfer code. We expect the emission of the most abundant molecules to be optically thick, and therefore we used the less abundant isotopologues in our calculations when possible.

        \subsection{Molecular column densities in TMC 1-C}
        
        Following the procedure described in Sections \ref{sec:colDensHfs} and \ref{sec:colDensNoHfs}, we obtained the column densities in TMC 1-C of the species shown in Table \ref{tab:columnDensitiesTMC1} assuming two different values of the hydrogen nuclei number density: $4\times 10^{5}$ and $10^{6}$ cm$^{-3}$. The first density value for TMC 1-C, $n_{\rm H} = 4\times 10^{5}$ cm$^{-3}$, was previously obtained by \citet{Navarro2021} in a multi-transition analysis with deuterated molecules. This offers the opportunity to find how the column densities estimated with the new collisional coefficients compare with those calculated there.   
        
            \begin{table}
	        \centering
	        \caption{Column densities N(X) (cm$^{-2}$) at 8 K assuming different densities.}
	        \label{tab:columnDensitiesTMC1}
	        \begin{tabular}{lcc}
		    \toprule
		{Molecule} & $n_{\rm H}=4\times 10^{5}$ cm$^{-3}$ & $n_{\rm H}=10^{6}$ cm$^{-3}$\\ \midrule
		        D$^{13}$CN & $<5.66\times 10^{10}$ & $<3.79\times 10^{10}$\\
				DCN & $(3.24\pm 0.81)\times 10^{12}$ & $(2.25\pm 0.56)\times 10^{12}$\\
				DN$^{13}$C & $(1.50\pm 0.38)\times 10^{11}$ & $(1.40\pm 0.35)\times 10^{11}$\\
				DNC & $(4.51\pm 1.13)\times 10^{12}$ & $(3.73\pm 0.93)\times 10^{12}$\\
				HC$^{15}$N & $(2.90\pm 0.73)\times 10^{11}$ & $(1.70\pm 0.43)\times 10^{11}$\\
				H$^{13}$CN & $(2.12\pm 0.53)\times 10^{12}$ & $(1.22\pm 0.31)\times 10^{12}$\\
				HN$^{13}$C & $(1.60\pm 0.40)\times 10^{12}$ & $(1.30\pm 0.33)\times 10^{12}$\\
				HCN & $(4.32\pm 1.08)\times 10^{13}$ & $(2.03\pm 0.51)\times 10^{13}$\\
				H$^{15}$NC & $(4.50\pm 1.13)\times 10^{11}$ & $(3.90\pm 0.98)\times 10^{11}$\\
				HNC & $(4.30\pm 1.08)\times 10^{12}$ & $(4.20\pm 1.05)\times 10^{12}$ \\
    		\bottomrule
	        \end{tabular}
        \end{table}
        
       Our estimation of the DNC column density with the new collisional coefficients (Table \ref{tab:columnDensitiesTMC1}) is a factor of $1.5$ times higher than the previous value obtained by \citet{Navarro2021}, but still compatible within the uncertainties. The DCN column density calculated here (Table \ref{tab:columnDensitiesTMC1}) shows a similar behavior, as it is a factor of $1.3$ times higher than that obtained in \citet{Navarro2021}. The column density of the DNC and DN$^{13}$C isotopologues computed in \citet{Navarro2021} relies on the assumption of a $^{12}{\rm C}/{^{13}{\rm C}}$ ratio of $\sim 60$. Here, we estimated the DN$^{13}$C column density free of assumptions on the $^{12}{\rm C}/{^{13}{\rm C}}$ ratio, obtaining a column density N(DN$^{13}$C) that is twice as high as the value tabulated there. The resulting isotopic ratio ${\rm DNC}/{{\rm DN}^{13}{\rm C}} = 21.60\pm 10.71$ (see Table \ref{tab:ratiosTMC1}) is thus lower than the previously assumed value of $^{12}{\rm C}/{^{13}{\rm C}} \sim 60$. We found a similar $^{12}{\rm C}/{^{13}{\rm C}}$ ratio from the fraction ${\rm HCN}/{{\rm H}^{13}{\rm CN}} = 20.38\pm 7.20$ (see Table \ref{tab:ratiosTMC1}). Finally, the ${\rm HNC}/{{\rm H}^{13}{\rm NC}} = 2.69\pm 0.95$ (Table \ref{tab:ratiosTMC1}) ratio is even lower than our previous results. These discrepancies with the commonly accepted value for the local interstellar medium $^{12}{\rm C}/{^{13}{\rm C}} = 68$ \citep{Milam2005} might be due to the underestimation of the HNC, HCN, and DNC column densities. As other studies on the subject have pointed out \citep[see, e.g.,][]{Daniel2013}, this underestimation of the column densities of the $^{12}$C-bearing species might be due to the high opacities of the emission and the subsequent absorption by external layers of the cloud \citep{Daniel2013, Navarro2021}. For HCN, this hypothesis is demonstrated by the fitting of the hyperfine splitting of the $1\rightarrow 0$ line, which provided $\tau_{\rm HCN} \gg 1$ for the weakest component (see Table \ref{tab:lineResultsTMC1C}). Since the hyperfine splitting is not resolved for the HNC and DCN $1\rightarrow 0$ lines, we need to follow a different strategy. We can estimate the opacity required to reproduce the low HNC/HN$^{13}$C and DNC/DN$^{13}$C ratios using the ratio of integrated intensities. To do so, we assumed optically thin emission from the HN$^{13}$C $1\rightarrow 0$ and DN$^{13}$C $1\rightarrow 0$ lines, an isotopic ratio of $^{12}{\rm C}/{^{13}{\rm C}} = 68$, and equal excitation temperatures. Under these assumptions, we obtained the following: W(HNC)/W(HN$^{13}$C) $\approx 68\times(1-\exp(-\tau_{\rm HNC}))/\tau_{\rm HNC})$, resulting in $\tau_{\rm HNC}\sim 29$.  This optical depth is comparable with the total optical depth of the HCN 1$\rightarrow$0 line obtained with the HFS fitting, which is reasonable as long as the HCN/HNC column density ratio is $\sim$1. Likewise, for DNC we obtained $\tau_{\rm DNC}\sim$ 2.9. This high opacity would imply a deuteration fraction of $\sim$0.1.
    
    \begin{table}
	\centering
	\caption{Column density ratios in TMC 1-C.}
	\label{tab:ratiosTMC1}
	
	\begin{tabular}{lcc}
		\toprule
		& $n_{\rm H}=4\times 10^{5}$ cm$^{-3}$ & $n_{\rm H}=10^{6}$ cm$^{-3}$\\ \midrule
		        \multicolumn{3}{c}{Deuterium fraction}\\ \midrule
		       D$^{13}$CN/H$^{13}$CN & $<2.67\times 10^{-2}$ & $<3.11\times 10^{-2}$\\
			   DN$^{13}$C/HN$^{13}$C & $(9.38\pm 3.34)\times 10^{-2}$ & $0.108\pm 0.038$\\
			   DCN/H$^{13}$CN & $1.53\pm 0.54$ & $1.84\pm 0.66$ \\
			   DNC/HN$^{13}$C & $2.82\pm 1.00$ & $2.87\pm 1.02$ \\
			   DCN/HCN & $(7.50\pm 2.65)\times 10^{-2}$ & $0.111\pm 0.039$ \\
			   DNC/HNC & $1.05\pm 0.37$ & $0.89\pm 0.31$ \\
			   \midrule
		       \multicolumn{3}{c}{C and N isotopic ratios}\\ \midrule
			   DCN/D$^{13}$CN & $> 57.24$ & $> 59.37$\\
			   DNC/DN$^{13}$C & $21.60\pm 10.71$ & $16.07\pm 9.41$\\
			   HCN/H$^{13}$CN & $20.38\pm 7.20$ & $16.64\pm 5.95$\\
			   HNC/HN$^{13}$C & $2.69\pm 0.95$ & $3.23\pm 1.15$\\
			   HCN/HC$^{15}$N & $148.97\pm 52.85$ & $119.41\pm 42.57$\\
			   HNC/H$^{15}$NC & $9.56\pm 3.39$ & $10.80\pm 3.82$\\
			   H$^{13}$CN/HC$^{15}$N & $7.31\pm 2.59$ & $7.18\pm 2.57$ \\
			   HN$^{13}$C/H$^{15}$NC  & $3.56\pm 1.26$ & $3.33\pm 1.19$ \\ \midrule
		        \multicolumn{3}{c}{Isomeric ratios}\\ \midrule
		        H$^{13}$CN/HN$^{13}$C & $1.33\pm 0.47$ & $0.94\pm 0.34$ \\
		        HC$^{15}$N/H$^{15}$NC & $0.64\pm 0.23$ & $0.44\pm 0.16$ \\
		        HCN/HNC & $10.05\pm 3.56$ & $4.83\pm 1.71$ \\
		        DCN/DNC & $0.72\pm 0.25$ & $0.60\pm 0.21$ \\
    		\bottomrule
	\end{tabular}
	
\end{table}
    
        The new collisional coefficients also allowed us to study the nitrogen fractionation in TMC 1-C, of great interest in molecular clouds due to its puzzling origin. A lot of effort has been done to elucidate the problem of the nitrogen fractionation \citep[see, e.g.,][]{Roueff2015, Loison2019}, which current chemical models are unable to reproduce. The measured HCN/HC$^{15}$N isotopic ratio in TMC 1-C is HCN/HC$^{15}$N$=148.97\pm 52.85$ ( Table \ref{tab:ratiosTMC1}), in agreement with the value obtained in Barnard 1b \citep{Daniel2013}, and much lower than the predicted value in chemical models \citep{Roueff2015}. Again, this may indicate an underestimation of the HCN column density due to high optical depth. A greater underestimation of the HNC column density might lead to the even lower nitrogen fractionation found in the isotopic ratio HNC/H$^{15}$NC $=9.56\pm 3.39$. The problem of high opacity is circumvented using the less abundant isotopologues H$^{13}$CN and HN$^{13}$C. Indeed, the isotopic ratios H$^{13}$CN/HC$^{15}$N $=7.31\pm 2.59$ and HN$^{13}$C/H$^{15}$NC $=3.56\pm1.26$ (Table \ref{tab:ratiosTMC1}) are in good agreement with the model predictions in \citet{Roueff2015} and observational results in TMC 1 \citep{Hily-Blant2013}. Indeed, \citet{Roueff2015} predicted H$^{13}$CN/HC$^{15}$N and HN$^{13}$C/H$^{15}$NC ratios between $\sim$3 and $\sim$7 for typical conditions of starless cores. Moreover, both ratios are expected to vary following the same trend with time evolution. The different values obtained for HCN and HNC isotopologues suggest that each molecule is probing different regions within dense cores.
        
        We also derived the isomeric ratios in TMC 1-C. Isomeric ratios involving HCN, HNC, and isotopologues, have been claimed to have a great potential as thermal probes due to their high temperature-dependent formation and destruction pathways \citep{Graninger2014, Hacar2020}. Variations in this isomer have also been linked to shocks and sputtering of material locked in icy grain surfaces \citep{Lefloch2021}. In cold core conditions, this ratio is expected to be close to 1 \citep{Daniel2013, Lefloch2021}, while it is expected to increase at higher temperatures. Owing again to the underestimation of the HNC column density, the measured HCN/HNC ratio is well above the expected value of 1. The expected value for this ratio in cold cores is recovered when we consider the less abundant isotopologues. Indeed, the H$^{13}$CN/HN$^{13}$C and DCN/DNC ratios are compatible and $\sim 1$. The HC$^{15}$N/H$^{15}$NC ratio is, however, sightly lower than 1.
        
        Finally, we derived the deuterium fraction of HCN, HNC, and their isotopologues. As discussed previously, the deuteration fraction of a prestellar or starless core is linked to its chemical age. Our results on deuteration are found to be similar to those computed in \citet{Daniel2013}, pointing to an evolved chemistry close to that of first hydrostatic cores (FHSCs) and YSOs. In fact, the ratio DCN/HCN $\sim 0.075\pm 0.027$ is in agreement with that obtained in Barnard 1b $\sim 0.05$. Similarly, the DCN/H$^{13}$CN $= 1.53\pm 0.54$ ratio is also found to be compatible with that derived in Barnard 1b $\sim 1.5$. Interestingly, the value  DNC/HN$^{13}$C = $2.82\pm1.0$ is slightly higher than that of DCN/H$^{13}$CN, suggesting that the HNC isomer comes from a colder region. However, this difference could be also the consequence of the high opacity of the DNC $1 \rightarrow 0$ line.
    
\subsection{Molecular column densities in NGC 1333}

    The procedures presented in Sections \ref{sec:colDensHfs} and \ref{sec:colDensNoHfs} were also applied to calculate the column density of the species in Table \ref{tab:summarylines}. In our calculations, we considered two density scenarios: for the NGC 1333-C7-2 to NGC 1333-C7-5 positions at a temperature of 15 K, whereas for NGC1333-C7-1, we used the physical properties obtained by \citet{RodriguezBaras2021}. The results are shown in Table \ref{tab:columnDensitiesNGC1333}.

    \begin{table*}
	\centering
	\caption{Column densities at the NGC1333-C7 positions assuming different densities.}
	\label{tab:columnDensitiesNGC1333}
	\resizebox{\textwidth}{!}{
	\begin{tabular}{lccccccccc}
		\toprule
		\multicolumn{10}{c}{{Column densities}} \\ \cmidrule{1-10}
		{Position} & {C7-1} & \multicolumn{2}{c}{C7-2} & \multicolumn{2}{c}{C7-3} & \multicolumn{2}{c}{C7-4} & \multicolumn{2}{c}{C7-5} \\ \cmidrule(lr){2-2}\cmidrule(lr){3-4}\cmidrule(lr){5-6}\cmidrule(lr){7-8}\cmidrule(lr){9-10}
		$n_{\rm H}$ (cm$^{-3}$) & $1.4\times 10^{5}$ & $4\times 10^{5}$ & $10^{6}$ & $4\times 10^{5}$ & $10^{6}$ & $4\times 10^{5}$ & $10^{6}$ & $4\times 10^{5}$ & $10^{6}$ \\
		\midrule
		        D$^{13}$CN & $<2.16\times 10^{11}$ & $<1.69\times 10^{11}$ & $<1.24\times 10^{11}$ & $<5.44\times 10^{11}$ &  $<3.87\times 10^{11}$ & $<2.63\times 10^{11}$ & $<1.84\times 10^{11}$ & $<2.12\times 10^{11}$ & $<1.71\times 10^{11}$\\
				DCN & $(3.42\pm 0.86)\times 10^{12}$ & $(5.76\pm 1.44)\times 10^{11}$ & $(4.14\pm 1.04)\times 10^{11}$ & $(1.62\pm 0.41)\times 10^{12}$ & $(1.13\pm 0.28)\times 10^{12}$ & $(2.28\pm 0.58)\times 10^{12}$ & $(1.58\pm 0.40)\times 10^{12}$ & $(2.52\pm 0.63)\times 10^{12}$ & $(1.98\pm 0.50)\times 10^{12}$\\
				DN$^{13}$C & $(1.25\pm 0.31)\times 10^{11}$ & $<4.90\times 10^{10}$ & $<4.60\times 10^{10}$ & $<8.78\times 10^{10}$ & $<8.51\times 10^{10}$ & $<7.60\times 10^{10}$ & $<7.40\times 10^{10}$ & $(7.30\pm 1.83)\times 10^{10}$ & $(7.80\pm 1.95)\times 10^{10}$\\
				DNC & $(4.50\pm 1.13)\times 10^{12}$ & $(6.20\pm 1.55)\times 10^{11}$ & $(5.60\pm 1.40)\times 10^{11}$ & $(1.62\pm 1.40)\times 10^{12}$ & $(1.48\pm 0.37)\times 10^{12}$ & $(2.80\pm 0.70)\times 10^{12}$ & $(2.40\pm 0.60)\times 10^{12}$ & $(3.85\pm 0.96)\times 10^{12}$ & $(3.75\pm 0.94)\times 10^{12}$\\
				HC$^{15}$N & $(1.94\pm 0.49)\times 10^{11}$ & $(1.20\pm 0.30)\times 10^{11}$ & $(7.00\pm 1.75)\times 10^{10}$ & $(2.00\pm 0.50)\times 10^{11}$ & $(1.21\pm 0.30)\times 10^{11}$ & $(2.61\pm 0.65)\times 10^{11}$ & $(1.57\pm 0.39)\times 10^{11}$ & $(1.88\pm 0.47)\times 10^{11}$ & $(1.26\pm 0.32)\times 10^{11}$ \\
				H$^{13}$CN & $(1.52\pm 0.38)\times 10^{12}$ & $(8.37\pm 2.09)\times 10^{11}$ & $(5.13\pm 1.28)\times 10^{11}$ & $(9.00\pm 2.25)\times 10^{11}$ & $(5.40\pm 1.35)\times 10^{11}$ & $(1.21\pm 0.30)\times 10^{12}$ & $(7.29\pm 1.82)\times 10^{11}$ & $(9.27\pm 2.32)\times 10^{11}$ & $(6.26\pm 1.57)\times 10^{11}$ \\
				HN$^{13}$C & $(1.31\pm 0.33)\times 10^{12}$ & $(2.40\pm 0.60)\times 10^{11}$ & $(2.16\pm 0.54)\times 10^{11}$ & $(5.40\pm 0.14)\times 10^{11}$ & $(4.75\pm 1.19)\times 10^{11}$ & $(8.78\pm 2.20)\times 10^{11}$ & $(7.68\pm 1.92)\times 10^{11}$ & $(1.09\pm 0.27)\times 10^{12}$ & $(1.06\pm 0.27)\times 10^{12}$\\
				HCN & $(1.08\pm 0.27)\times 10^{14}$ & $(4.44\pm 1.10)\times 10^{13}$ & $(2.32\pm 0.58)\times 10^{13}$ & $(7.92\pm 1.98)\times 10^{13}$ & $(3.69\pm 0.92)\times 10^{13}$ & $(1.44\pm 0.36)\times 10^{14}$ & $(5.85\pm 1.46)\times 10^{13}$ & $(5.63\pm 1.41)\times 10^{13}$ & $(2.97\pm 0.74)\times 10^{13}$\\
				H$^{15}$NC & $(3.38\pm 0.85)\times 10^{11}$ & $(7.58\pm 1.90)\times 10^{10}$ & $(7.05\pm 1.76)\times 10^{10}$ & $(1.65\pm 0.41)\times 10^{11}$ & $(1.48\pm 0.37)\times 10^{11}$ & $(2.27\pm 0.57)\times 10^{11}$ & $(2.09\pm 0.52)\times 10^{11}$ & $(2.55\pm 0.64)\times 10^{11}$ & $(2.57\pm 0.64)\times 10^{11}$\\
				HNC & $(1.56\pm 0.39)\times 10^{13}$ &$-$ & $-$ & $-$ & $-$ & $-$ & $-$ & $-$ & $-$ \\
    		\bottomrule
	\end{tabular}
	}
\end{table*}

    We first analyzed the Carbon isotopic ratio. We detected DN$^{13}$C $1\rightarrow0$ emission in only two positions belonging to NGC 1333. In these positions, NGC1333-C7-1 and NGC 1333-C7-5, the DNC/DN$^{13}$C ratio is similar to that of TMC 1-C (see Table \ref{tab:ratiosNGC1333}) and below the commonly used value of $^{12}{\rm C}/^{13}{\rm C} = 68$. Assuming the DN$^{13}$C $1\rightarrow 0$ line as optically thin, a $^{12}{\rm C}/^{13}{\rm C} = 68$ ratio, and equal excitation temperatures, we found that W(DNC)/W(DN$^{13}$C) $\approx 68\times(1-\exp(-\tau_{\rm DNC}))/\tau_{\rm DNC})$, resulting in $\tau_{\rm DNC}\sim 2.10$ in NGC 1333-C7-1 and $\tau_{\rm DNC}\sim 0.89$ in NGC1333-C7-5. These moderately high opacities may hinder a reliable estimation of the DNC column density, hence, the abnormal DNC/DN$^{13}$C ratio we obtained. This is not, however, the case of the HCN/H$^{13}$CN, which is found to be in good agreement with the standard value of 68 at all positions, as a result of the lower opacities we found for the HCN $1\rightarrow0$ lines (see Tables \ref{tab:lineResultsNGC1333_C7_1} to \ref{tab:lineResultsNGC1333_C7_5}), compared to TMC 1-C. Finally, like in the case of TMC 1-C, we underestimated the $^{12}{\rm C}/^{13}{\rm C}$ ratio with the HNC/HN$^{13}$C fraction (see Table \ref{tab:ratiosNGC1333}) in NGC 1333-C7-1. The opacity of the HNC $1\rightarrow0$ line, assuming the same conditions as our previous estimations, is $\tau_{\rm DNC}\sim 10$, making this line optically thick and, consequently, leading to an underestimation of the HNC column density. While the $^{12}{\rm C}/^{13}{\rm C}$ isotopic ratio calculated using DNC and DN$^{13}$C is, in general, lower than that calculated using HCN and H$^{13}$CN, they are compatible within the uncertainties. The differences between estimations of the $^{12}{\rm C}/^{13}{\rm C}$ isotopic ratio may arise from the different layers of gas traced with HCN and DNC, as DNC traces denser and colder gas than HCN.

    The nitrogen isotopic ratio in the positions observed here are shown in Table \ref{tab:ratiosNGC1333}. The HCN/HC$^{15}$N ratios in NGC1333-C7 are, in general, higher than those derived in TMC 1-C. The lower optical depths of the transitions help constrain the column densities better, and therefore our estimations across the NGC1333-C7 positions is in better agreement with model predictions \citep{Roueff2015}. The H$^{13}{\rm C}^{14}{\rm N}/{\rm H}^{12}{\rm C}^{15}{\rm N}$ and ${\rm H^{14}{\rm N}^{13}{\rm C}}/{\rm H^{15}{\rm N}^{12}{\rm C}}$ ratios found in NGC1333-C7 are also in good agreement with model predictions \citep{Roueff2015}, Barnard 1b \citep{Daniel2013}, and previous TMC 1 observations \citep{Liszt2012}. {As in the case of TMC 1-C, there is a trend with the  ${\rm H^{14}{\rm N}^{13}{\rm C}}/{\rm H^{15}{\rm N}^{12}{\rm C}}$ ratio being generally lower than the H$^{13}{\rm C}^{14}{\rm N}/{\rm H}^{12}{\rm C}^{15}{\rm N}$ one.}
    
    As discussed previously, the isomeric ratio HCN/HNC and the possible isotopic variations of it are indicators of the gas temperature in a region. For cold, starless or prestellar cores, this ratio is $\sim 1$, and higher for warmer gas. Our results for the H$^{13}$CN/HN$^{13}$C ratio across the observed positions show a general agreement with a $\sim 1$ ratio, indicating cold temperatures of the cores. This ratio in NGC1333-C7-2 is, however, up to three times higher than in the rest of the positions, suggesting higher gas temperatures in this core. We found a similar behavior in the HC$^{15}$N/H$^{15}$NC ratio, although the values are always lower than those of the H$^{13}$CN/HN$^{13}$C ratio.
    
    We last discuss the deuterium fraction at the different positions observed here. On one hand, our analysis shows that deuteration is higher in the cores NGC1333-C7-1 and NGC1333-C7-5. This is consistent with the detection of DN$^{13}$C in these cores, which indeed indicates a high deuteration. On the other hand, the lowest deuteration is found in NGC1333-C7-2. As commented below, the higher gas temperatures found at this position using the H$^{13}$CN/HN$^{13}$C ratio may lead to a less efficient deuteration route thus reducing the deuterium fraction \citep{Roueff2007}.
    
\begin{table*}
	\centering
	\caption{Column density ratios across the positions observed in NGC1333-C7.}
	\label{tab:ratiosNGC1333}
	\resizebox{\textwidth}{!}{
	\begin{tabular}{lccccccccc}
		\toprule
		{Position} & {C7-1} & \multicolumn{2}{c}{C7-2} & \multicolumn{2}{c}{C7-3} & \multicolumn{2}{c}{C7-4} & \multicolumn{2}{c}{C7-5} \\ \cmidrule(lr){2-2}\cmidrule(lr){3-4}\cmidrule(lr){5-6}\cmidrule(lr){7-8}\cmidrule(lr){9-10}
		$n_{\rm H}$ (cm$^{-3}$) & $1.4\times 10^{5}$ & $4\times 10^{5}$ & $10^{6}$ & $4\times 10^{5}$ & $10^{6}$ & $4\times 10^{5}$ & $10^{6}$ & $4\times 10^{5}$ & $10^{6}$ \\
		\midrule
		\multicolumn{10}{c}{Deuterium fraction}\\ \midrule
		        D$^{13}$CN/H$^{13}$CN & $<0.142$ & $<0.202$ & $<0.242$ & $<0.604$ & $<0.717$ & $<0.217$ & $<0.252$ & $<0.229$ & $<0.273$ \\
		        DN$^{13}$C/HN$^{13}$C & $0.095\pm 0.034$ & $<0.204$ & $<0.213$ & $<0.163$ & $<0.179$ & $<0.087$ & $<0.096$ & $0.067\pm 0.024$ & $0.074\pm 0.026$\\
		        DCN/H$^{13}$CN & $2.25\pm 0.80$ & $0.69\pm 0.24$ & $0.81\pm 0.29$ & $1.80\pm 0.64$ & $2.09\pm 0.74$ & $1.88\pm 0.67$ & $2.17\pm 0.77$ & $2.72\pm 0.96$ & $3.16\pm 1.12$\\
		        DNC/HN$^{13}$C & $3.44\pm 1.21$ & $2.58\pm 0.91$ & $2.59\pm 0.92$ & $3.00\pm 1.06$ & $3.12\pm 1.10$ & $3.19\pm 1.13$ & $3.13\pm 1.10$ & $3.53\pm 1.25$ & $3.54\pm 1.25$\\
		        DCN/HCN & $(3.17\pm 1.11)\times 10^{-2}$ & $(1.30\pm 0.46)\times 10^{-2}$ & $(1.78\pm 0.63)\times 10^{-2}$ & $(2.05\pm 0.72)\times 10^{-2}$ & $(3.06\pm 0.11)\times 10^{-2}$ & $(1.58\pm 0.56)\times 10^{-2}$ & $(2.70\pm 0.95)\times 10^{-2}$ & $(4.48\pm 1.58)\times 10^{-2}$ & $(6.67\pm 0.24)\times 10^{-2}$\\
		        DNC/HNC & $0.29\pm 0.10$ & $-$ & $-$ & $-$ & $-$ & $-$ & $-$ & $-$ & $-$\\
		        \midrule
		        	\multicolumn{10}{c}{C and N isotopic ratios}\\ \midrule
		        DCN/D$^{13}$CN & $>15.83$ & $>3.40$ & $>3.33$ & $>2.98$ & $>2.92$ & $>8.68$ & $>8.61$ & $>11.89$ & $>11.58$ \\
		        DNC/DN$^{13}$C & $27.36\pm 12.73$ & $>11.76$ & $>9.00$ & $>18.45$ & $>13.28$ & $>30.00$ & $>21.35$ & $34.52\pm 18.65$ & $25.39\pm 17.00$ \\
		        HCN/H$^{13}$CN & $71.05\pm 25.12$ & $53.05\pm 18.76$ & $45.22\pm 15.99$ & $88.00\pm 31.11$ & $68.33\pm 24.16$ & $119.01\pm 42.08$ & $80.25\pm 28.37$ & $60.73\pm 21.47$ & $47.44\pm 17.78$\\ 
		        HNC/HN$^{13}$C & $11.91\pm 4.21$ & $-$ & $-$ & $-$ & $-$ & $-$ & $-$ & $-$ & $-$ \\
		        HCN/HC$^{15}$N & $556.70\pm 196.82$ & $370.00\pm 130.82$ & $331.43\pm 117.18$ & $396.00\pm 140.01$ & $304.96\pm 107.82$ & $551.72\pm 195.06$ & $372.61\pm 131.74$ & $299.47\pm 105.88$ & $235.71\pm 83.34$ \\
		        HNC/H$^{15}$NC & $46.15\pm 16.32$ & $-$ & $-$ & $-$ & $-$ & $-$ & $-$ & $-$ & $-$ \\
		        H$^{13}$CN/HC$^{15}$N & $7.84\pm 2.77$ & $6.98\pm 2.47$ & $7.33\pm 2.59$ & $4.50\pm 1.59$ & $4.46\pm 1.58$ & $4.64\pm 1.64$ & $4.64\pm 1.64$ & $4.93\pm 1.74$ & $4.97\pm 1.76$ \\
		        HN$^{13}$C/H$^{15}$NC & $3.88\pm 1.37$ & $3.17\pm 1.12$ & $3.06\pm 1.08$ & $3.27\pm 1.16$ & $3.21\pm 1.14$ & $3.87\pm 1.37$ & $3.68\pm 1.30$ & $4.28\pm 1.51$ & $4.13\pm 1.46$ \\
		        \midrule
		        \multicolumn{10}{c}{Isomeric ratios}\\ \midrule
		        H$^{13}$CN/HN$^{13}$C & $1.16\pm 0.41$ & $3.49\pm 1.23$ & $2.38\pm 0.84$ & $1.67\pm 0.59$ & $1.14\pm 0.40$ & $1.38\pm 0.49$ & $0.95\pm 0.34$ & $0.85\pm 0.30$ & $0.59\pm 0.21$ \\
		        HC$^{15}$N/H$^{15}$NC & $0.57\pm 0.20$ & $1.58\pm 0.56$ & $0.99\pm 0.35$ & $1.21\pm 0.43$ & $0.82\pm 0.29$ & $1.15\pm 0.41$ & $0.75\pm 0.27$ & $0.74\pm 0.26$ & $0.49\pm 0.17$\\
		        HCN/HNC & $6.92\pm 2.45$ & $-$ & $-$ & $-$ & $-$ & $-$ & $-$ & $-$ & $-$ \\
		        DCN/DNC & $0.76\pm 0.27$ & $0.93\pm 0.33$ & $0.74\pm 0.26$ & $1.00\pm 0.35$ & $0.76\pm 0.27$ & $0.81\pm 0.29$ & $0.66\pm 0.23$ & $0.66\pm 0.23$ & $0.53\pm 0.19$ \\ 
    		\bottomrule
	\end{tabular}
	}
\end{table*}

\section{Continuum emission, dust opacity, and the spectral index}

As discussed above, large deuteration fractions are often linked to a more evolved stage of prestellar and starless cores. The CO freeze out and consequent ice mantle growth that leads to an enhancement of deuterated molecules beyond the elemental D/H ratio in cold cores is also responsible for grain coagulation and changes in the properties of grains \citep{Whittet1988, Ossenkopf1993, Ormel2009}. Dust sizes play important roles in star and planet formation. In the low-density gas of molecular clouds, dust size is involved in the synthesis of molecules that radiatively cool the gas \citep{Draine2011}. At higher densities $n>10^{5}$ cm$^{-3}$, dust dominates the gas cooling \citep{Goldsmith2001}. In the subsequent stages of the protostellar evolution, dust size is a key property in the formation of protostellar disks and, eventually, planets \citep{Zhao2016}.

In dense cloud cores there is evidence of grain growth unveiled by the study of infrared radiation at different wavelengths. Previous works focused on the study of extended dust emission known as coreshine. For instance, dust continuum observations at 3.6 $\mu$m with the Spitzer Infrared Array Camera (IRAC) instrument revealed emission patterns only possible by radiation scattering by large grains \citep{Pagani2010, Steinacker2010}. Further observations at 3.6 and 4.5 $\mu$m with the same instrument also helped constrain the grain sizes \citep{Steinacker2015} and, at 8 $\mu$m, \citep{Lefevre2016} showed that uncoagulated grains are unable to reproduce the observations. 

\subsection{Dust emissivity spectral index}

Here, we examine the behavior of the dust emissivity spectral index, $\beta$, using  our 3 mm MUSTANG-2 images and previous data. The value of $\beta$ is a quantity that directly depends on the grain size distribution \citep[see, e.g.,][]{Schnee2014, Chacon2017, Chacon2019}. From now on, we assume that the mm wavelength emission of dust is described by a modified blackbody. Given that dust emission is optically thin at the {\em Herschel} and MUSTANG-2 observation frequencies:

\begin{equation}
    \label{eq:modBlackbody}
    I_{\nu} \simeq B_{\nu}(T_{d})\kappa_{\nu}\mu_{\rm H_{2}}m_{\rm H}N(H_{2}) = B_{\nu}(T_{d})\tau_{\nu},
\end{equation}
where $B_{\nu}(T_{d})$ is the blackbody function at a temperature, T$_{d}$, while $\kappa_{\nu}$ is the dust opacity at a given frequency, $\nu$, and $\mu_{\rm H_{2}}$ is the molecular weight per hydrogen molecule, $m_{\rm H}$ is the atomic hydrogen mass, $N(H_{2})$ is the molecular hydrogen column density, and $\tau_{\nu}$ is the optical depth at the frequency $\nu$. We assume that the dust opacity $\kappa_{\nu}$ can be approximated by a power law at millimeter wavelengths, so that $\kappa_{\nu}\sim \nu^{\ \beta}$, where $\beta$ is the spectral index \citep{Hildebrand1983}. Therefore, computing the dust opacity at two different wavelengths provides the spectral index $\beta$ in that wavelength range. Throughout this paper, we compare the values of $\tau_{3\rm mm}$ and $\tau_{0.85\rm mm}$ to compute $\beta$:

\begin{equation}
    \label{eq:beta}
    \beta_{0.85mm-3mm} = \frac{\log{\left(\tau_{3\rm mm}/\tau_{0.85 \rm mm}\right)}}{\log{\left(\nu_{3\rm mm}/\nu_{0.85\rm mm}\right)}},
\end{equation}
with the spectral index $\beta$ typically taking values in the range $1.5\leq\beta\leq2.5$ \citep[see, e.g.,][]{Schnee2010, Sadavoy2013}. Shallower spectral indexes point at the presence of large (mm) grains \citep[see, e.g.,][]{Schnee2010, Schnee2014, Sadavoy2013}. 

As discussed above, grain growth is expected in evolved prestellar and protostellar structures \citep[see, e.g.,][]{Chacon2019, Silsbee2020, Caselli2022}. Since the deuterium fraction has been used as a chemical clock of prestellar and starless cores and is highly dependent on the temperature, in the next section we explore the possible correlations between the spectral index, deuteration, and temperature, linking chemistry, dust properties, and evolution.

\subsection{Spectral index using MUSTANG-2 data}

\begin{figure*}
    \centering
    \begin{subfigure}[b]{0.49\textwidth}\includegraphics[width=\textwidth,keepaspectratio]{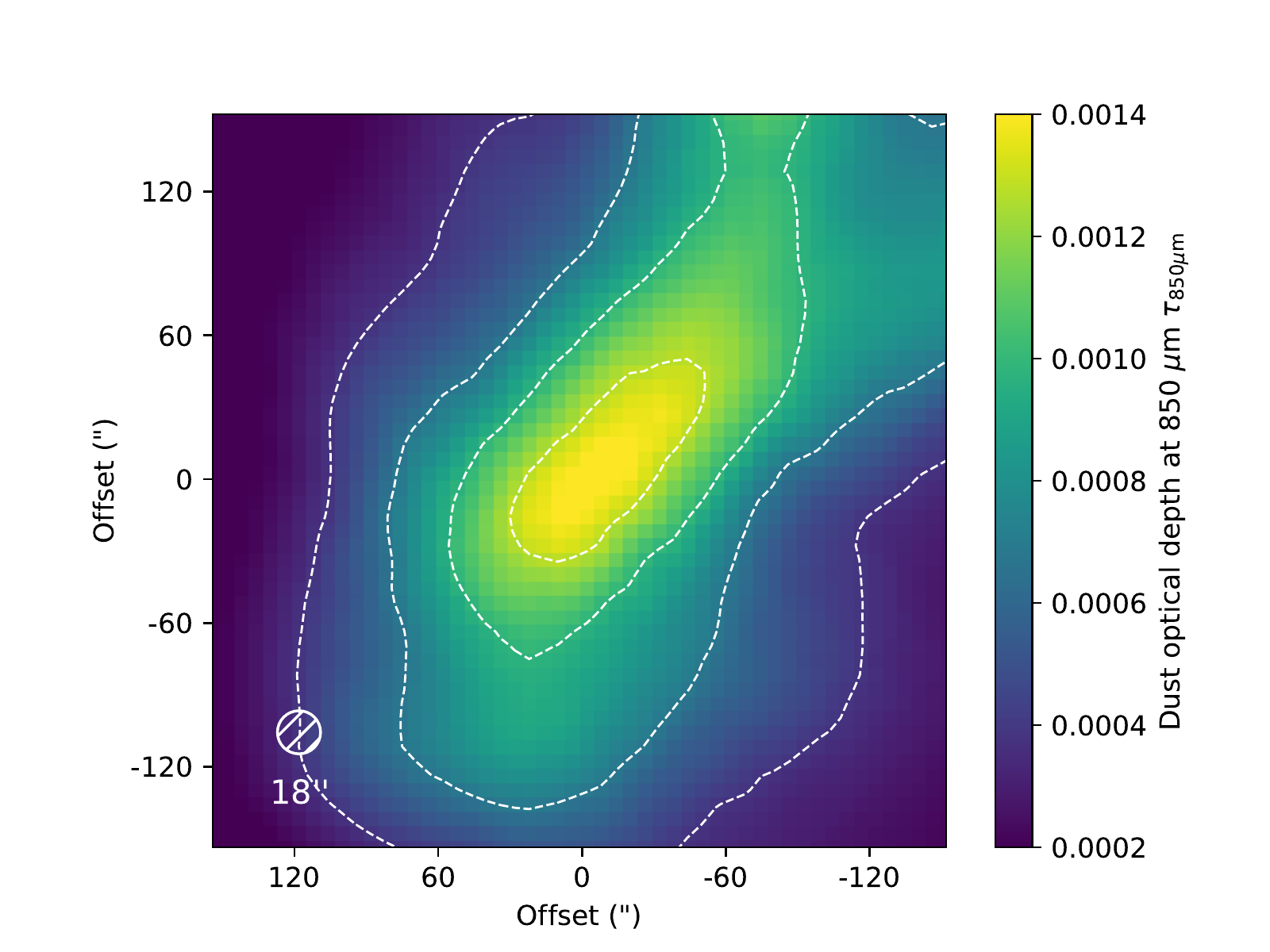}
    \end{subfigure}
    \begin{subfigure}[b]{0.49\textwidth}\includegraphics[width=\textwidth,keepaspectratio]{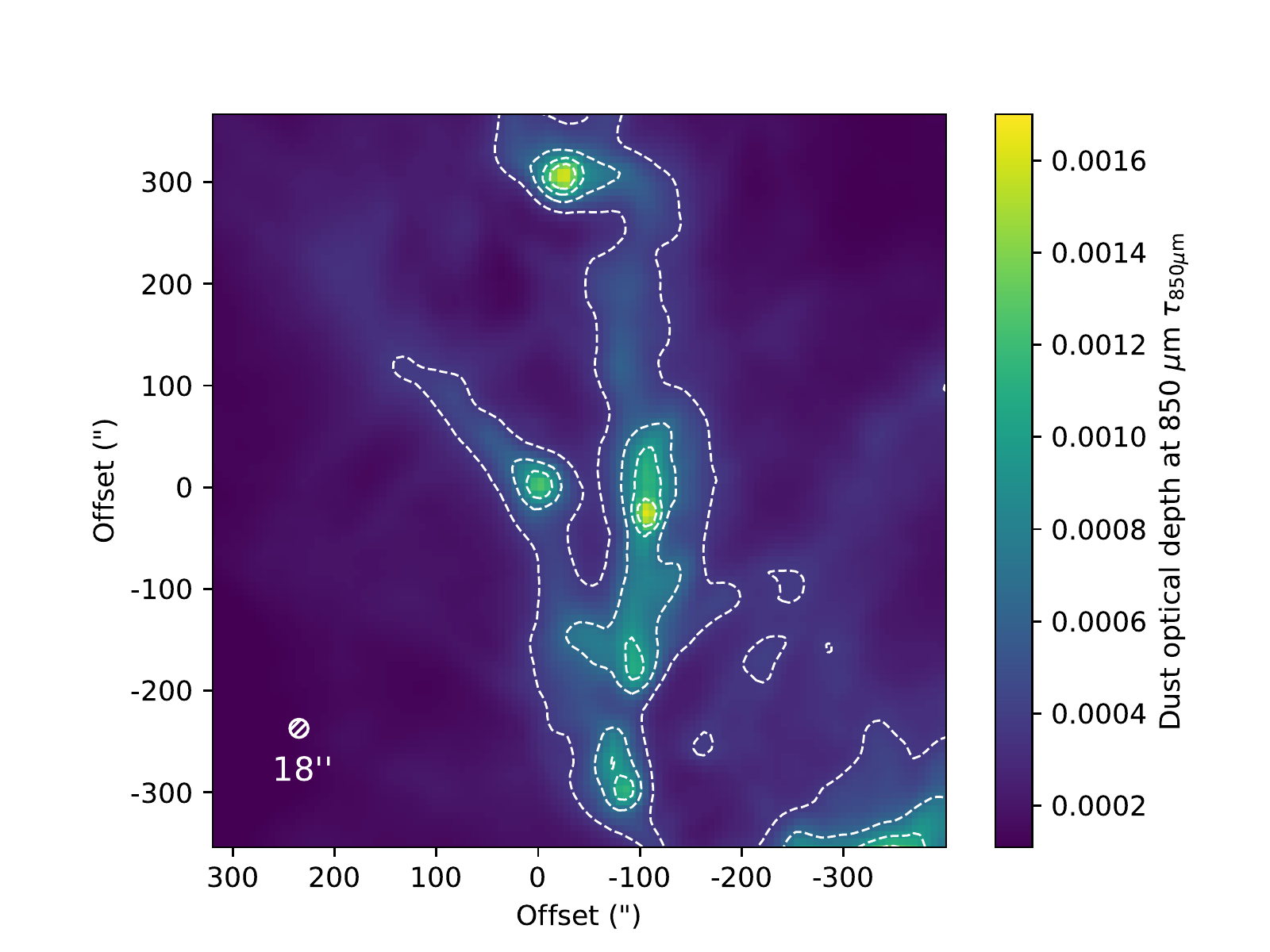}
    \end{subfigure}
		    \caption{Dust optical depth maps at 850 $\mu$m from the {\it Herschel} data (Kirk et al. in prep.) used to compute the dust emissivity spectral index $\beta_{0.85mm-3mm}$. {\it Left panel:} dust optical depth at 850 $\mu$m in TMC 1-C. Contours correspond to levels $\tau_{850{\mu\rm m}} = 5\sigma\times[5, 9, 13, 17]$ with $\sigma = 1.5\times 10^{-5}$ (Kirk et al. in prep.). {\it Right panel:} dust optical depth at 850 $\mu$m in the northern sector of NGC 1333 \citep{Zari2016}. Contours correspond to levels $\tau_{850{\mu\rm m}} = 5\sigma\times[5, 9, 13, 17]$ with $\sigma = 1.5\times 10^{-5}$. The origin (0,0) of these maps is the same as in Figure \ref{fig:mapsTMC1NGC1333}.}
		    \label{fig:mapsOpacity850}
\end{figure*}

Figure \ref{fig:mapsOpacity850} shows the dust optical depth maps at 850 $\mu$m, $\tau_{0.85\rm mm}$, obtained with the spectral energy distribution (SED) fitting of the dust continuum emission images provided by the {\em Herschel} space telescope. The TMC 1-C map was created following the procedure described in \citet{Kirk2013}, in which a SED fitting is done pixel by pixel, where the dust opacity, $\kappa_\nu$, is parameterized as $\kappa_\nu \propto \nu^\beta$, with $\beta=2$ and the reference value is 0.1 cm$^2$ g$^{-1}$ at 1 THz \citep{Beckwith1990}. In this fitting, the {\em Herschel} maps at 70 $\mu$m, 160 $\mu$m, 250 $\mu$m, 350 $\mu$m, and 500 $\mu$m were processed following the procedure described in \citet{Lombardi2014} to obtain the dust parameters at an angular resolution of 18$"$. The value of $\tau_{850\mu\rm m}$ is therefore a result of the fitting. In the case of NGC 1333 we used the dust temperature and dust optical depth images provided by \citet{Zari2016} following a similar procedure.

\begin{figure*}
    \centering
    \begin{subfigure}[b]{0.49\textwidth}\includegraphics[width=\textwidth,keepaspectratio]{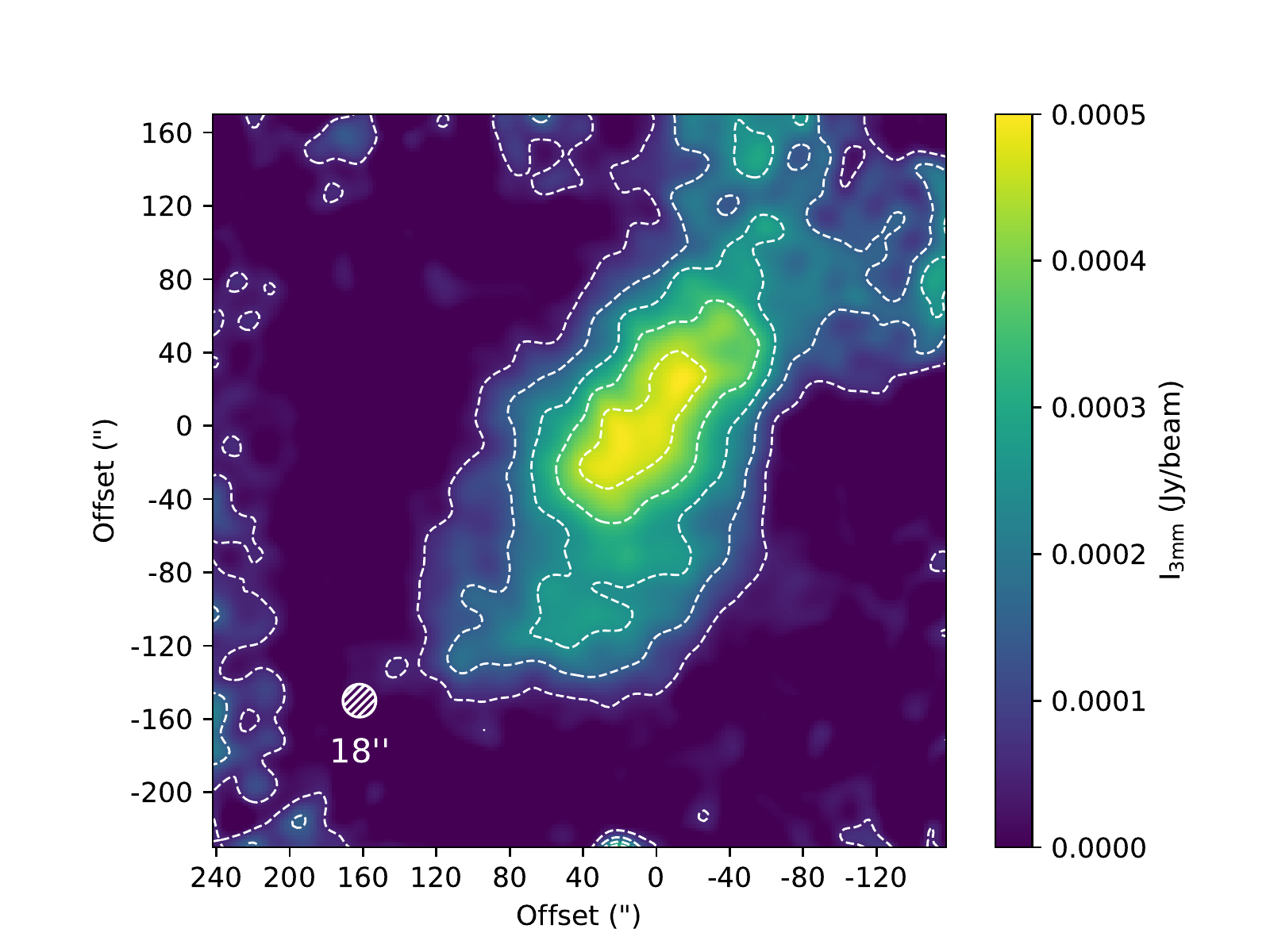}
    \end{subfigure}
    \begin{subfigure}[b]{0.49\textwidth}\includegraphics[width=\textwidth,keepaspectratio]{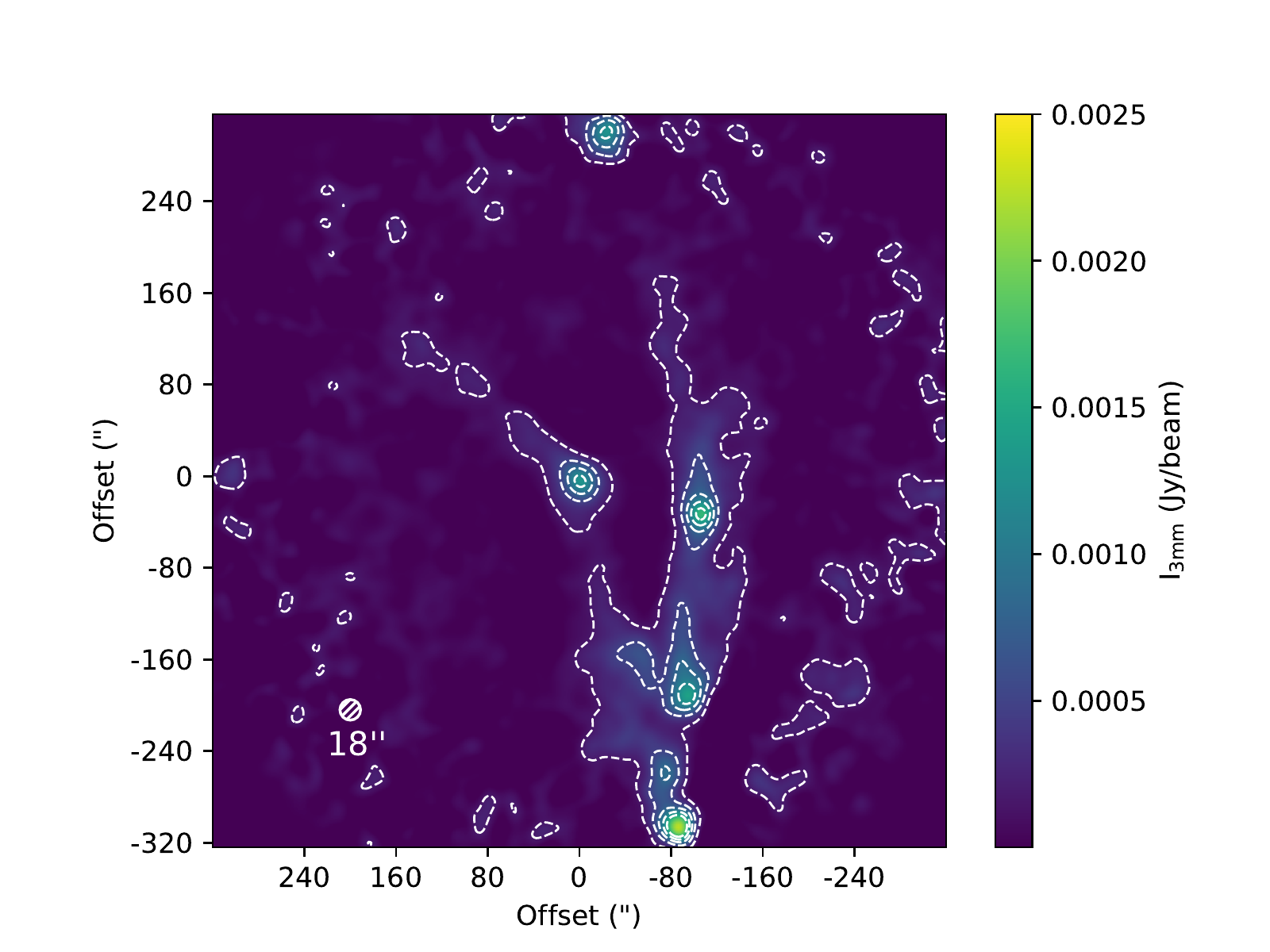}
    \end{subfigure}
		    \caption{Continuum flux maps at 3 mm obtained with the MUSTANG-2 bolometer used to compute the dust emissivity spectral index $\beta_{0.85mm-3mm}$. {\it Left panel:} Flux at 3 mm in the TMC 1-C prestellar core. Contours correspond to levels $5\sigma\times[1,3,5,7,9]$, with $\sigma = 10^{-5}$ Jy/beam. {\it Right panel:} Flux at 3 mm in the NGC 1333-C7 sector. Contours correspond to levels $5\sigma\times[1,3,5,7,9]$, with $\sigma = 3.4\times 10^{-5}$ Jy/beam. The fluxes were obtained using the MUSTANG-2 bolometer of the Green Bank Telescope. The origin (0,0) of these maps is the same as in Figure \ref{fig:mapsTMC1NGC1333}.}
		    \label{fig:mapsOpacity3mm}
\end{figure*}
\begin{figure*}
    \centering
    \begin{subfigure}[b]{0.495\textwidth}\includegraphics[width=\textwidth,keepaspectratio]{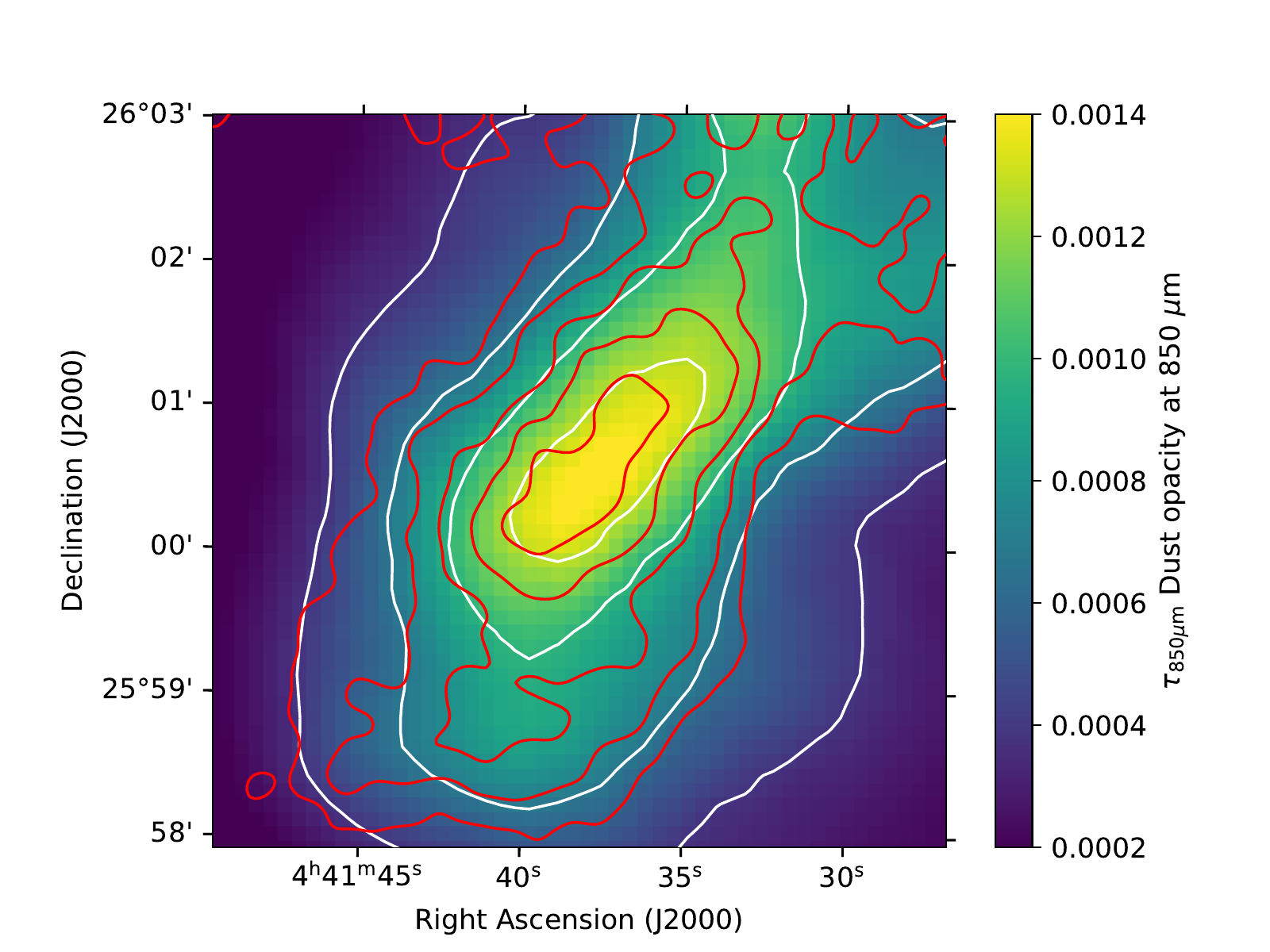}
    \end{subfigure}
    \begin{subfigure}[b]{0.495\textwidth}\includegraphics[width=\textwidth,keepaspectratio]{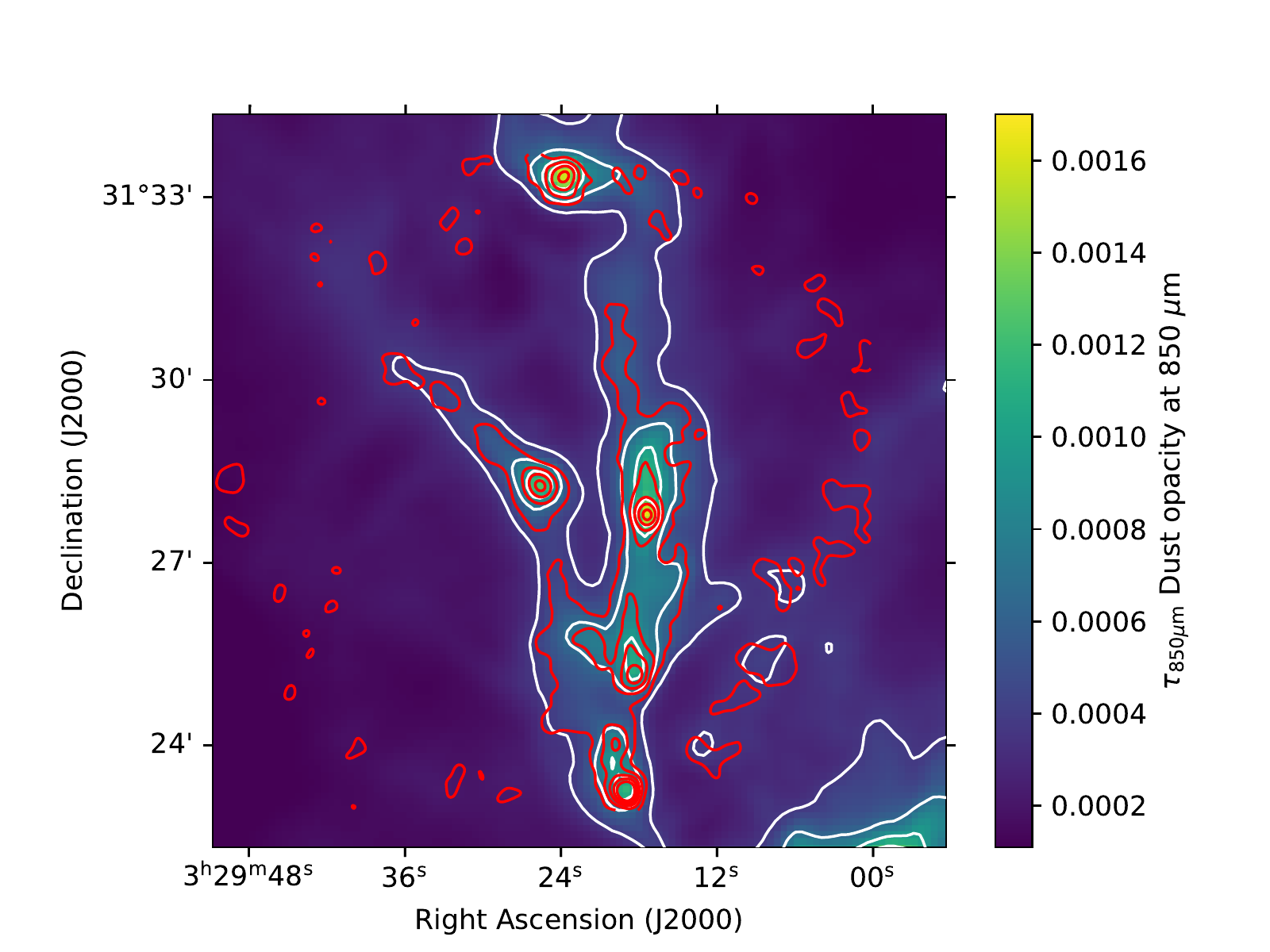}
    \end{subfigure}
		    \caption{Overlay of the 3 mm contours from Figure \ref{fig:mapsOpacity3mm} over the maps and contours from Figure \ref{fig:mapsOpacity850}. {\it Left panel:} TMC 1-C. {\it Right panel:} northern sector of NGC 1333.}
		    \label{fig:overlay}
\end{figure*}

With the MUSTANG-2 instrument, we observed the continuum emission intensity at 3 mm in TMC 1-C and NGC~1333-C7 with an angular resolution of $\sim 9-10"$. We convolved the 3 mm maps to match the angular resolution of the T$_d$ and $\tau_{850\mu\rm m}$ maps derived from the {\em Herschel} data (see Figure \ref{fig:mapsOpacity3mm}). Introducing the dust temperature, $T_{d}$, in the modified blackbody model (Eq. \ref{eq:modBlackbody}), we obtain the dust optical depth at 3 mm. The resulting values are used to derive the dust emissivity spectral index, $\beta_{0.85mm-3mm}$, using Eq. \ref{eq:beta} and the values of $\tau_{850\mu\rm m}$. The obtained values range between $\beta_{0.85mm-3mm}\sim 1.7$ and 2.1, as shown in Table~\ref{tab:continuumData}. We are aware that these values might be upper limits to the real value of $\beta_{0.85mm-3mm}$ if the 3mm continuum emission have an extended component in spatial scales larger than $\sim$ 4$'$, which can be filtered by MUSTANG-2 observations. In Figure \ref{fig:overlay}, we plotted the overlay between the continuum maps at 850 $\mu$m and the 3 mm contours from Figure \ref{fig:mapsOpacity3mm} to assess the impact of extended emission filtering in our measurements. The morphology of the MUSTANG-2 map towards NGC~1333 is similar to that of the $\tau_{850\mu\rm m}$ maps, taking into account the difference in sensitivity. Therefore, despite the expectation of a loss of large-scale emission, strong filtering effects ($>25 \%$) are not supposed to affect the 3 mm fluxes measured toward the dense core positions in NGC 1333. However, TMC 1-C, due to its proximity, displays a more extended emission that might be filtered. Comparing the ratio between the extended and peak values in the  850 $\mu$m image, and assuming that the flux at 3 mm is proportional to the N(H$_2$) (i.e., neglecting any effect due to changes in the dust temperature) we estimated that the filtered flux is $<40 \%$ at the peak emission in TMC~1-C. This means that $\beta_{0.85mm-3mm}\sim1.7-2.1$ in this prestellar core.

We would generally expect that the dust emissivity spectral index decreases towards the center of the cores due to the larger sizes of the grains in the densest regions. In order to check this possibility, we used the continuum emission at 850 $\mu$m detected with SCUBA on the JCMT \citep{Hatchell2005} to derive  $\beta_{0.85mm-3mm}$ towards the  NGC~1333 cores at a higher angular resolution of 14$"$. For this calculation, we convolved the MUSTANG-2 images to match angular resolution of JCMT 850 $\mu m$ observations and assumed the dust temperature obtained from the {\em Herschel} maps, as before. The values of $\beta_{0.85mm-850 \mu m}$ thus obtained are shown in Table~\ref{tab:continuumData_2}. The values of $\beta_{0.85mm-3mm}$ at 14$"$ are fully consistent with  those derived from the {\em Herschel} products towards the Class 0 object NGC 1333-C7-1 and the starless core NGC 1333-C7-3. However, we obtained lower values of $\beta_{0.85mm-3mm}$ towards the Class 0/I sources NGC~1333-C7-2 and NGC~1333-C7-4, which is in line with the interpretation of grain growth in the densest regions of these protostars. It is only towards NGC 1333-C7-5 that the value of $\beta_{0.85mm-3mm}$ at 14$"$ is higher than that obtained at 18$"$. It should be noted that this core is at the edge of the MUSTANG-2 field of view which could introduce uncertainties once the image is convolved at lower angular resolutions. One main result is that the trend found in $\beta_{0.85mm-3mm}$ among the sampled cores in NGC~1333 using {\em Herschel} data remains the same using the JCMT fluxes, which reinforces the existence of an evolutionary sequence in these cores, with the lowest value of $\beta$, namely, the largest grain sizes towards the Class I protostar NGC~1333-C7-2.

The value of $\beta_{0.85mm-3mm} \sim 1.7 - 2.1$ estimated using our data towards TMC 1-C is consistent with that derived by \citet{Schnee2010} using maps at 160, 450, 850, 1200 and 2100 $\mu$m at an angular resolution of 14$"$.  These authors proposed that the value of $\beta$ remains close to $\sim 2$, between 1.5 $-$ 2.5, across this starless core.  Uncertainties in the continuum fluxes and dust temperature across the core make a more reliable estimate of $\beta$ difficult.

\begin{table*}
	        \centering
	        \caption{Continuum 3 mm flux, dust temperature, optical depth at 850 $\mu$m and 3 mm, and the spectral index.}
            \label{tab:continuumData}
	        \begin{tabular}{lccccccccc}
		        \toprule
		        \multirow{2}{*}{{Position}} & $T_{\rm d}$ & Flux F${_{3 \rm mm}}^{(a)}$ & $\tau_{3\rm mm}\times 10^4$ & $\tau_{850\mu\rm m}\times 10^3$ & $\beta_{0.85mm-3mm}$  \\
		        & (K) & (mJy/beam $18''$) & (18$''$)$^{(a)}$ & (18$''$)$^{(b)}$ & (18$''$) & \\\midrule
				TMC 1-C & 11 & $0.49\pm 0.03$ & $0.82\pm 0.02$ & $1.43\pm 0.14$ & $2.09\pm 0.08$  \\
				NGC1333-C7-1& 17 & $1.29\pm 0.04$ & $1.35\pm 0.04$ & $1.26\pm 0.13$ & $1.63\pm 0.09$ \\
				NGC1333-C7-2 & 21 & $2.23\pm 0.04$ & $1.85\pm 0.03$ & $1.15\pm 0.12$ & $1.34\pm 0.08$\\
				NGC1333-C7-3 & 20 & $1.44\pm 0.04$ & $1.26\pm 0.04$ & $1.09\pm 0.11$ & $1.56\pm 0.09$\\
				NGC1333-C7-4 & 18 & $1.68\pm 0.04$ & $1.65\pm 0.04$ & $1.64\pm 0.16$ & $1.68\pm 0.09$\\
				NGC1333-C7-5 & 15 & $1.27\pm 0.04$ & $1.54\pm 0.05$ & $1.57\pm 0.16$ & $1.70\pm 0.09$\\
    		    \bottomrule
	        \end{tabular}
	        \flushleft
			    {\small
			\ \ \ \ $^{{(\rm a)}}$ Flux and optical depth derived from the MUSTANG-2 data after smoothing to achieve a resolution of 18$''$.\\
			\ \ \ \ $^{{(\rm b)}}$ Optical depth from the NGC 1333 maps in \citet{Zari2016}.\\
			}
\end{table*}

\begin{table*}
	        \centering
	        \caption{Flux, optical depth, and spectral index in NGC 1333 at different frequencies and resolutions from the literature.}
            \label{tab:continuumData_2}
	        \begin{tabular}{lcccccccccc}
		        \toprule
		        \multirow{2}{*}{{Position}} & $T_{\rm d}$ & Flux F${_{3 \rm mm}}^{(a)}$ & $\tau_{3\rm mm}\times 10^4$ & Flux F${_{850 \mu\rm m}}^{(b)}$ & $\tau_{850\mu\rm m}\times 10^3$ & $\beta_{0.85mm-3mm}$ \\
		        & (K) & (mJy/beam $14''$) & (14$''$)$^{(a)}$ & (mJy/beam $14''$) & (14$''$)$^{(b)}$ & (14$''$)\\\midrule
				NGC1333-C7-1& 17 & $1.62\pm 0.05$ & $1.70\pm 0.05$ & $292\pm 35$ & $1.53\pm 0.18$ & $1.61\pm 0.08$\\
				NGC1333-C7-2 & 21 & $2.82\pm 0.05$ & $2.33\pm 0.05$ & $308\pm 35$ & $1.17\pm 0.14$ & $1.18\pm 0.08$ \\
				NGC1333-C7-3 & 20 & $1.72\pm 0.05$ & $1.50\pm 0.05$ & $376\pm 35$ & $1.54\pm 0.18$ & $1.70\pm 0.08$ \\
				NGC1333-C7-4 & 18 & $2.16\pm 0.05$ & $2.12\pm 0.05$ & $349\pm 35$ & $1.67\pm 0.20$ & $1.51\pm 0.08$ \\
				NGC1333-C7-5 & 15 & $1.46\pm 0.05$ & $1.78\pm 0.05$ & $371\pm 35$ & $2.38\pm 0.29$ & $1.90\pm 0.08$ \\
    		    \bottomrule
	        \end{tabular}
	        \flushleft
			    {\small
			\ \ \ \ $^{{(\rm a)}}$ Flux and optical depth derived from the MUSTANG-2 data after smoothing to achieve a resolution of 14$''$.\\
			\ \ \ \ $^{{(\rm b)}}$ Flux and optical depth from \citet{Hatchell2005}.\\
			}
\end{table*}

\begin{figure*}
    \centering
    \begin{subfigure}[b]{0.49\textwidth}\includegraphics[width=\textwidth,keepaspectratio]{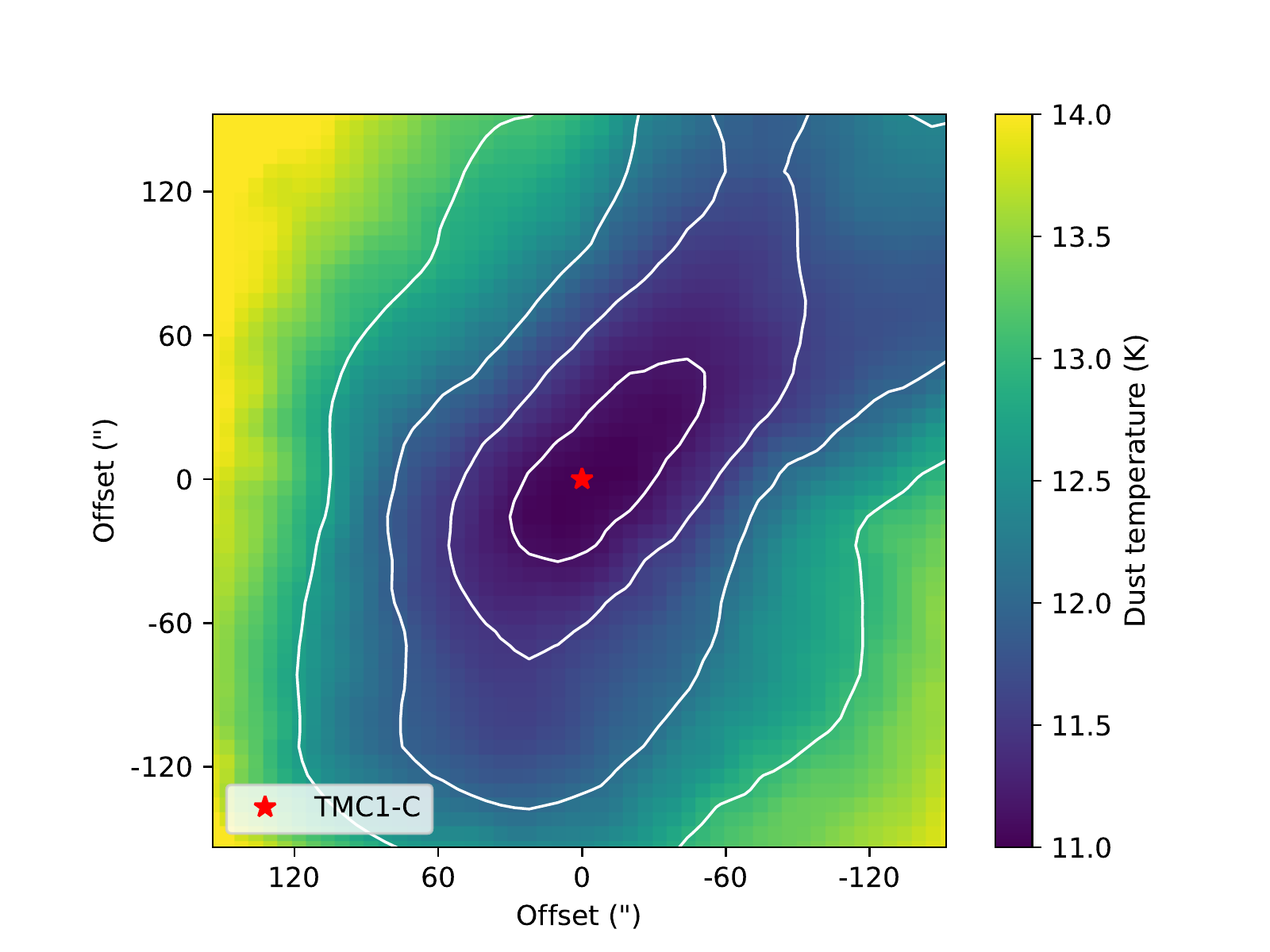}
    \end{subfigure}
    \begin{subfigure}[b]{0.49\textwidth}\includegraphics[width=\textwidth,keepaspectratio]{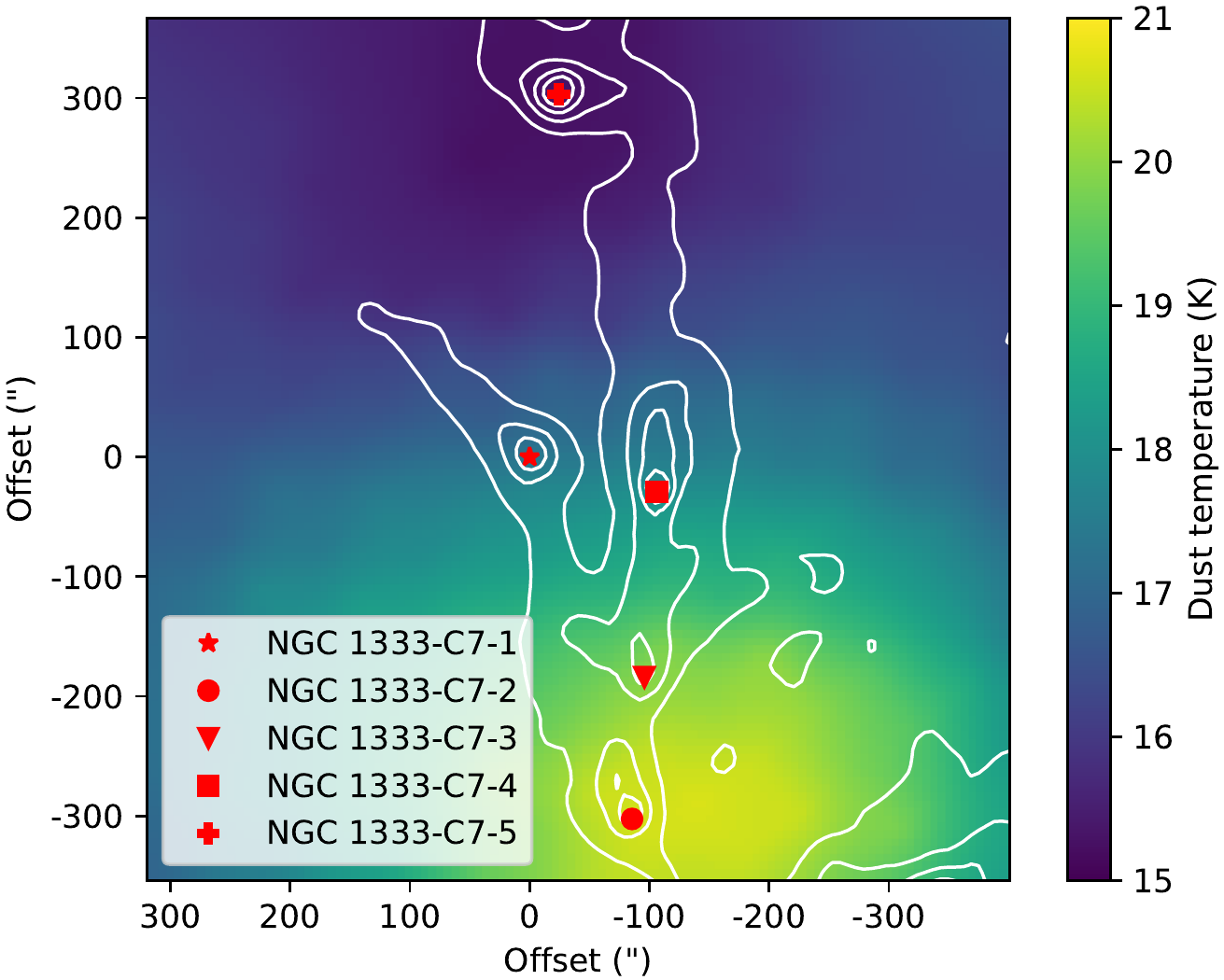}
    \end{subfigure}
		    \caption{Dust temperature resulting from SED fitting (Kirk et al. in prep.) in TMC 1-C (left panel) and NGC 1333-C7 (right panel). The contours are the same as in Figure \ref{fig:mapsOpacity850}.}
		    \label{fig:dustTemp}
\end{figure*}

\section{Discussion}

So far, we analyzed the chemical composition and the continuum emission from different positions located in two star-forming regions that feature two different formation regimes: isolated, low-mass star formation in TMC 1-C, and clustered, intermediate-mass star formation in NGC 1333. Our goal is studying the evolutionary stage of the different positions and investigate the effects of potential stellar feedback on the chemistry and evolution of these objects. 

\subsection{Isotopic and isomeric HCN ratios as chemical clocks}

As discussed above, deuterium fraction is enhanced in the cold and dense conditions of prestellar and starless cores, increasing by around three to four orders of magnitude. This is due to the fact that the exothermic reaction in Eq. \ref{eq:deuteration} is highly favored at these low temperatures. As such, the deuterium fraction and, in particular, the deuterium fraction of HCN \citep{Fontani2015}, are tightly related to the gas temperature. The HCN/HNC ratio is known to strongly depend on the kinetic temperature \citep{Schilke1992} and therefore has been used as a chemical thermometer \citep{Hacar2020} to explore the physical and chemical conditions of star-forming regions. The isotopic and isomeric ratios of nitriles are also dependent on the chemical time and are potential tracers of the chemical evolution of starless cores and cold protostellar envelopes \citep{Roueff2015}. The H$^{12}$CN/H$^{13}$CN and  HN$^{12}$C/HN$^{13}$C ratio may reach values of around two times larger than the canonical $^{12}$C/$^{13}$C isotopic ratios in evolved ($t\sim1$ Myr) cores. Similarly, the $^{14}$N/$^{15}$N, and $^{13}$C $^{14}$N/$^{12}$C $^{15}$N ratios also present significant variation (larger than a factor of 2) with the time evolution, providing a tool to determine the chemical age of the dense gas.

\begin{figure*}
    \centering
    \includegraphics[width=\textwidth,keepaspectratio]{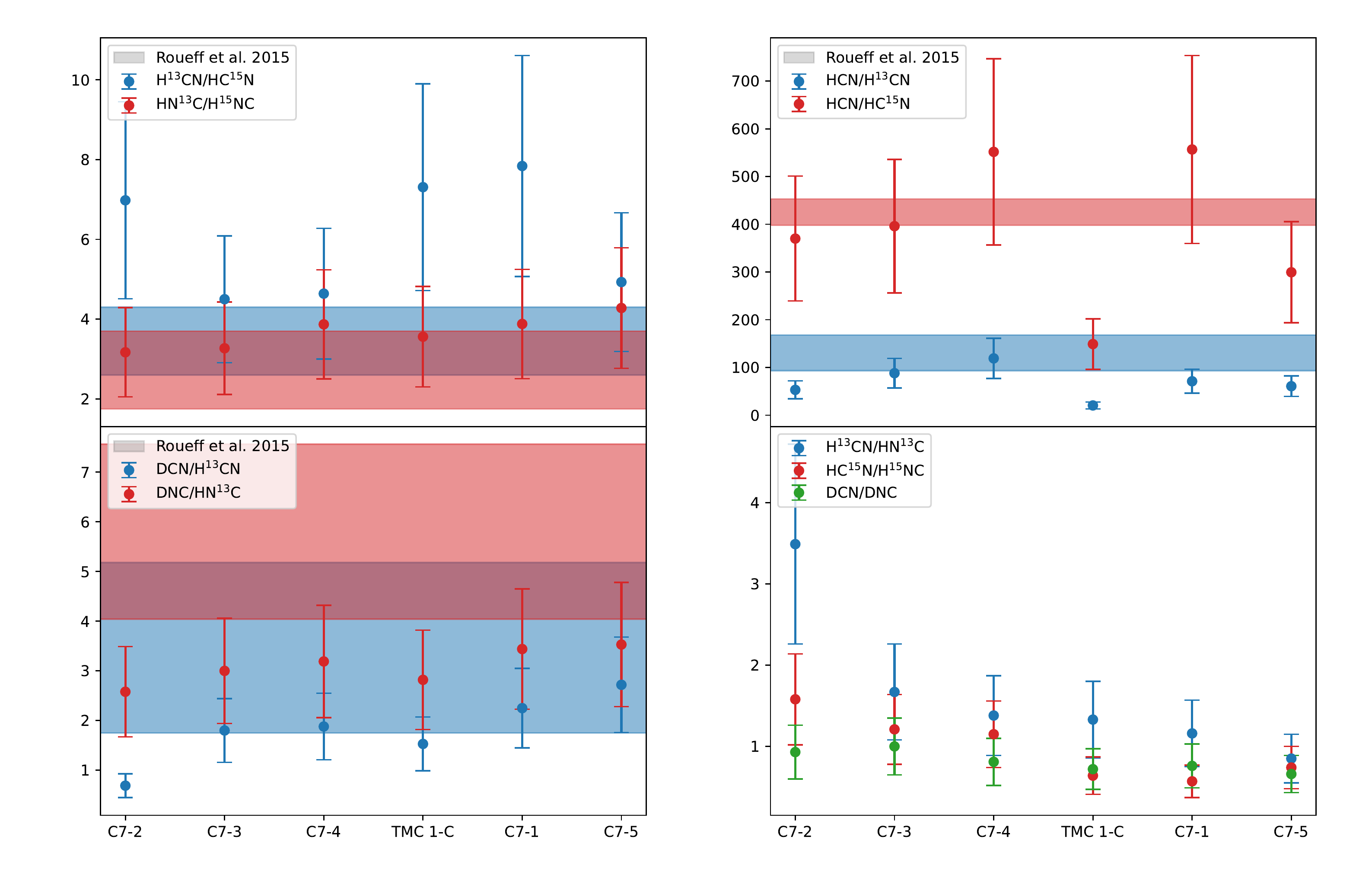}
    \caption{Blue, red, and green dots represent the different observed molecular ratios across the positions analyzed in this paper. The horizontal bands span from the minimum to the maximum values for the ratios obtained in the models of \citet{Roueff2015}. The sources are ordered by decreasing dust temperature.}
	\label{fig:ratios}
\end{figure*}

From millimeter observations and using the collisional coefficients presented in this paper, we carried out a chemical characterization of HCN, HNC, and isotopologues, computing the column density of these species, the $^{13}$C/$^{12}$C and $^{15}$N/$^{14}$N isotopic ratios, the HCN and HNC deuterium fractions, and the isomeric ratio HCN/HNC. The ratios involving the most abundant isotopologues, specially HNC, are not reliable due to high optical depths. Figure \ref{fig:ratios} shows the H$^{13}$CN/ HC$^{15}$N, HN$^{13}$C/ H$^{15}$NC, H$^{13}$CN/ HN$^{13}$C, HC$^{15}$N/H$^{15}$NC, DNC/HN$^{13}$C, and DCN/H$^{13}$CN ratios. We found a differentiated behavior between the two isomers. The isotopic ratios involving HNC, the HN$^{13}$C/H$^{15}$NC, and DNC/HN$^{13}$C present uniform values across the sample. However, we detected variations in those of HCN, in particular DCN/H$^{13}$CN, H$^{13}$CN/HN$^{13}$C, and HC$^{15}$N/H$^{15}$NC show systematic changes with core evolution. In the following, we discuss these ratios in more detail.

The HN$^{13}$C/H$^{15}$NC ratio present values of 3-4 along our sample without any trend of variation among the different cores within the uncertainties of our estimates. Using a time-dependent gas-phase chemical model, \citet{Roueff2015} predicted the evolution of the nitriles and isonitriles isotopic ratios for the typical conditions of a dense cold core. The varying hydrogen number density and initial chemical abundances of these models were set to cover typical properties for dense cores. The resulting isotopic ratios were given at the steady state and at an age of 1 Myr. In Figure \ref{fig:ratios}, we compared our results with the range of ratios these models provided. The derived isotopic H$^{13}$CN/HC$^{15}$N ratio is in good agreement with these chemical model predictions. The values of H$^{13}$CN/HC$^{15}$N ratio ranges from $\sim$4 to $\sim$7. In each core, the H$^{13}$CN/H$^{15}$NC ratio is systematically higher than the  HN$^{13}$C/H$^{15}$NC one. Moreover, it is systematically higher than the values predicted by \citet{Roueff2015}.

As previously commented, the HCN/HNC isomeric ratio has been used as a proxy of the kinetic temperature. The H$^{13}$CN/HN$^{13}$C ratio we computed across our sample shows a trend that can be interpreted as a progressive increase in the average gas temperature from the coldest core starless cores NGC1333-C7-5 and TMC 1-C to the Class I source NGC1333-C7-2. This trend is also observed in the HC$^{15}$N/H$^{15}$NC ratio, which demonstrates that is not affected by strong optical depth effects. However, the value of the  HC$^{15}$N/H$^{15}$NC ratio is systematically lower than that of the  H$^{13}$CN/HN$^{13}$C one, suggesting that these isotopologues are tracing different regions within the prestellar or proto-stellar envelopes.

\begin{figure}
    \centering
    \includegraphics[width=0.48\textwidth,keepaspectratio]{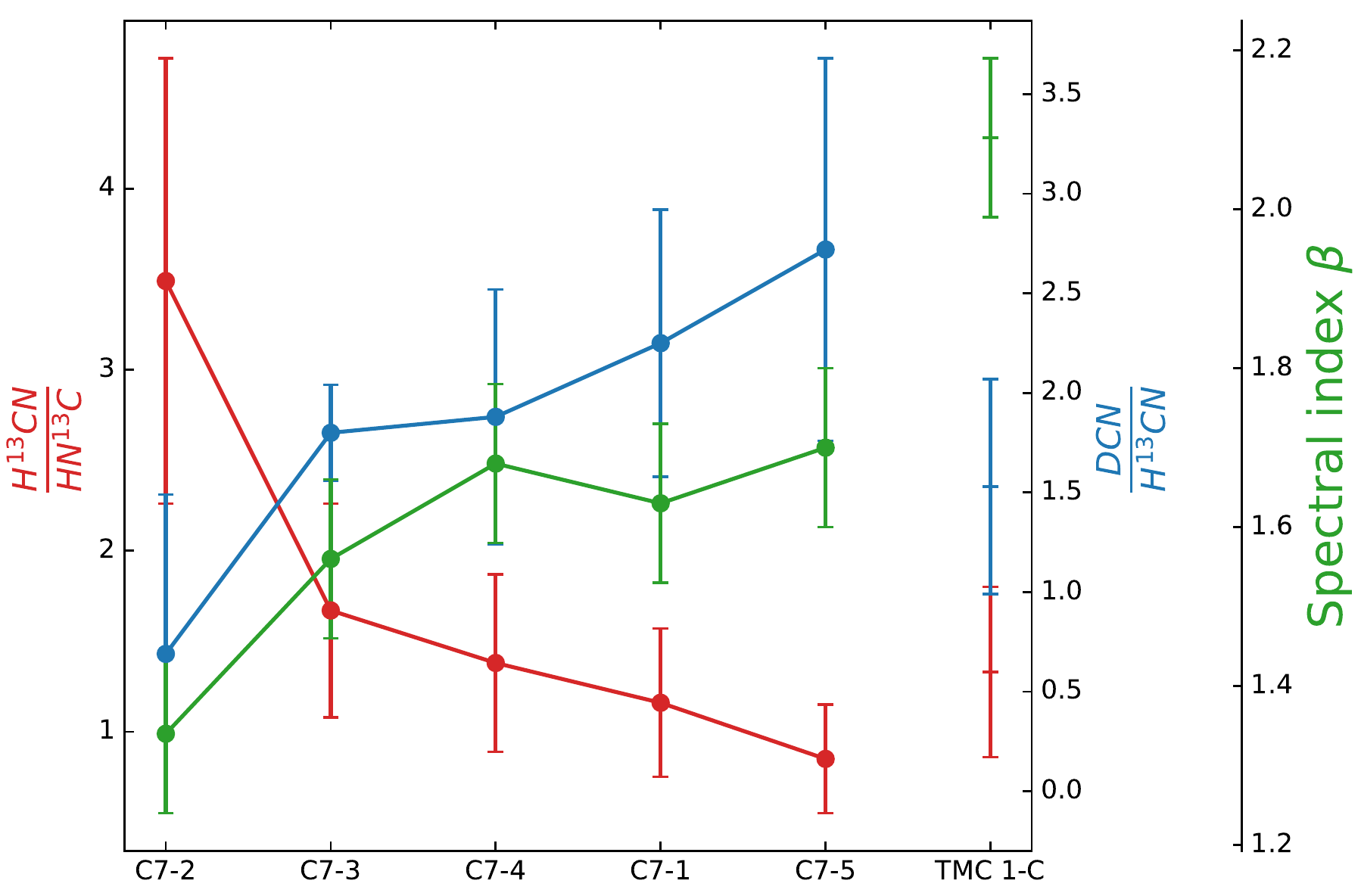}
    \caption{Values of the H$^{13}$CN/HN$^{13}$C (red) and DCN/H$^{13}$CN (blue) ratios, and the spectral index (green) across the sources of our sample.}
	\label{fig:trends1}
\end{figure}

The deuterium fraction of HCN is, however, an evolutionary tracer past the prestellar phase since it decreases as heating from the nascent star increases the gas and dust temperature. Following this reasoning, the HCN/HNC ratio and the deuterium fraction are expected to be anticorrelated. This anticorrelation is present across our dataset. In Figure \ref{fig:trends1}, we plotted the H$^{13}$CN/HN$^{13}$C and DCN/H$^{13}$CN ratios at the observed positions, where we see that the H$^{13}$CN/HN$^{13}$C is indeed anticorrelated with the deuterium fraction DCN/H$^{13}$CN. As the H$^{13}$CN/HN$^{13}$C ratio approaches 1, the gas temperature is expected to decrease, promoting the formation of deuterated compounds at the expense of their hydrogenated counterparts. The H$^{13}$CN/HN$^{13}$C ratio is, moreover, correlated with the dust temperature (see Table \ref{tab:continuumData}). Interestingly, the deuterium fraction of HNC remains quite uniform across the sample. Furthermore, the deuterium fraction of HNC is always higher than that of HCN. Opacity effects are expected to be more important in the HNC isotopologues than in those of HCN and could produce an underestimation of the HNC deuterium fraction towards the cores, TMC~1, NGC~1333-C7-1, and  NGC~1333-C7-5 (see Table~\ref{tab:lineResultsTMC1C}, \ref{tab:lineResultsNGC1333_C7_1},  and \ref{tab:lineResultsNGC1333_C7_5}). It should be noted that DN$^{13}$C has been detected in these objects. This could explain the absence  of variation  of the HNC deuterium fraction with the evolutionary stage. However, opacity effects are not expected to play a role in   NGC~1333-C7-2 and  NGC~1333-C7-3, where the difference in deuterium fraction between the two isomers remains. Gradients in the HCN/HNC ratio along the line of sight towards these cores are needed to be invoked to account for these differences.

The deuterium fraction has also been used as a chemical clock of prestellar and starless cores \citep[see, e.g.,][]{Pagani2013, Kong2015, Bianchi2017} so that evolved cores feature a higher deuterium fraction as a consequence of Eq. \ref{eq:deuteration}. However, this statement only applies to prestellar and starless cores. As the core collapses and evolves into later stages, heating from the incipient star or its outflow activity reduce the deuterium fraction of gas-phase molecules like N$_{2}$H$^{+}$ \citep{Emprechtinger2009}. Instead, high abundances of singly, doubly or even triply deuterated molecules have been measured \citep[see, e.g.,][]{Ceccarelli1998, Ceccarelli2007} as a consequence of the sublimation of the inherited chemical content present in icy mantles. The high deuterium fraction observed in hot cores and corinos is therefore found in molecules that are efficiently formed on grain surfaces like CH$_{3}$OH, H$_{2}$CO, H$_{2}$S, or H$_{2}$CS \citep[see, e.g.,][]{Ceccarelli1998, Vastel2003, Parise2004, Marcelino2005, Parise2006}. Our study shows that the deuterium fraction of HCN increases in the prestellar and starless phase and would decrease, past the prestellar borderline, due to increasing temperatures. This trend is not evident in the case of HNC due to the opacity effects and the HCN/HNC gradients along the core. This suggests that HCN is a better tracer of the protostellar evolution. To better assess the evolutionary stage of the targets in our sample, together with the chemical information we obtained, we also need hints on the grain size of the object.

\subsection{Evolutionary stages: Grain growth}

Grain growth is considered as an evolutionary tracer since it has been observed in dense, evolved areas inside molecular clouds \citep[see, e.g.,][]{Weingartner2001, Pagani2010, Steinacker2010, Miotello2014, Wong2016}. Grain growth has an impact on the SED, measurable via the dust emissivity spectral index $\beta$, the slope of such a distribution between two different wavelengths. We used our 3mm continuum images obtained with MUSTANG-2 to investigate possible variations of $\beta$ within our sample of two starless cores, three Class 0 and  one Class I objects. Our results (Tables \ref{tab:continuumData} and  \ref{tab:continuumData_2}) show, within the uncertainties, little variations of the spectral index among the starless cores and Class 0 objects, with values of $\beta_{0.85mm-3mm}$ ranging from $\sim$1.5 to $\sim$2.1 (see Figure \ref{fig:trends1}). However, the Class I object NGC1333-C7-2 presents the lower spectral index, $\beta_{0.85mm-3mm} \sim 1.1 - 1.3$, consistent with the presence of large grains in its protostellar disk. The low spectral index found in NGC1333-C7-3, comparable to those of the Class 0 objects present in our sample, suggests again that this source might be more evolved than a starless core.

These values are consistent with those derived in previous works. \citet{Shirley2011} determined $\beta$ towards a sample of Class 0 and I protostars using dust continuum emission images at 0.862 mm, 1.25 mm, and MUSTANG images, deriving values of $\beta$ ranging from 1.0 to 2.4, with the lowest values for the most evolved Class I sources.
\citet{Bracco2017} mapped a chain of 4 prestellar and Class 0/I protostellar cores in the B 213 filament using the NIKA camera on the IRAM 30m telescope. Based on these observations, they estimated values of $\beta_{1mm-2mm}$ between $\sim$0.6 and $\sim$1.1  for the protostellar cores, and $\sim$2 for prestellar ones using a 1D dust emission model. \citet{Chacon2017, Chacon2019} imaged the prototypical prestellar core L1544 using the NIKA camera at 1mm and 2mm. They derived values of $\beta_{1mm-2mm}$ ranging from $\sim$2 in the densest inner region to $\sim$2.4 in the outer layers. 

Putting all together, we can conclude that the value of $\beta$ at millimeter wavelengths is useful to discern starless cores and Class 0 objects from Class I protostars, pointing to the existence of grain growth in the Class I protostellar disks. The large uncertainties of our values of $\beta$ in starless cores and Class 0 objects make the detection of evidences of grain growth in these early stages of star formation difficult and therefore gas chemistry would be a better tool to date these objects. An important source of uncertainty in our calculations is the loss of large spatial scale emission. The MUSTANG-2 instrument has a field-of-view of $4.25’$ (Section 4.3) and is therefore not sensitive to detect larger scales of the two regions presented here. In addition, the observed MUSTANG-2 maps are only $5’$ in diameter (Section 4.3). The Herschel maps and the resulting map of optical extinctions (Figure \ref{fig:mapsTMC1NGC1333}) show that these regions extend far beyond a few arcminutes. This leads to a loss of the larger spatial scales beyond about $4’$, as confirmed by a detailed analysis of the MUSTANG-2 transfer function (Romero, priv. comm.). The SCUBA maps do probably suffer from a similar issue. These caveats will have an impact on the dust emissivities index $\beta$ estimated here. But a detailed discussion is beyond the scope of the present paper.

\subsection{Chemistry and large-scale feedback}

We investigated the H$^{13}$CN/HN$^{13}$C ratio, a proxy of the gas temperature, across our sample and how it affects other chemical quantities like the deuterium fraction, finding that they are anticorrelated. Dust temperature follows the same trend, as in the case of the gas temperature. In Figure \ref{fig:dustTemp}, we show the spatial distribution of the dust temperature in NGC 1333, which is increasing from north to south. This is most likely due to the heating produced by the active star forming region located in the central part of this molecular cloud. Indeed, the feedback of the main NGC 1333 cluster has been the subject of several works in literature. Extensive outflows and jets have been described in this region that even escape the cluster core \citep{Bally2001, Walawender2008}, playing a fundamental role in its energetic balance \citep{Sandell2001}. This suggests that at this resolution, cloud-scale and environmental properties influences the chemistry of this region. 

On the other hand, there is a south-to-north gradient in dust size and evolutionary stage of the objects considered in this sample, with later classes (NGC1333-C7-2 as a Class I object) located toward the south, closer to the central protostellar cluster, and the youngest classes (starless and Class 0 objects) towards the north of the region. This trend in the evolutionary stage indicate that star formation in the northern sector of NGC~1333 could be regulated or induced by the stellar feedback coming from the main cluster in the south. Evidence of outflow activity triggering star formation in this region was found by, for instance, \citet{Sandell2001}. NGC 1333 has been documented to host a large amount of Herbig-Haro objects associated with cavities revealing interactions between outflows and surrounding gas \citep[see, e.g.,][]{Bachiller1990}. These injections of new material may trigger core collapse, inducing star formation \citep[see, e.g.,][]{Chen2016, Dhabal2019, DeSimone2022}. Since the feedback of the active protostellar cluster would be producing both the heating of the interstellar material and the induced star formation, it is difficult to discern to what extent the environment is responsible for determining the isotopic and isomeric ratios measured in the targeted NGC~1333 cores. A larger sample, with dozens of cores located in different environments, would be required to discern the effects of the environment on the chemistry of starless cores and YSOs.


\section{Summary and conclusions}
\label{sec:summ}

We performed a chemical study of two different star-forming regions with updated collisional rates to investigate their chemical and evolutionary characteristics:

\begin{itemize}
    \item We present new collisional coefficients of HCN, HNC, and their C, N, and H isotopologues that provided the most up-to-date information about the chemistry of these compounds, offering an updated view of the isotopic ratios, deuterium fraction, isomeric ratios, and evolution. We used these new collisional coefficients to study the chemistry of two star-forming regions with different properties in terms of the stellar masses and the presence of stellar feedback from nearby sources: 1) TMC 1-C is an isolated prestellar core in the TMC 1 cloud in Taurus, a low-mass star forming region; 2) and five positions in the northern sector of NGC 1333, a proto-cluster located in Perseus, an intermediate-mass star-forming region.
    \item Using millimeter observations and the collisional rates presented in this paper, we carried out a chemical characterization of HCN, HNC, and their isotopologues, computing the column density of these species, the $^{13}$C/$^{12}$C and $^{15}$N/$^{14}$N isotopic ratios, the HCN and HNC deuterium fractions, and isomeric ratios HCN/HNC. The isotopic ratios between the least abundant species show good agreement with current chemical models \citep{Roueff2015}, with $^{13}$CN/C$^{15}$N $\sim 4.50-7.84$ and N$^{13}$C/$^{15}$NC $\sim 3.06-4.28$. The most abundant isotopic species display unreliable ratios due to high optical depths, especially in the case of TMC 1-C, where C/$^{13}$C $\sim 20$ and N/$^{15}$N $\sim 150$. The ratios between the two isomers show different variability across our sample, suggesting a different origin of their emission.
    \item The deuterium fraction was calculated to study the evolutionary stage of the members of our sample. We found that the deuterium fraction of HCN, measured as the column density ratio DCN/H$^{13}$CN, is mainly driven by the reaction in Eq. \ref{eq:deuteration}, since it is anticorrelated with dust temperature and the H$^{13}$CN/HN$^{13}$C ratio, a proxy of the kinetic temperature. The deuterium fraction of HCN is an evolutionary tracer past the prestellar phase since it decreases as heating from the nascent star increases the gas and dust temperature. Therefore, the H$^{13}$CN/HN$^{13}$C and DCN/H$^{13}$CN ratios behave as chemical clocks in our sample.
    \item We analyzed the continuum emission of our sample at two different wavelengths, 850 $\mu$m and 3 mm, to estimate the spectral index. Variations in the spectral index are thought to be a consequence of grain growth, another evolutionary tracer. The dust spectral index $\beta_{0.85mm-3mm}$ is found to be as low as $\beta_{0.85mm-3mm}=1.34$ for the most evolved member of our sample, and rises up to $\beta_{0.85mm-3mm}=2.09$ in TMC 1-C, a prestellar core. Combining the dust spectral index $\beta_{0.85mm-3mm}$ with the chemical information previously obtained, we find a tentative correlation between the spectral index and the HCN deuterium fraction. This is consistent with the presence of grain growth in evolved objects past the prestellar phase as the dust and gas temperatures increase, thus reducing the HCN deuterium fraction.
    \item At the resolution of our data, the NGC 1333 region shows a south-to-north gradient in the spectral index, spectral class, and dust temperature. The cloud-scale properties turns out to be a key driver of the chemistry in this region in contrast with TMC 1-C. This gradient suggests that the formation of the sources present in our sample may be induced by cloud-scale feedback coming from the main NGC 1333 cloud at the south of the observed region. High-resolution and sensitivity observations are needed to further study the effect of the stellar feedback and the environment in the star formation process in clusters.
\end{itemize}

The new collisional coefficients presented in this paper allowed us to perform the most up-to-date chemical characterization of cores from two star-forming regions based on the set of molecules that includes HCN, HNC, and their isotopologues. Quantities of great astrochemical interest such as the isotopic fractions, deuterium fractions, and isomer ratios were derived and compared to other results from the literature. The study of the dust continuum emission helped us find connections between the nature of the dust emission, the chemical processes occurring in the areas under consideration, the evolution of YSOs, and the effects of feedback from nearby star formation.

\begin{acknowledgements}

    DNA acknowledges funding support from the Ram\'on Areces Foundation through their international postdoc grant program. We thank the Spanish MICINN for funding support from PID2019-106235GB-I00. We also acknowledge financial support from the European Research Council (Consolidator Grant COLLEXISM, Grant Agreement No. 811363). F. Lique acknowledges financial support from the Institut Universitaire de France and the Programme National `Physique et Chimie du Milieu Interstellaire' (PCMI) of INSU-CNRS with INC/INP cofunded by CEA and CNES.

\end{acknowledgements}



\bibliography{colCoeff.bib}

\begin{thebibliography}{153}
\expandafter\ifx\csname natexlab\endcsname\relax\def\natexlab#1{#1}\fi

\bibitem[{{Aalto} {et~al.}(2012){Aalto}, {Garcia-Burillo}, {Muller}, {Winters},
  {van der Werf}, {Henkel}, {Costagliola}, \& {Neri}}]{Aalto2012}
{Aalto}, S., {Garcia-Burillo}, S., {Muller}, S., {et~al.} 2012, \aap, 537, A44

\bibitem[{{Aalto} {et~al.}(2007){Aalto}, {Monje}, \& {Mart{\'\i}n}}]{Aalto2007}
{Aalto}, S., {Monje}, R., \& {Mart{\'\i}n}, S. 2007, \aap, 475, 479

\bibitem[{{Ag{\'u}ndez} \& {Wakelam}(2013)}]{Agundez2013}
{Ag{\'u}ndez}, M. \& {Wakelam}, V. 2013, Chemical Reviews, 113, 8710

\bibitem[{{Al{\'e}on}(2010)}]{Aleon2010}
{Al{\'e}on}, J. 2010, \apj, 722, 1342

\bibitem[{{Aspin}(2003)}]{Aspin2003}
{Aspin}, C. 2003, \aj, 125, 1480

\bibitem[{{Aspin} {et~al.}(1994){Aspin}, {Sandell}, \& {Russell}}]{Aspin1994}
{Aspin}, C., {Sandell}, G., \& {Russell}, A.~P.~G. 1994, \aaps, 106, 165

\bibitem[{{Baan} {et~al.}(2010){Baan}, {Loenen}, \& {Spaans}}]{Baan2010}
{Baan}, W.~A., {Loenen}, A.~F., \& {Spaans}, M. 2010, \aap, 516, A40

\bibitem[{{Bachiller} {et~al.}(1990){Bachiller}, {Cernicharo},
  {Martin-Pintado}, {Tafalla}, \& {Lazareff}}]{Bachiller1990}
{Bachiller}, R., {Cernicharo}, J., {Martin-Pintado}, J., {Tafalla}, M., \&
  {Lazareff}, B. 1990, \aap, 231, 174

\bibitem[{{Bally} \& {Reipurth}(2001)}]{Bally2001}
{Bally}, J. \& {Reipurth}, B. 2001, \apj, 546, 299

\bibitem[{{Beckwith} {et~al.}(1990){Beckwith}, {Sargent}, {Chini}, \&
  {Guesten}}]{Beckwith1990}
{Beckwith}, S. V.~W., {Sargent}, A.~I., {Chini}, R.~S., \& {Guesten}, R. 1990,
  \aj, 99, 924

\bibitem[{{Bianchi} {et~al.}(2017){Bianchi}, {Codella}, {Ceccarelli},
  {Fontani}, {Testi}, {Bachiller}, {Lefloch}, {Podio}, \&
  {Taquet}}]{Bianchi2017}
{Bianchi}, E., {Codella}, C., {Ceccarelli}, C., {et~al.} 2017, \mnras, 467,
  3011

\bibitem[{{Bracco} {et~al.}(2017){Bracco}, {Palmeirim}, {Andr{\'e}}, {Adam},
  {Ade}, {Bacmann}, {Beelen}, {Beno{\^\i}t}, {Bideaud}, {Billot}, {Bourrion},
  {Calvo}, {Catalano}, {Coiffard}, {Comis}, {D'Addabbo}, {D{\'e}sert},
  {Didelon}, {Doyle}, {Goupy}, {K{\"o}nyves}, {Kramer}, {Lagache}, {Leclercq},
  {Mac{\'\i}as-P{\'e}rez}, {Maury}, {Mauskopf}, {Mayet}, {Monfardini}, {Motte},
  {Pajot}, {Pascale}, {Peretto}, {Perotto}, {Pisano}, {Ponthieu},
  {Rev{\'e}ret}, {Rigby}, {Ritacco}, {Rodriguez}, {Romero}, {Roy}, {Ruppin},
  {Schuster}, {Sievers}, {Triqueneaux}, {Tucker}, \& {Zylka}}]{Bracco2017}
{Bracco}, A., {Palmeirim}, P., {Andr{\'e}}, P., {et~al.} 2017, \aap, 604, A52

\bibitem[{{Cabezas} {et~al.}(2022){Cabezas}, {Ag{\'u}ndez}, {Marcelino},
  {Tercero}, {Endo}, {Fuentetaja}, {Pardo}, {de Vicente}, \&
  {Cernicharo}}]{Cabezas2022}
{Cabezas}, C., {Ag{\'u}ndez}, M., {Marcelino}, N., {et~al.} 2022, \aap, 657, L4

\bibitem[{{Cambr{\'e}sy}(1999)}]{Cambresy1999}
{Cambr{\'e}sy}, L. 1999, \aap, 345, 965

\bibitem[{{Caselli}(2002)}]{Caselli2002}
{Caselli}, P. 2002, \planss, 50, 1133

\bibitem[{{Caselli} {et~al.}(2022){Caselli}, {Pineda}, {Sipil{\"a}}, {Zhao},
  {Redaelli}, {Spezzano}, {Maureira}, {Alves}, {Bizzocchi}, {Bourke},
  {Chac{\'o}n-Tanarro}, {Friesen}, {Galli}, {Harju}, {Jim{\'e}nez-Serra},
  {Keto}, {Li}, {Padovani}, {Schmiedeke}, {Tafalla}, \& {Vastel}}]{Caselli2022}
{Caselli}, P., {Pineda}, J.~E., {Sipil{\"a}}, O., {et~al.} 2022, \apj, 929, 13

\bibitem[{{Caselli} {et~al.}(1999){Caselli}, {Walmsley}, {Tafalla}, {Dore}, \&
  {Myers}}]{Caselli1999}
{Caselli}, P., {Walmsley}, C.~M., {Tafalla}, M., {Dore}, L., \& {Myers}, P.~C.
  1999, \apjl, 523, L165

\bibitem[{{Ceccarelli} {et~al.}(2014){Ceccarelli}, {Caselli},
  {Bockel{\'e}e-Morvan}, {Mousis}, {Pizzarello}, {Robert}, \&
  {Semenov}}]{Ceccarelli2014}
{Ceccarelli}, C., {Caselli}, P., {Bockel{\'e}e-Morvan}, D., {et~al.} 2014, in
  Protostars and Planets VI, ed. H.~{Beuther}, R.~S. {Klessen}, C.~P.
  {Dullemond}, \& T.~{Henning}, 859

\bibitem[{{Ceccarelli} {et~al.}(2007){Ceccarelli}, {Caselli}, {Herbst},
  {Tielens}, \& {Caux}}]{Ceccarelli2007}
{Ceccarelli}, C., {Caselli}, P., {Herbst}, E., {Tielens}, A.~G.~G.~M., \&
  {Caux}, E. 2007, in Protostars and Planets V, ed. B.~{Reipurth}, D.~{Jewitt},
  \& K.~{Keil}, 47

\bibitem[{{Ceccarelli} {et~al.}(1998){Ceccarelli}, {Castets}, {Loinard},
  {Caux}, \& {Tielens}}]{Ceccarelli1998}
{Ceccarelli}, C., {Castets}, A., {Loinard}, L., {Caux}, E., \& {Tielens},
  A.~G.~G.~M. 1998, \aap, 338, L43

\bibitem[{{Cernicharo} {et~al.}(2022){Cernicharo}, {Ag{\'u}ndez}, {Cabezas},
  {Marcelino}, {Tercero}, {Pardo}, {Fuentetaja}, \& {de
  Vicente}}]{Cernicharo2022}
{Cernicharo}, J., {Ag{\'u}ndez}, M., {Cabezas}, C., {et~al.} 2022, in European
  Physical Journal Web of Conferences, Vol. 265, European Physical Journal Web
  of Conferences, 00041

\bibitem[{{Cernicharo} {et~al.}(2021{\natexlab{a}}){Cernicharo}, {Ag{\'u}ndez},
  {Kaiser}, {Cabezas}, {Tercero}, {Marcelino}, {Pardo}, \& {de
  Vicente}}]{Cernicharo2021b}
{Cernicharo}, J., {Ag{\'u}ndez}, M., {Kaiser}, R.~I., {et~al.}
  2021{\natexlab{a}}, \aap, 655, L1

\bibitem[{{Cernicharo} {et~al.}(2021{\natexlab{b}}){Cernicharo}, {Cabezas},
  {Endo}, {Ag{\'u}ndez}, {Tercero}, {Pardo}, {Marcelino}, \& {de
  Vicente}}]{Cernicharo2021a}
{Cernicharo}, J., {Cabezas}, C., {Endo}, Y., {et~al.} 2021{\natexlab{b}}, \aap,
  650, L14

\bibitem[{{Cernicharo} \& {Guelin}(1987)}]{Cernicharo1987}
{Cernicharo}, J. \& {Guelin}, M. 1987, \aap, 176, 299

\bibitem[{{Chac{\'o}n-Tanarro} {et~al.}(2017){Chac{\'o}n-Tanarro}, {Caselli},
  {Bizzocchi}, {Pineda}, {Harju}, {Spaans}, \& {D{\'e}sert}}]{Chacon2017}
{Chac{\'o}n-Tanarro}, A., {Caselli}, P., {Bizzocchi}, L., {et~al.} 2017, \aap,
  606, A142

\bibitem[{{Chac{\'o}n-Tanarro} {et~al.}(2019){Chac{\'o}n-Tanarro}, {Pineda},
  {Caselli}, {Bizzocchi}, {Gutermuth}, {Mason}, {G{\'o}mez-Ruiz}, {Harju},
  {Devlin}, {Dicker}, {Mroczkowski}, {Romero}, {Sievers}, {Stanchfield},
  {Offner}, \& {S{\'a}nchez-Arg{\"u}elles}}]{Chacon2019}
{Chac{\'o}n-Tanarro}, A., {Pineda}, J.~E., {Caselli}, P., {et~al.} 2019, \aap,
  623, A118

\bibitem[{{Chen} {et~al.}(2015){Chen}, {Liu}, {Chen}, {Huang}, \&
  {Wu}}]{Chen2015}
{Chen}, H.-F., {Liu}, M.-C., {Chen}, S.-C., {Huang}, T.-P., \& {Wu}, Y.-J.
  2015, \apj, 804, 36

\bibitem[{{Chen} {et~al.}(2011){Chen}, {Liu}, {Su}, \& {Wang}}]{Chen2011}
{Chen}, H.-R., {Liu}, S.-Y., {Su}, Y.-N., \& {Wang}, M.-Y. 2011, \apj, 743, 196

\bibitem[{{Chen} {et~al.}(2016){Chen}, {Di Francesco}, {Johnstone}, {Sadavoy},
  {Hatchell}, {Mottram}, {Kirk}, {Buckle}, {Berry}, {Broekhoven-Fiene},
  {Currie}, {Fich}, {Jenness}, {Nutter}, {Pattle}, {Pineda}, {Quinn}, {Salji},
  {Tisi}, {Hogerheijde}, {Ward-Thompson}, {Bastien}, {Bresnahan}, {Butner},
  {Chrysostomou}, {Coude}, {Davis}, {Drabek-Maunder}, {Duarte-Cabral}, {Fiege},
  {Friberg}, {Friesen}, {Fuller}, {Graves}, {Greaves}, {Gregson}, {Holland},
  {Joncas}, {Kirk}, {Knee}, {Mairs}, {Marsh}, {Matthews}, {Moriarty-Schieven},
  {Mowat}, {Pezzuto}, {Rawlings}, {Richer}, {Robertson}, {Rosolowsky},
  {Rumble}, {Schneider-Bontemps}, {Thomas}, {Tothill}, {Viti}, {White},
  {Wouterloot}, {Yates}, \& {Zhu}}]{Chen2016}
{Chen}, M. C.-Y., {Di Francesco}, J., {Johnstone}, D., {et~al.} 2016, \apj,
  826, 95

\bibitem[{{Churchwell} {et~al.}(1984){Churchwell}, {Nash}, \&
  {Walmsley}}]{Churchwell1984}
{Churchwell}, E., {Nash}, A.~G., \& {Walmsley}, C.~M. 1984, \apj, 287, 681

\bibitem[{{Colzi} {et~al.}(2018){Colzi}, {Fontani}, {Caselli}, {Ceccarelli},
  {Hily-Blant}, \& {Bizzocchi}}]{Colzi2018}
{Colzi}, L., {Fontani}, F., {Caselli}, P., {et~al.} 2018, \aap, 609, A129

\bibitem[{{Colzi} {et~al.}(2020){Colzi}, {Sipil{\"a}}, {Roueff}, {Caselli}, \&
  {Fontani}}]{Colzi2020}
{Colzi}, L., {Sipil{\"a}}, O., {Roueff}, E., {Caselli}, P., \& {Fontani}, F.
  2020, \aap, 640, A51

\bibitem[{{Daniel} {et~al.}(2013){Daniel}, {G{\'e}rin}, {Roueff}, {Cernicharo},
  {Marcelino}, {Lique}, {Lis}, {Teyssier}, {Biver}, \&
  {Bockel{\'e}e-Morvan}}]{Daniel2013}
{Daniel}, F., {G{\'e}rin}, M., {Roueff}, E., {et~al.} 2013, \aap, 560, A3

\bibitem[{{De Simone} {et~al.}(2022){De Simone}, {Codella}, {Ceccarelli},
  {L{\'o}pez-Sepulcre}, {Neri}, {Rivera-Ortiz}, {Busquet}, {Caselli},
  {Bianchi}, {Fontani}, {Lefloch}, {Oya}, \& {Pineda}}]{DeSimone2022}
{De Simone}, M., {Codella}, C., {Ceccarelli}, C., {et~al.} 2022, \mnras, 512,
  5214

\bibitem[{Denis-Alpizar {et~al.}(2013)Denis-Alpizar, Kalugina, Stoecklin, Vera,
  \& Lique}]{denis2013new}
Denis-Alpizar, O., Kalugina, Y., Stoecklin, T., Vera, M.~H., \& Lique, F. 2013,
  The Journal of chemical physics, 139, 224301

\bibitem[{Denis-Alpizar {et~al.}(2015)Denis-Alpizar, Stoecklin, \&
  Halvick}]{denis2015isotopic}
Denis-Alpizar, O., Stoecklin, T., \& Halvick, P. 2015, Monthly Notices of the
  Royal Astronomical Society, 453, 1317

\bibitem[{{Dhabal} {et~al.}(2019){Dhabal}, {Mundy}, {Chen}, {Teuben}, \&
  {Storm}}]{Dhabal2019}
{Dhabal}, A., {Mundy}, L.~G., {Chen}, C.-y., {Teuben}, P., \& {Storm}, S. 2019,
  \apj, 876, 108

\bibitem[{{Dicker} {et~al.}(2014){Dicker}, {Ade}, {Aguirre}, {Brevik}, {Cho},
  {Datta}, {Devlin}, {Dober}, {Egan}, {Ford}, {Ford}, {Hilton}, {Hubmayr},
  {Irwin}, {Mason}, {Marganian}, {Mello}, {McMahon}, {Mroczkowski}, {Romero},
  {Stanchfield}, {Tucker}, {Vale}, {White}, {Whitehead}, \&
  {Young}}]{Dicker2014}
{Dicker}, S.~R., {Ade}, P.~A.~R., {Aguirre}, J., {et~al.} 2014, in Society of
  Photo-Optical Instrumentation Engineers (SPIE) Conference Series, Vol. 9153,
  Millimeter, Submillimeter, and Far-Infrared Detectors and Instrumentation for
  Astronomy VII, ed. W.~S. {Holland} \& J.~{Zmuidzinas}, 91530J

\bibitem[{{Draine}(2011)}]{Draine2011}
{Draine}, B.~T. 2011, {Physics of the Interstellar and Intergalactic Medium}
  (Princeton University Press)

\bibitem[{Dumouchel {et~al.}(2011)Dumouchel, K{\l}os, \&
  Lique}]{dumouchel2011rotational}
Dumouchel, F., K{\l}os, J., \& Lique, F. 2011, Physical Chemistry Chemical
  Physics, 13, 8204

\bibitem[{Dumouchel {et~al.}(2017)Dumouchel, Lique, Spielfiedel, \&
  Feautrier}]{dumouchel2017hyperfine}
Dumouchel, F., Lique, F., Spielfiedel, A., \& Feautrier, N. 2017, Monthly
  Notices of the Royal Astronomical Society, 471, 1849

\bibitem[{{Emprechtinger} {et~al.}(2009){Emprechtinger}, {Caselli}, {Volgenau},
  {Stutzki}, \& {Wiedner}}]{Emprechtinger2009}
{Emprechtinger}, M., {Caselli}, P., {Volgenau}, N.~H., {Stutzki}, J., \&
  {Wiedner}, M.~C. 2009, \aap, 493, 89

\bibitem[{{Endres} {et~al.}(2016){Endres}, {Schlemmer}, {Schilke}, {Stutzki},
  \& {M{\"u}ller}}]{Endres2016}
{Endres}, C.~P., {Schlemmer}, S., {Schilke}, P., {Stutzki}, J., \&
  {M{\"u}ller}, H. S.~P. 2016, Journal of Molecular Spectroscopy, 327, 95

\bibitem[{{Enoch} {et~al.}(2008){Enoch}, {Evans}, {Sargent}, {Glenn},
  {Rosolowsky}, \& {Myers}}]{Enoch2008}
{Enoch}, M.~L., {Evans}, Neal~J., I., {Sargent}, A.~I., {et~al.} 2008, \apj,
  684, 1240

\bibitem[{{Enoch} {et~al.}(2006){Enoch}, {Young}, {Glenn}, {Evans}, {Golwala},
  {Sargent}, {Harvey}, {Aguirre}, {Goldin}, {Haig}, {Huard}, {Lange},
  {Laurent}, {Maloney}, {Mauskopf}, {Rossinot}, \& {Sayers}}]{Enoch2006}
{Enoch}, M.~L., {Young}, K.~E., {Glenn}, J., {et~al.} 2006, \apj, 638, 293

\bibitem[{{Esplugues} {et~al.}(2022){Esplugues}, {Fuente}, {Navarro-Almaida},
  {Rodr{\'\i}guez-Baras}, {Majumdar}, {Caselli}, {Wakelam}, {Roueff},
  {Bachiller}, {Spezzano}, {Rivi{\`e}re-Marichalar},
  {Mart{\'\i}n-Dom{\'e}nech}, \& {Mu{\~n}oz Caro}}]{Esplugues2022}
{Esplugues}, G., {Fuente}, A., {Navarro-Almaida}, D., {et~al.} 2022, \aap, 662,
  A52

\bibitem[{{Feh{\'e}r} {et~al.}(2016){Feh{\'e}r}, {T{\'o}th}, {Ward-Thompson},
  {Kirk}, {Kraus}, {Pelkonen}, {Pint{\'e}r}, \& {Zahorecz}}]{Feher2016}
{Feh{\'e}r}, O., {T{\'o}th}, L.~V., {Ward-Thompson}, D., {et~al.} 2016, \aap,
  590, A75

\bibitem[{Flower \& Lique(2015)}]{flower2015excitation}
Flower, D. \& Lique, F. 2015, Monthly Notices of the Royal Astronomical
  Society, 446, 1750

\bibitem[{{Fontani} {et~al.}(2015){Fontani}, {Busquet}, {Palau}, {Caselli},
  {S{\'a}nchez-Monge}, {Tan}, \& {Audard}}]{Fontani2015}
{Fontani}, F., {Busquet}, G., {Palau}, A., {et~al.} 2015, \aap, 575, A87

\bibitem[{{Fuente} {et~al.}(2019){Fuente}, {Navarro}, {Caselli}, {Gerin},
  {Kramer}, {Roueff}, {Alonso-Albi}, {Bachiller}, {Cazaux}, {Commercon},
  {Friesen}, {Garc{\'\i}a-Burillo}, {Giuliano}, {Goicoechea}, {Gratier},
  {Hacar}, {Jim{\'e}nez-Serra}, {Kirk}, {Lattanzi}, {Loison}, {Malinen},
  {Marcelino}, {Mart{\'\i}n-Dom{\'e}nech}, {Mu{\~n}oz-Caro}, {Pineda},
  {Tafalla}, {Tercero}, {Ward-Thompson}, {Trevi{\~n}o-Morales},
  {Rivi{\'e}re-Marichalar}, {Roncero}, {Vidal}, \& {Ballester}}]{Fuente2019}
{Fuente}, A., {Navarro}, D.~G., {Caselli}, P., {et~al.} 2019, \aap, 624, A105

\bibitem[{{Fuentetaja} {et~al.}(2022){Fuentetaja}, {Ag{\'u}ndez}, {Cabezas},
  {Tercero}, {Marcelino}, {Pardo}, {de Vicente}, \&
  {Cernicharo}}]{Fuentetaja2022}
{Fuentetaja}, R., {Ag{\'u}ndez}, M., {Cabezas}, C., {et~al.} 2022, \aap, 667,
  L4

\bibitem[{{Gerlich} {et~al.}(2002){Gerlich}, {Herbst}, \&
  {Roueff}}]{Gerlich2002}
{Gerlich}, D., {Herbst}, E., \& {Roueff}, E. 2002, \planss, 50, 1275

\bibitem[{{Godard} {et~al.}(2010){Godard}, {Falgarone}, {Gerin}, {Hily-Blant},
  \& {de Luca}}]{Godard2010}
{Godard}, B., {Falgarone}, E., {Gerin}, M., {Hily-Blant}, P., \& {de Luca}, M.
  2010, \aap, 520, A20

\bibitem[{{Goldsmith}(2001)}]{Goldsmith2001}
{Goldsmith}, P.~F. 2001, \apj, 557, 736

\bibitem[{{Goldsmith} {et~al.}(2008){Goldsmith}, {Heyer}, {Narayanan}, {Snell},
  {Li}, \& {Brunt}}]{Goldsmith2008}
{Goldsmith}, P.~F., {Heyer}, M., {Narayanan}, G., {et~al.} 2008, \apj, 680, 428

\bibitem[{{Goodman} {et~al.}(1998){Goodman}, {Barranco}, {Wilner}, \&
  {Heyer}}]{Goodman1998}
{Goodman}, A.~A., {Barranco}, J.~A., {Wilner}, D.~J., \& {Heyer}, M.~H. 1998,
  \apj, 504, 223

\bibitem[{{Graninger} {et~al.}(2014){Graninger}, {Herbst}, {{\"O}berg}, \&
  {Vasyunin}}]{Graninger2014}
{Graninger}, D.~M., {Herbst}, E., {{\"O}berg}, K.~I., \& {Vasyunin}, A.~I.
  2014, \apj, 787, 74

\bibitem[{{Gratier} {et~al.}(2016){Gratier}, {Majumdar}, {Ohishi}, {Roueff},
  {Loison}, {Hickson}, \& {Wakelam}}]{Gratier2016}
{Gratier}, P., {Majumdar}, L., {Ohishi}, M., {et~al.} 2016, \apjs, 225, 25

\bibitem[{{Greissl} {et~al.}(2007){Greissl}, {Meyer}, {Wilking}, {Fanetti},
  {Schneider}, {Greene}, \& {Young}}]{Greissl2007}
{Greissl}, J., {Meyer}, M.~R., {Wilking}, B.~A., {et~al.} 2007, \aj, 133, 1321

\bibitem[{{Hacar} {et~al.}(2020){Hacar}, {Bosman}, \& {van
  Dishoeck}}]{Hacar2020}
{Hacar}, A., {Bosman}, A.~D., \& {van Dishoeck}, E.~F. 2020, \aap, 635, A4

\bibitem[{{Harvey} {et~al.}(1984){Harvey}, {Wilking}, \& {Joy}}]{Harvey1984}
{Harvey}, P.~M., {Wilking}, B.~A., \& {Joy}, M. 1984, \apj, 278, 156

\bibitem[{{Hatchell} {et~al.}(2007{\natexlab{a}}){Hatchell}, {Fuller}, \&
  {Richer}}]{Hatchell2007b}
{Hatchell}, J., {Fuller}, G.~A., \& {Richer}, J.~S. 2007{\natexlab{a}}, \aap,
  472, 187

\bibitem[{{Hatchell} {et~al.}(2007{\natexlab{b}}){Hatchell}, {Fuller},
  {Richer}, {Harries}, \& {Ladd}}]{Hatchell2007a}
{Hatchell}, J., {Fuller}, G.~A., {Richer}, J.~S., {Harries}, T.~J., \& {Ladd},
  E.~F. 2007{\natexlab{b}}, \aap, 468, 1009

\bibitem[{{Hatchell} {et~al.}(2005){Hatchell}, {Richer}, {Fuller},
  {Qualtrough}, {Ladd}, \& {Chandler}}]{Hatchell2005}
{Hatchell}, J., {Richer}, J.~S., {Fuller}, G.~A., {et~al.} 2005, \aap, 440, 151

\bibitem[{Hern{\'a}ndez~Vera {et~al.}(2017)Hern{\'a}ndez~Vera, Lique,
  Dumouchel, Hily-Blant, \& Faure}]{hernandez2017rotational}
Hern{\'a}ndez~Vera, M., Lique, F., Dumouchel, F., Hily-Blant, P., \& Faure, A.
  2017, Monthly Notices of the Royal Astronomical Society, 468, 1084

\bibitem[{{Hildebrand}(1983)}]{Hildebrand1983}
{Hildebrand}, R.~H. 1983, \qjras, 24, 267

\bibitem[{{Hily-Blant} {et~al.}(2013){Hily-Blant}, {Bonal}, {Faure}, \&
  {Quirico}}]{Hily-Blant2013}
{Hily-Blant}, P., {Bonal}, L., {Faure}, A., \& {Quirico}, E. 2013, \icarus,
  223, 582

\bibitem[{{Hily-Blant} {et~al.}(2010){Hily-Blant}, {Walmsley}, {Pineau Des
  For{\^e}ts}, \& {Flower}}]{Hily2010}
{Hily-Blant}, P., {Walmsley}, M., {Pineau Des For{\^e}ts}, G., \& {Flower}, D.
  2010, \aap, 513, A41

\bibitem[{{Hirota} {et~al.}(1998){Hirota}, {Yamamoto}, {Mikami}, \&
  {Ohishi}}]{Hirota1998}
{Hirota}, T., {Yamamoto}, S., {Mikami}, H., \& {Ohishi}, M. 1998, \apj, 503,
  717

\bibitem[{Hutson \& Green(1994)}]{hutson1994molscat}
Hutson, J. \& Green, S. 1994, Collaborative computational project

\bibitem[{{Imai} {et~al.}(2018){Imai}, {Sakai}, {L{\'o}pez-Sepulcre},
  {Higuchi}, {Zhang}, {Oya}, {Watanabe}, {Sakai}, {Ceccarelli}, {Lefloch}, \&
  {Yamamoto}}]{Imai2018}
{Imai}, M., {Sakai}, N., {L{\'o}pez-Sepulcre}, A., {et~al.} 2018, \apj, 869, 51

\bibitem[{{Jennings} {et~al.}(1987){Jennings}, {Cameron}, {Cudlip}, \&
  {Hirst}}]{Jennings1987}
{Jennings}, R.~E., {Cameron}, D.~H.~M., {Cudlip}, W., \& {Hirst}, C.~J. 1987,
  \mnras, 226, 461

\bibitem[{{Jin} {et~al.}(2015){Jin}, {Lee}, \& {Kim}}]{Jin2015}
{Jin}, M., {Lee}, J.-E., \& {Kim}, K.-T. 2015, \apjs, 219, 2

\bibitem[{{J{\o}rgensen} {et~al.}(2005){J{\o}rgensen}, {Bourke}, {Myers},
  {Sch{\"o}ier}, {van Dishoeck}, \& {Wilner}}]{Jorgensen2005}
{J{\o}rgensen}, J.~K., {Bourke}, T.~L., {Myers}, P.~C., {et~al.} 2005, \apj,
  632, 973

\bibitem[{{J{\o}rgensen} {et~al.}(2008){J{\o}rgensen}, {Johnstone}, {Kirk},
  {Myers}, {Allen}, \& {Shirley}}]{Jorgensen2008}
{J{\o}rgensen}, J.~K., {Johnstone}, D., {Kirk}, H., {et~al.} 2008, \apj, 683,
  822

\bibitem[{{J{\o}rgensen} {et~al.}(2004){J{\o}rgensen}, {Sch{\"o}ier}, \& {van
  Dishoeck}}]{Jorgensen2004}
{J{\o}rgensen}, J.~K., {Sch{\"o}ier}, F.~L., \& {van Dishoeck}, E.~F. 2004,
  \aap, 416, 603

\bibitem[{{Kirk} {et~al.}(2006){Kirk}, {Johnstone}, \& {Di
  Francesco}}]{Kirk2006}
{Kirk}, H., {Johnstone}, D., \& {Di Francesco}, J. 2006, \apj, 646, 1009

\bibitem[{{Kirk} {et~al.}(2007){Kirk}, {Johnstone}, \& {Tafalla}}]{Kirk2007}
{Kirk}, H., {Johnstone}, D., \& {Tafalla}, M. 2007, \apj, 668, 1042

\bibitem[{{Kirk} {et~al.}(2013){Kirk}, {Ward-Thompson}, {Palmeirim},
  {Andr{\'e}}, {Griffin}, {Hargrave}, {K{\"o}nyves}, {Bernard}, {Nutter},
  {Sibthorpe}, {Di Francesco}, {Abergel}, {Arzoumanian}, {Benedettini},
  {Bontemps}, {Elia}, {Hennemann}, {Hill}, {Men'shchikov}, {Motte},
  {Nguyen-Luong}, {Peretto}, {Pezzuto}, {Rygl}, {Sadavoy}, {Schisano},
  {Schneider}, {Testi}, \& {White}}]{Kirk2013}
{Kirk}, J.~M., {Ward-Thompson}, D., {Palmeirim}, P., {et~al.} 2013, \mnras,
  432, 1424

\bibitem[{{Kong} {et~al.}(2015){Kong}, {Caselli}, {Tan}, {Wakelam}, \&
  {Sipil{\"a}}}]{Kong2015}
{Kong}, S., {Caselli}, P., {Tan}, J.~C., {Wakelam}, V., \& {Sipil{\"a}}, O.
  2015, \apj, 804, 98

\bibitem[{{Lada} {et~al.}(1996){Lada}, {Alves}, \& {Lada}}]{Lada1996}
{Lada}, C.~J., {Alves}, J., \& {Lada}, E.~A. 1996, \aj, 111, 1964

\bibitem[{{Lef{\`e}vre} {et~al.}(2016){Lef{\`e}vre}, {Pagani}, {Min}, {Poteet},
  \& {Whittet}}]{Lefevre2016}
{Lef{\`e}vre}, C., {Pagani}, L., {Min}, M., {Poteet}, C., \& {Whittet}, D.
  2016, \aap, 585, L4

\bibitem[{{Lefloch} {et~al.}(2021){Lefloch}, {Busquet}, {Viti}, {Vastel},
  {Mendoza}, {Benedettini}, {Codella}, {Podio}, {Schutzer}, {Rivera-Ortiz},
  {L{\'e}pine}, \& {Bachiller}}]{Lefloch2021}
{Lefloch}, B., {Busquet}, G., {Viti}, S., {et~al.} 2021, \mnras, 507, 1034

\bibitem[{{Lefloch} {et~al.}(1998){Lefloch}, {Castets}, {Cernicharo}, {Langer},
  \& {Zylka}}]{Lefloch1998}
{Lefloch}, B., {Castets}, A., {Cernicharo}, J., {Langer}, W.~D., \& {Zylka}, R.
  1998, \aap, 334, 269

\bibitem[{{Liszt} \& {Lucas}(2001)}]{Liszt2001}
{Liszt}, H. \& {Lucas}, R. 2001, \aap, 370, 576

\bibitem[{{Liszt} \& {Ziurys}(2012)}]{Liszt2012}
{Liszt}, H.~S. \& {Ziurys}, L.~M. 2012, \apj, 747, 55

\bibitem[{{Liu} {et~al.}(2013){Liu}, {Wang}, \& {Xu}}]{Liu2013}
{Liu}, X.-L., {Wang}, J.-J., \& {Xu}, J.-L. 2013, \mnras, 431, 27

\bibitem[{{Loison} {et~al.}(2019){Loison}, {Wakelam}, {Gratier}, \&
  {Hickson}}]{Loison2019}
{Loison}, J.-C., {Wakelam}, V., {Gratier}, P., \& {Hickson}, K.~M. 2019,
  \mnras, 484, 2747

\bibitem[{{Loison} {et~al.}(2014){Loison}, {Wakelam}, \&
  {Hickson}}]{Loison2014}
{Loison}, J.-C., {Wakelam}, V., \& {Hickson}, K.~M. 2014, \mnras, 443, 398

\bibitem[{{Lombardi} {et~al.}(2014){Lombardi}, {Bouy}, {Alves}, \&
  {Lada}}]{Lombardi2014}
{Lombardi}, M., {Bouy}, H., {Alves}, J., \& {Lada}, C.~J. 2014, \aap, 566, A45

\bibitem[{{Long} {et~al.}(2021){Long}, {Bosman}, {Cazzoletti}, {van Dishoeck},
  {{\"O}berg}, {Facchini}, {Tazzari}, {Guzm{\'a}n}, \& {Testi}}]{Long2021}
{Long}, F., {Bosman}, A.~D., {Cazzoletti}, P., {et~al.} 2021, \aap, 647, A118

\bibitem[{{Majumdar} {et~al.}(2017){Majumdar}, {Gratier}, {Ruaud}, {Wakelam},
  {Vastel}, {Sipil{\"a}}, {Hersant}, {Dutrey}, \& {Guilloteau}}]{Majumdar2017}
{Majumdar}, L., {Gratier}, P., {Ruaud}, M., {et~al.} 2017, \mnras, 466, 4470

\bibitem[{{Marcelino} {et~al.}(2005){Marcelino}, {Cernicharo}, {Roueff},
  {Gerin}, \& {Mauersberger}}]{Marcelino2005}
{Marcelino}, N., {Cernicharo}, J., {Roueff}, E., {Gerin}, M., \&
  {Mauersberger}, R. 2005, \apj, 620, 308

\bibitem[{{Maureira} {et~al.}(2020){Maureira}, {Arce}, {Dunham}, {Mardones},
  {Guzm{\'a}n}, {Pineda}, \& {Bourke}}]{Maureira2020}
{Maureira}, M.~J., {Arce}, H.~G., {Dunham}, M.~M., {et~al.} 2020, \mnras, 499,
  4394

\bibitem[{{Milam} {et~al.}(2005){Milam}, {Savage}, {Brewster}, {Ziurys}, \&
  {Wyckoff}}]{Milam2005}
{Milam}, S.~N., {Savage}, C., {Brewster}, M.~A., {Ziurys}, L.~M., \& {Wyckoff},
  S. 2005, \apj, 634, 1126

\bibitem[{{Miotello} {et~al.}(2014){Miotello}, {Testi}, {Lodato}, {Ricci},
  {Rosotti}, {Brooks}, {Maury}, \& {Natta}}]{Miotello2014}
{Miotello}, A., {Testi}, L., {Lodato}, G., {et~al.} 2014, \aap, 567, A32

\bibitem[{{Mizuno} {et~al.}(1995){Mizuno}, {Onishi}, {Yonekura}, {Nagahama},
  {Ogawa}, \& {Fukui}}]{Mizuno1995}
{Mizuno}, A., {Onishi}, T., {Yonekura}, Y., {et~al.} 1995, \apjl, 445, L161

\bibitem[{{M{\"u}ller} {et~al.}(2005){M{\"u}ller}, {Schl{\"o}der}, {Stutzki},
  \& {Winnewisser}}]{Muller2005}
{M{\"u}ller}, H. S.~P., {Schl{\"o}der}, F., {Stutzki}, J., \& {Winnewisser}, G.
  2005, Journal of Molecular Structure, 742, 215

\bibitem[{{M{\"u}ller} {et~al.}(2001){M{\"u}ller}, {Thorwirth}, {Roth}, \&
  {Winnewisser}}]{Muller2001}
{M{\"u}ller}, H.~S.~P., {Thorwirth}, S., {Roth}, D.~A., \& {Winnewisser}, G.
  2001, \aap, 370, L49

\bibitem[{{Narayanan} {et~al.}(2008){Narayanan}, {Heyer}, {Brunt}, {Goldsmith},
  {Snell}, \& {Li}}]{Narayanan2008}
{Narayanan}, G., {Heyer}, M.~H., {Brunt}, C., {et~al.} 2008, \apjs, 177, 341

\bibitem[{{Navarro-Almaida} {et~al.}(2021){Navarro-Almaida}, {Fuente},
  {Majumdar}, {Wakelam}, {Caselli}, {Rivi{\`e}re-Marichalar},
  {Trevi{\~n}o-Morales}, {Cazaux}, {Jim{\'e}nez-Serra}, {Kramer},
  {Chac{\'o}n-Tanarro}, {Kirk}, {Ward-Thompson}, \& {Tafalla}}]{Navarro2021}
{Navarro-Almaida}, D., {Fuente}, A., {Majumdar}, L., {et~al.} 2021, \aap, 653,
  A15

\bibitem[{{Navarro-Almaida} {et~al.}(2020){Navarro-Almaida}, {Le Gal},
  {Fuente}, {Rivi{\`e}re-Marichalar}, {Wakelam}, {Cazaux}, {Caselli}, {Laas},
  {Alonso-Albi}, {Loison}, {Gerin}, {Kramer}, {Roueff}, {Bachiller},
  {Commer{\c{c}}on}, {Friesen}, {Garc{\'\i}a-Burillo}, {Goicoechea},
  {Giuliano}, {Jim{\'e}nez-Serra}, {Kirk}, {Lattanzi}, {Malinen}, {Marcelino},
  {Mart{\'\i}n-Dom{\`e}nech}, {Mu{\~n}oz Caro}, {Pineda}, {Tercero},
  {Trevi{\~n}o-Morales}, {Roncero}, {Hacar}, {Tafalla}, \&
  {Ward-Thompson}}]{Navarro2020}
{Navarro-Almaida}, D., {Le Gal}, R., {Fuente}, A., {et~al.} 2020, \aap, 637,
  A39

\bibitem[{{Nikolic} {et~al.}(2007){Nikolic}, {Prestage}, {Balser}, {Chandler},
  \& {Hills}}]{Nikolic2007}
{Nikolic}, B., {Prestage}, R.~M., {Balser}, D.~S., {Chandler}, C.~J., \&
  {Hills}, R.~E. 2007, \aap, 465, 685

\bibitem[{{Onishi} {et~al.}(1996){Onishi}, {Mizuno}, {Kawamura}, {Ogawa}, \&
  {Fukui}}]{Onishi1996}
{Onishi}, T., {Mizuno}, A., {Kawamura}, A., {Ogawa}, H., \& {Fukui}, Y. 1996,
  \apj, 465, 815

\bibitem[{{Ormel} {et~al.}(2011){Ormel}, {Min}, {Tielens}, {Dominik}, \&
  {Paszun}}]{Ormel2011}
{Ormel}, C.~W., {Min}, M., {Tielens}, A.~G.~G.~M., {Dominik}, C., \& {Paszun},
  D. 2011, \aap, 532, A43

\bibitem[{{Ormel} {et~al.}(2009){Ormel}, {Paszun}, {Dominik}, \&
  {Tielens}}]{Ormel2009}
{Ormel}, C.~W., {Paszun}, D., {Dominik}, C., \& {Tielens}, A.~G.~G.~M. 2009,
  \aap, 502, 845

\bibitem[{{Ortiz-Le{\'o}n} {et~al.}(2018){Ortiz-Le{\'o}n}, {Loinard}, {Dzib},
  {Galli}, {Kounkel}, {Mioduszewski}, {Rodr{\'\i}guez}, {Torres}, {Hartmann},
  {Boden}, {Evans}, {Brice{\~n}o}, \& {Tobin}}]{OrtizLeon2018}
{Ortiz-Le{\'o}n}, G.~N., {Loinard}, L., {Dzib}, S.~A., {et~al.} 2018, \apj,
  865, 73

\bibitem[{{Ossenkopf}(1993)}]{Ossenkopf1993}
{Ossenkopf}, V. 1993, \aap, 280, 617

\bibitem[{{Padoan} {et~al.}(2002){Padoan}, {Cambr{\'e}sy}, \&
  {Langer}}]{Padoan2002}
{Padoan}, P., {Cambr{\'e}sy}, L., \& {Langer}, W. 2002, \apjl, 580, L57

\bibitem[{{Pagani} {et~al.}(2007){Pagani}, {Bacmann}, {Cabrit}, \&
  {Vastel}}]{Pagani2007}
{Pagani}, L., {Bacmann}, A., {Cabrit}, S., \& {Vastel}, C. 2007, \aap, 467, 179

\bibitem[{{Pagani} {et~al.}(2013){Pagani}, {Lesaffre}, {Jorfi}, {Honvault},
  {Gonz{\'a}lez-Lezana}, \& {Faure}}]{Pagani2013}
{Pagani}, L., {Lesaffre}, P., {Jorfi}, M., {et~al.} 2013, \aap, 551, A38

\bibitem[{{Pagani} {et~al.}(2010){Pagani}, {Steinacker}, {Bacmann}, {Stutz}, \&
  {Henning}}]{Pagani2010}
{Pagani}, L., {Steinacker}, J., {Bacmann}, A., {Stutz}, A., \& {Henning}, T.
  2010, Science, 329, 1622

\bibitem[{{Parise} {et~al.}(2004){Parise}, {Castets}, {Herbst}, {Caux},
  {Ceccarelli}, {Mukhopadhyay}, \& {Tielens}}]{Parise2004}
{Parise}, B., {Castets}, A., {Herbst}, E., {et~al.} 2004, \aap, 416, 159

\bibitem[{{Parise} {et~al.}(2006){Parise}, {Ceccarelli}, {Tielens}, {Castets},
  {Caux}, {Lefloch}, \& {Maret}}]{Parise2006}
{Parise}, B., {Ceccarelli}, C., {Tielens}, A.~G.~G.~M., {et~al.} 2006, \aap,
  453, 949

\bibitem[{{Rebull}(2015)}]{Rebull2015}
{Rebull}, L.~M. 2015, \aj, 150, 17

\bibitem[{{Remijan} {et~al.}(2007){Remijan}, {Markwick-Kemper}, \& {ALMA
  Working Group on Spectral Line Frequencies}}]{Remijan2007}
{Remijan}, A.~J., {Markwick-Kemper}, A., \& {ALMA Working Group on Spectral
  Line Frequencies}. 2007, in American Astronomical Society Meeting Abstracts,
  Vol. 211, American Astronomical Society Meeting Abstracts, 132.11

\bibitem[{{Rodr{\'\i}guez-Baras} {et~al.}(2021){Rodr{\'\i}guez-Baras},
  {Fuente}, {Rivi{\'e}re-Marichalar}, {Navarro-Almaida}, {Caselli}, {Gerin},
  {Kramer}, {Roueff}, {Wakelam}, {Esplugues}, {Garc{\'\i}a-Burillo}, {Le Gal},
  {Spezzano}, {Alonso-Albi}, {Bachiller}, {Cazaux}, {Commercon}, {Goicoechea},
  {Loison}, {Trevi{\~n}o-Morales}, {Roncero}, {Jim{\'e}nez-Serra}, {Laas},
  {Hacar}, {Kirk}, {Lattanzi}, {Mart{\'\i}n-Dom{\'e}nech}, {Mu{\~n}oz-Caro},
  {Pineda}, {Tercero}, {Ward-Thompson}, {Tafalla}, {Marcelino}, {Malinen},
  {Friesen}, \& {Giuliano}}]{RodriguezBaras2021}
{Rodr{\'\i}guez-Baras}, M., {Fuente}, A., {Rivi{\'e}re-Marichalar}, P.,
  {et~al.} 2021, \aap, 648, A120

\bibitem[{{Romero} {et~al.}(2020){Romero}, {Sievers}, {Ghirardini}, {Dicker},
  {Giacintucci}, {Mroczkowski}, {Mason}, {Sarazin}, {Devlin}, {Gaspari},
  {Battaglia}, {Hilton}, {Bulbul}, {Lowe}, \& {Stanchfield}}]{Romero2020}
{Romero}, C.~E., {Sievers}, J., {Ghirardini}, V., {et~al.} 2020, \apj, 891, 90

\bibitem[{{Roueff} {et~al.}(2015){Roueff}, {Loison}, \& {Hickson}}]{Roueff2015}
{Roueff}, E., {Loison}, J.~C., \& {Hickson}, K.~M. 2015, \aap, 576, A99

\bibitem[{{Roueff} {et~al.}(2007){Roueff}, {Parise}, \& {Herbst}}]{Roueff2007}
{Roueff}, E., {Parise}, B., \& {Herbst}, E. 2007, \aap, 464, 245

\bibitem[{{Sadavoy} {et~al.}(2014){Sadavoy}, {Di Francesco}, {Andr{\'e}},
  {Pezzuto}, {Bernard}, {Maury}, {Men'shchikov}, {Motte}, {Nguyen-Lu'o'ng},
  {Schneider}, {Arzoumanian}, {Benedettini}, {Bontemps}, {Elia}, {Hennemann},
  {Hill}, {K{\"o}nyves}, {Louvet}, {Peretto}, {Roy}, \& {White}}]{Sadavoy2014}
{Sadavoy}, S.~I., {Di Francesco}, J., {Andr{\'e}}, P., {et~al.} 2014, \apjl,
  787, L18

\bibitem[{{Sadavoy} {et~al.}(2013){Sadavoy}, {Di Francesco}, {Johnstone},
  {Currie}, {Drabek}, {Hatchell}, {Nutter}, {Andr{\'e}}, {Arzoumanian},
  {Benedettini}, {Bernard}, {Duarte-Cabral}, {Fallscheer}, {Friesen},
  {Greaves}, {Hennemann}, {Hill}, {Jenness}, {K{\"o}nyves}, {Matthews},
  {Mottram}, {Pezzuto}, {Roy}, {Rygl}, {Schneider-Bontemps}, {Spinoglio},
  {Testi}, {Tothill}, {Ward-Thompson}, {White}, {JCMT}, \& {Herschel Gould Belt
  Survey Teams}}]{Sadavoy2013}
{Sadavoy}, S.~I., {Di Francesco}, J., {Johnstone}, D., {et~al.} 2013, \apj,
  767, 126

\bibitem[{{Sandell} \& {Knee}(2001)}]{Sandell2001}
{Sandell}, G. \& {Knee}, L. B.~G. 2001, \apjl, 546, L49

\bibitem[{{Sarrasin} {et~al.}(2010){Sarrasin}, {Abdallah}, {Wernli}, {Faure},
  {Cernicharo}, \& {Lique}}]{Sarrasin2010}
{Sarrasin}, E., {Abdallah}, D.~B., {Wernli}, M., {et~al.} 2010, \mnras, 404,
  518

\bibitem[{{Schilke} {et~al.}(1992){Schilke}, {Walmsley}, {Pineau Des Forets},
  {Roueff}, {Flower}, \& {Guilloteau}}]{Schilke1992}
{Schilke}, P., {Walmsley}, C.~M., {Pineau Des Forets}, G., {et~al.} 1992, \aap,
  256, 595

\bibitem[{{Schmalzl} {et~al.}(2010){Schmalzl}, {Kainulainen}, {Quanz}, {Alves},
  {Goodman}, {Henning}, {Launhardt}, {Pineda}, \&
  {Rom{\'a}n-Z{\'u}{\~n}iga}}]{Schmalzl2010}
{Schmalzl}, M., {Kainulainen}, J., {Quanz}, S.~P., {et~al.} 2010, \apj, 725,
  1327

\bibitem[{{Schnee} {et~al.}(2007){Schnee}, {Caselli}, {Goodman}, {Arce},
  {Ballesteros-Paredes}, \& {Kuchibhotla}}]{Schnee2007}
{Schnee}, S., {Caselli}, P., {Goodman}, A., {et~al.} 2007, \apj, 671, 1839

\bibitem[{{Schnee} {et~al.}(2010){Schnee}, {Enoch}, {Noriega-Crespo}, {Sayers},
  {Terebey}, {Caselli}, {Foster}, {Goodman}, {Kauffmann}, {Padgett}, {Rebull},
  {Sargent}, \& {Shetty}}]{Schnee2010}
{Schnee}, S., {Enoch}, M., {Noriega-Crespo}, A., {et~al.} 2010, \apj, 708, 127

\bibitem[{{Schnee} {et~al.}(2014){Schnee}, {Mason}, {Di Francesco}, {Friesen},
  {Li}, {Sadavoy}, \& {Stanke}}]{Schnee2014}
{Schnee}, S., {Mason}, B., {Di Francesco}, J., {et~al.} 2014, \mnras, 444, 2303

\bibitem[{{Shirley} {et~al.}(2011){Shirley}, {Mason}, {Mangum}, {Bolin},
  {Devlin}, {Dicker}, \& {Korngut}}]{Shirley2011}
{Shirley}, Y.~L., {Mason}, B.~S., {Mangum}, J.~G., {et~al.} 2011, \aj, 141, 39

\bibitem[{{Silsbee} {et~al.}(2020){Silsbee}, {Ivlev}, {Sipil{\"a}}, {Caselli},
  \& {Zhao}}]{Silsbee2020}
{Silsbee}, K., {Ivlev}, A.~V., {Sipil{\"a}}, O., {Caselli}, P., \& {Zhao}, B.
  2020, \aap, 641, A39

\bibitem[{{Sipil{\"a}} \& {Caselli}(2018)}]{Sipila2018}
{Sipil{\"a}}, O. \& {Caselli}, P. 2018, \aap, 615, A15

\bibitem[{{Sipil{\"a}} {et~al.}(2019){Sipil{\"a}}, {Caselli}, \&
  {Harju}}]{Sipila2019}
{Sipil{\"a}}, O., {Caselli}, P., \& {Harju}, J. 2019, \aap, 631, A63

\bibitem[{{Steinacker} {et~al.}(2015){Steinacker}, {Andersen}, {Thi},
  {Paladini}, {Juvela}, {Bacmann}, {Pelkonen}, {Pagani}, {Lef{\`e}vre},
  {Henning}, \& {Noriega-Crespo}}]{Steinacker2015}
{Steinacker}, J., {Andersen}, M., {Thi}, W.~F., {et~al.} 2015, \aap, 582, A70

\bibitem[{{Steinacker} {et~al.}(2010){Steinacker}, {Pagani}, {Bacmann}, \&
  {Guieu}}]{Steinacker2010}
{Steinacker}, J., {Pagani}, L., {Bacmann}, A., \& {Guieu}, S. 2010, \aap, 511,
  A9

\bibitem[{{Tafalla} {et~al.}(2006){Tafalla}, {Santiago-Garc{\'\i}a}, {Myers},
  {Caselli}, {Walmsley}, \& {Crapsi}}]{Tafalla2006}
{Tafalla}, M., {Santiago-Garc{\'\i}a}, J., {Myers}, P.~C., {et~al.} 2006, \aap,
  455, 577

\bibitem[{{Tennekes} {et~al.}(2006){Tennekes}, {Harju}, {Juvela}, \&
  {T{\'o}th}}]{Tennekes2006}
{Tennekes}, P.~P., {Harju}, J., {Juvela}, M., \& {T{\'o}th}, L.~V. 2006, \aap,
  456, 1037

\bibitem[{{Tercero} {et~al.}(2021){Tercero}, {L{\'o}pez-P{\'e}rez}, {Gallego},
  {Beltr{\'a}n}, {Garc{\'\i}a}, {Patino-Esteban}, {L{\'o}pez-Fern{\'a}ndez},
  {G{\'o}mez-Molina}, {Diez}, {Garc{\'\i}a-Carre{\~n}o}, {Malo}, {Amils},
  {Serna}, {Albo}, {Hern{\'a}ndez}, {Vaquero}, {Gonz{\'a}lez-Garc{\'\i}a},
  {Barbas}, {L{\'o}pez-Fern{\'a}ndez}, {Bujarrabal}, {G{\'o}mez-Garrido},
  {Pardo}, {Santander-Garc{\'\i}a}, {Tercero}, {Cernicharo}, \& {de
  Vicente}}]{Tercero2021}
{Tercero}, F., {L{\'o}pez-P{\'e}rez}, J.~A., {Gallego}, J.~D., {et~al.} 2021,
  \aap, 645, A37

\bibitem[{{Tobin} {et~al.}(2016){Tobin}, {Looney}, {Li}, {Chandler}, {Dunham},
  {Segura-Cox}, {Sadavoy}, {Melis}, {Harris}, {Kratter}, \&
  {Perez}}]{Tobin2016}
{Tobin}, J.~J., {Looney}, L.~W., {Li}, Z.-Y., {et~al.} 2016, \apj, 818, 73

\bibitem[{{Ungerechts} \& {Thaddeus}(1987)}]{Ungerechts1987}
{Ungerechts}, H. \& {Thaddeus}, P. 1987, \apjs, 63, 645

\bibitem[{{van der Tak} {et~al.}(2007){van der Tak}, {Black}, {Sch{\"o}ier},
  {Jansen}, \& {van Dishoeck}}]{VanDerTak2007}
{van der Tak}, F.~F.~S., {Black}, J.~H., {Sch{\"o}ier}, F.~L., {Jansen}, D.~J.,
  \& {van Dishoeck}, E.~F. 2007, \aap, 468, 627

\bibitem[{{Vastel} {et~al.}(2003){Vastel}, {Phillips}, {Ceccarelli}, \&
  {Pearson}}]{Vastel2003}
{Vastel}, C., {Phillips}, T.~G., {Ceccarelli}, C., \& {Pearson}, J. 2003,
  \apjl, 593, L97

\bibitem[{{Wakelam} {et~al.}(2005){Wakelam}, {Ceccarelli}, {Castets},
  {Lefloch}, {Loinard}, {Faure}, {Schneider}, \& {Benayoun}}]{Wakelam2005}
{Wakelam}, V., {Ceccarelli}, C., {Castets}, A., {et~al.} 2005, \aap, 437, 149

\bibitem[{{Walawender} {et~al.}(2008){Walawender}, {Bally}, {Francesco},
  {J{\o}rgensen}, \& {Getman}}]{Walawender2008}
{Walawender}, J., {Bally}, J., {Francesco}, J.~D., {J{\o}rgensen}, J., \&
  {Getman}, K.~. 2008, in Handbook of Star Forming Regions, Volume I, ed.
  B.~{Reipurth}, Vol.~4 (-), 346

\bibitem[{{Weingartner} \& {Draine}(2001)}]{Weingartner2001}
{Weingartner}, J.~C. \& {Draine}, B.~T. 2001, \apj, 548, 296

\bibitem[{{Whittet} {et~al.}(1988){Whittet}, {Bode}, {Longmore}, {Adamson},
  {McFadzean}, {Aitken}, \& {Roche}}]{Whittet1988}
{Whittet}, D.~C.~B., {Bode}, M.~F., {Longmore}, A.~J., {et~al.} 1988, \mnras,
  233, 321

\bibitem[{{Wong} {et~al.}(2016){Wong}, {Hirashita}, \& {Li}}]{Wong2016}
{Wong}, Y. H.~V., {Hirashita}, H., \& {Li}, Z.-Y. 2016, \pasj, 68, 67

\bibitem[{{Wu} {et~al.}(2014){Wu}, {Chuang}, {Chen}, \& {Huang}}]{Wu2014}
{Wu}, Y.-J., {Chuang}, S.-J., {Chen}, S.-C., \& {Huang}, T.-P. 2014, \apjs,
  212, 7

\bibitem[{{Wu} {et~al.}(2012){Wu}, {Wu}, {Chou}, {Lin}, {Lu}, {Lo}, \&
  {Cheng}}]{Wu2012}
{Wu}, Y.-J., {Wu}, C.~Y.~R., {Chou}, S.-L., {et~al.} 2012, \apj, 746, 175

\bibitem[{{Xu} {et~al.}(2016){Xu}, {Li}, {Yue}, \& {Goldsmith}}]{Xu2016}
{Xu}, D., {Li}, D., {Yue}, N., \& {Goldsmith}, P.~F. 2016, \apj, 819, 22

\bibitem[{{Yan} {et~al.}(2019){Yan}, {Zhang}, {Xu}, {Guo}, {Macquart}, {Tang},
  \& {Walsh}}]{Yan2019}
{Yan}, Q.-Z., {Zhang}, B., {Xu}, Y., {et~al.} 2019, \aap, 624, A6

\bibitem[{{Zari} {et~al.}(2016){Zari}, {Lombardi}, {Alves}, {Lada}, \&
  {Bouy}}]{Zari2016}
{Zari}, E., {Lombardi}, M., {Alves}, J., {Lada}, C.~J., \& {Bouy}, H. 2016,
  \aap, 587, A106

\bibitem[{{Zhao} {et~al.}(2016){Zhao}, {Caselli}, {Li}, {Krasnopolsky},
  {Shang}, \& {Nakamura}}]{Zhao2016}
{Zhao}, B., {Caselli}, P., {Li}, Z.-Y., {et~al.} 2016, \mnras, 460, 2050

\end{thebibliography}
\bibliographystyle{aa}





\begin{appendix}
\onecolumn
\section{TMC 1-C line properties}

    \begin{figure}[h!]
	    \centering
	    \includegraphics[width=0.98\textwidth,keepaspectratio]{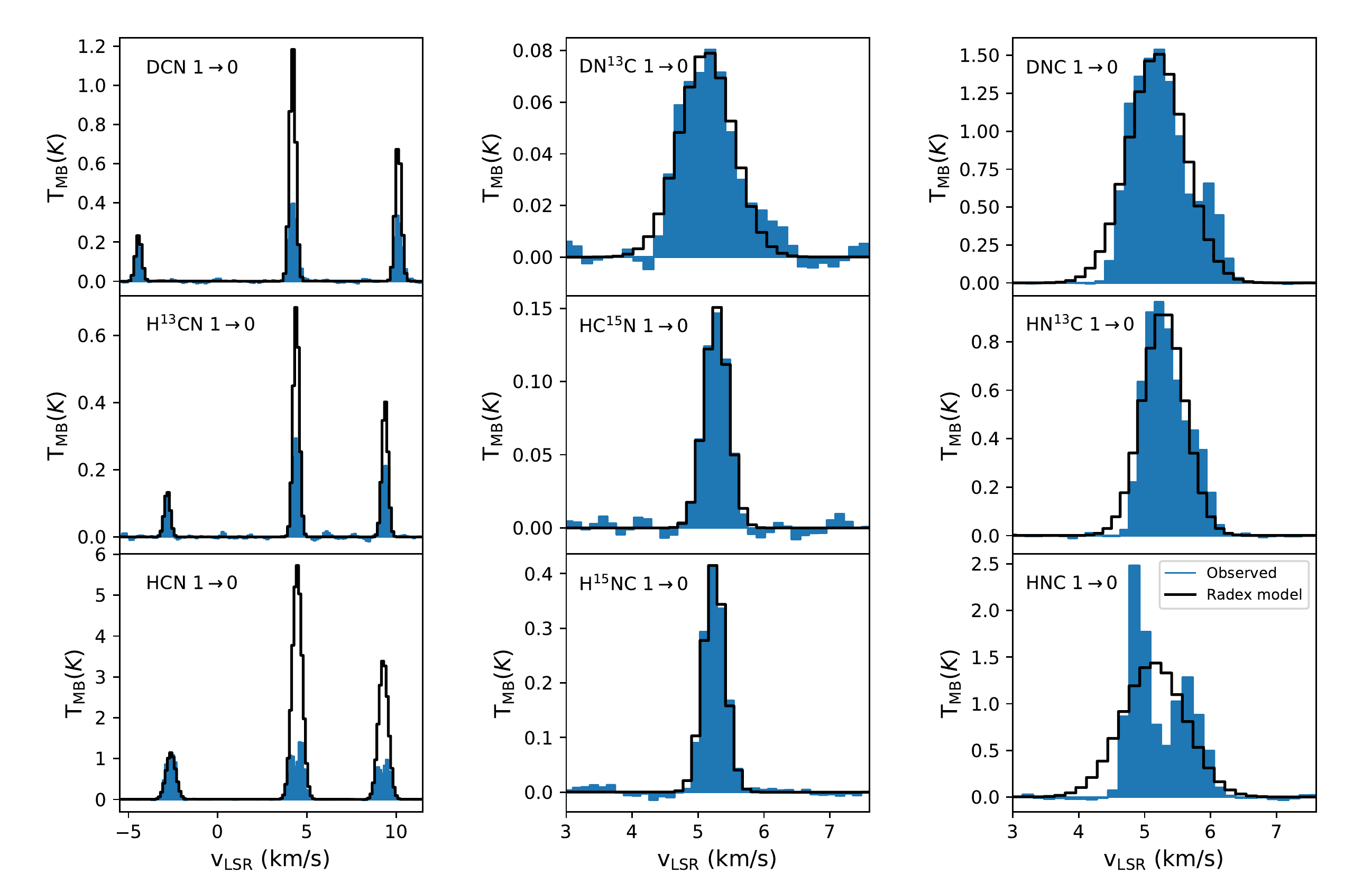}
	    \caption{Line spectra of the rotational transitions considered in this paper observed towards the TMC 1-C position. The frequency taken as reference in each panel is given in Table \ref{tab:summarylines}. The solid black lines represent the RADEX fitting over each line to compute column densities.}
	    \label{fig:spectraTMC1C}
    \end{figure}

    \begin{table}[h!]
		\centering
		\caption{Properties, main beam temperatures, and integrated intensities of the spectral lines in TMC~1-C. We include the opacities of the resolved hyperfine structure components.}
		\resizebox{0.98\textwidth}{!}{
			\begin{tabular}{lccccrrcccc}
				\toprule
				\multirow{2}{*}{{Species}} &  \multirow{2}{*}{{Transition}} &  {Frequency} & \multirow{2}{*}{{E$_{\rm up}$} {(K)}} & \multirow{2}{*}{{log(A$_{\rm{ij}}$)}} & \multirow{2}{*}{Peak position (km s$^{-1}$)} & \multirow{2}{*}{Width (km s$^{-1}$)} & & \multirow{2}{*}{{T}$_{\rm MB}$ {(K)}} & $\int${T}$_{\rm MB}\ dv$ & \multirow{2}{*}{${\tau}$}\\
					&  & {(MHz)} & & & & & & & {(K km s}$^{{-1}}${)} & \\
				 \midrule\midrule
					 D$^{13}$CN & $1\rightarrow 0$ & 71175.07 & 3.4 & -4.90 & $-$ & $-$ & & \multicolumn{2}{c}{$\rm rms = 7.77\times 10^{-3}$ K} & \\
					\midrule
					\multirow{3}{*}{DCN} & $1_{1}\rightarrow 0_{1}$   & 72413.50 & \multirow{3}{*}{3.5} & \multirow{3}{*}{-4.88} & $10.122\pm 0.158$ & $0.523\pm 0.158$ & & $0.339\pm 0.009$ & $0.189\pm 0.011$ & $2.32\pm 0.12$ \\
					 & $1_{2}\rightarrow 0_{1}$ & 72414.93 & & & $4.202\pm 0.158$ & $0.567\pm 0.158$ & & $0.412\pm 0.009$ & $0.249\pm 0.011$ & $3.87\pm 0.20$\\
					 & $1_{0}\rightarrow 0_{1}$ & 72417.03 & & & $-4.428\pm 0.158$ & $0.414\pm 0.158$ & & $0.238\pm 0.009$ & $0.105\pm 0.011$ & $0.77\pm 0.04$\\\midrule
					 DN$^{13}$C & $1\rightarrow 0$ & 73367.75 & 3.5 & -4.90 & $5.106\pm 0.027$ & $0.908\pm 0.061$ & & $0.077\pm 0.008$ & $0.073\pm 0.004$\\
					\midrule
					DNC & $1\rightarrow 0$ & 76305.70 & 3.7 & -4.80 & $5.181\pm 0.002$ & $1.020\pm 0.004$ & & $1.514\pm 0.011$ & $1.643\pm 0.006$ \\\midrule
					HC$^{15}$N & $1\rightarrow 0$ & 86054.97 & 4.1 & -4.66 & $5.277\pm 0.008$ & $0.452\pm 0.019$ & & $1.535\pm 0.008$ & $0.074\pm 0.003$ \\\midrule
					\multirow{3}{*}{H$^{13}$CN} &  $1_{1}\rightarrow 0_{1}$ & 86338.74 & \multirow{3}{*}{4.1} & \multirow{3}{*}{-4.65} & $9.376\pm 0.005$ & $0.456\pm 0.011$ & &  $0.220\pm 0.006$ & $0.107\pm 0.002$ & $1.28\pm 0.12$\\
					&  $1_{2}\rightarrow 0_{1}$ & 86340.17 & & & $4.401\pm 0.004$ & $0.502\pm 0.008$ & & $0.303\pm 0.006$ & $0.162\pm 0.002$ & $2.14\pm 0.20$\\
					&  $1_{0}\rightarrow 0_{1}$ & 86342.25 & & & $-2.825\pm 0.007$ & $0.401\pm 0.017$ & & $0.138\pm 0.006$ & $0.059\pm 0.002$ & $0.43\pm 0.04$\\\midrule
					HN$^{13}$C & $1\rightarrow 0$ & 87090.83 & 4.2 & -4.62 & $5.282\pm 0.002$ & $0.747\pm 0.004$ & & $0.923\pm 0.009$ & $0.735\pm 0.004$ \\\midrule
					\multirow{3}{*}{HCN} &  $1_{1}\rightarrow 0_{1}$ & 88630.42 & \multirow{3}{*}{4.2} & \multirow{3}{*}{-4.62} & $9.286\pm 0.003$ & $0.862\pm 0.005$ & &  $0.943\pm 0.008$ & $0.866\pm 0.005$ & $6.72\pm 0.03$\\
					&  $1_{2}\rightarrow 0_{1}$ & 88631.85 & & & $4.457\pm 0.000$ & $0.915\pm 0.003$ & & $1.305\pm 0.008$ & $1.270\pm 0.002$ & $11.20\pm 0.05$\\
					&  $1_{0}\rightarrow 0_{1}$ & 88633.94 & & & $-2.623\pm 0.000$ & $0.664\pm 0.004$ & & $1.146\pm 0.008$ & $0.810\pm 0.004$ & $2.24\pm 0.01$\\
					\midrule
					H$^{15}$NC & $1\rightarrow 0$ & 88865.69 & 4.2 & -4.70 & $5.246\pm 0.004$ & $0.419\pm 0.009$ & & $0.434\pm 0.010$ & $0.193\pm 0.004$ \\\midrule
					HNC$^{(a)}$ & $1\rightarrow 0$ & 90663.57 & 4.3 & -4.57 & $5.138\pm 0.000$ & $1.127\pm 0.006$ & & $1.440\pm 0.014$ & $1.727\pm 0.009$ \\
    			\bottomrule
			\end{tabular}
			}
			\label{tab:lineResultsTMC1C}
            \flushleft
			    {\small
			\ \ \ \ $^{{(\rm a)}}$ HNC $1\rightarrow0$ features self-absorption. The parameters shown here were calculated fitting one gaussian profile to the spectra.
			}
        \end{table}
        
\newpage
\section{NGC 1333 line properties}
    \subsection{NGC1333-C7-1}
        
        \begin{figure}[h!]
	        \centering
	        \includegraphics[width=0.95\textwidth,keepaspectratio]{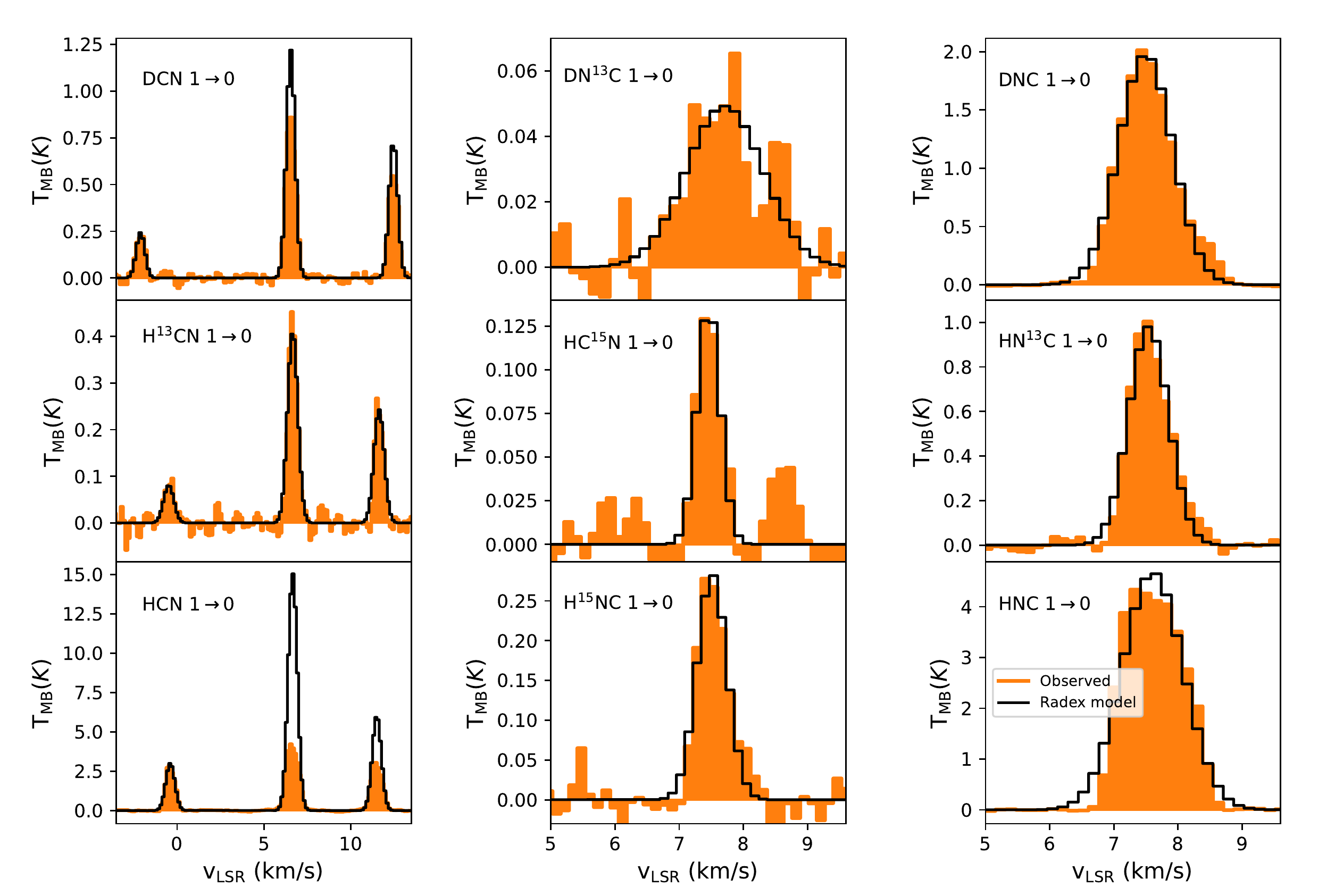}
	        \caption{Line spectra of the rotational transitions considered in this paper observed towards the NGC1333-C7-1 position. The frequency taken as reference in each panel is given in Table \ref{tab:summarylines}. The solid black lines represent the RADEX fitting over each line to compute column densities.}
	        \label{fig:spectraNGC1333_C7_1}
        \end{figure}
        
        \begin{table}[h!]
		\centering
		\caption{Properties, main beam temperatures, and integrated intensities of the spectral lines in NGC1333-C7-1. We include the opacities of the resolved hyperfine structure components.}
		\resizebox{0.95\textwidth}{!}{
			\begin{tabular}{lccccrrcccc}
				\toprule
				\multirow{2}{*}{{Species}} &  \multirow{2}{*}{{Transition}} &  {Frequency} & \multirow{2}{*}{{E$_{\rm up}$} {(K)}} & \multirow{2}{*}{{log(A$_{\rm{ij}}$)}} & \multirow{2}{*}{Peak position (km s$^{-1}$)} & \multirow{2}{*}{ Width (km s$^{-1}$)} & & \multirow{2}{*}{{T}$_{\rm MB}$ {(K)}} & $\int${T}$_{\rm MB}\ dv$ & \multirow{2}{*}{${\tau}$}\\
					&  & {(MHz)} & & & & & & & {(K km s}$^{{-1}}${)} & \\
				 \midrule\midrule
					 D$^{13}$CN & $1\rightarrow 0$ & 71175.07 & 3.4 & -4.90 & $-$ & $-$ & & \multicolumn{2}{c}{$\rm rms = 2.51\times 10^{-2}$ K} & \\
					\midrule
					\multirow{3}{*}{DCN} & $1_{1}\rightarrow 0_{1}$   & 72413.50 & \multirow{3}{*}{3.5} & \multirow{3}{*}{-4.88} & $12.461\pm 0.008$ & $0.624\pm 0.017$ & & $0.581\pm 0.020$ & $0.386\pm 0.010$ & $0.77\pm 0.04$ \\
					 & $1_{2}\rightarrow 0_{1}$ & 72414.93 & & & $6.544\pm 0.005$ & $0.652\pm 0.012$ & & $0.871\pm 0.020$ & $0.605\pm 0.010$ & $0.84\pm 0.14$\\
					 & $1_{0}\rightarrow 0_{1}$ & 72417.03 & & & $-2.108\pm 0.017$ & $0.590\pm 0.038$ & & $0.244\pm 0.020$ & $0.153\pm 0.010$ & $0.50\pm 0.10$ \\\midrule
					 DN$^{13}$C & $1\rightarrow 0$ & 73367.75 & 3.5 & -4.90 & $7.447\pm 0.090$ & $1.343\pm 0.277$ & & $0.052\pm 0.014$ & $0.074\pm 0.011$\\
					\midrule
					DNC & $1\rightarrow 0$ & 76305.70 & 3.7 & -4.80 & $7.500\pm 0.000$ & $1.004\pm 0.005$ & & $2.010\pm 0.015$ & $2.148\pm 0.008$ \\\midrule
					HC$^{15}$N & $1\rightarrow 0$ & 86054.97 & 4.1 & -4.66 & $7.471\pm 0.018$ & $0.475\pm 0.039$ & & $0.133\pm 0.013$ & $0.068\pm 0.005$ \\\midrule
					\multirow{3}{*}{H$^{13}$CN} &  $1_{1}\rightarrow 0_{1}$ & 86338.74 & \multirow{3}{*}{4.1} & \multirow{3}{*}{-4.65} & $11.633\pm 0.012$ & $0.600\pm 0.027$ & & $0.272\pm 0.016$ & $0.174\pm 0.007$ & $0.07\pm 0.17$\\
					&  $1_{2}\rightarrow 0_{1}$ & 86340.17 & & & $6.658\pm 0.007$ & $0.574\pm 0.016$ & & $0.447\pm 0.016$ & $0.273\pm 0.007$ & $0.12\pm 0.28$\\
					&  $1_{0}\rightarrow 0_{1}$ & 86342.25 & & &  $-0.485\pm 0.045$ & $0.695\pm 0.100$ & & $0.081\pm 0.016$ & $0.060\pm 0.007$ & $0.02\pm 0.07$\\\midrule
					HN$^{13}$C & $1\rightarrow 0$ & 87090.83 & 4.2 & -4.62 & $7.544\pm 0.004$ & $0.738\pm 0.009$ & & $1.037\pm 0.017$ & $0.814\pm 0.008$ \\\midrule
					\multirow{3}{*}{HCN} &  $1_{1}\rightarrow 0_{1}$ & 88630.42 & \multirow{3}{*}{4.2} & \multirow{3}{*}{-4.62} & $11.500\pm 0.031$ & $0.746\pm 0.074$ & & $3.146\pm 0.315$ & $2.592\pm 0.146$ & $3.25\pm 0.59$\\
					&  $1_{2}\rightarrow 0_{1}$ & 88631.85 & & & $6.659\pm 0.022$ & $0.830\pm 0.054$ & & $4.480\pm 0.315$ & $4.079\pm 0.219$ & $5.42\pm 0.98$\\
					&  $1_{0}\rightarrow 0_{1}$ & 88633.94 & & &  $-0.394\pm 0.029$ & $0.624\pm 0.066$ & & $3.014\pm 0.315$ & $2.127\pm 0.185$ & $1.08\pm 0.20$\\
					\midrule
					H$^{15}$NC & $1\rightarrow 0$ & 88865.69 & 4.2 & -4.70 & $7.515\pm 0.012$ & $0.609\pm 0.027$ & & $0.287\pm 0.018$ & $0.186\pm 0.007$ \\\midrule
					HNC & $1\rightarrow 0$ & 90663.57 & 4.3 & -4.57 & $7.607\pm 0.001$ & $1.118\pm 0.002$ & & $4.674\pm 0.016$ & $5.563\pm 0.010$\\
    			\bottomrule
			\end{tabular}
			}
			\label{tab:lineResultsNGC1333_C7_1}
        \end{table}

\newpage
    \subsection{NGC1333-C7-2}
    
        \begin{figure}[h!]
	        \centering
	        \includegraphics[width=\textwidth,keepaspectratio]{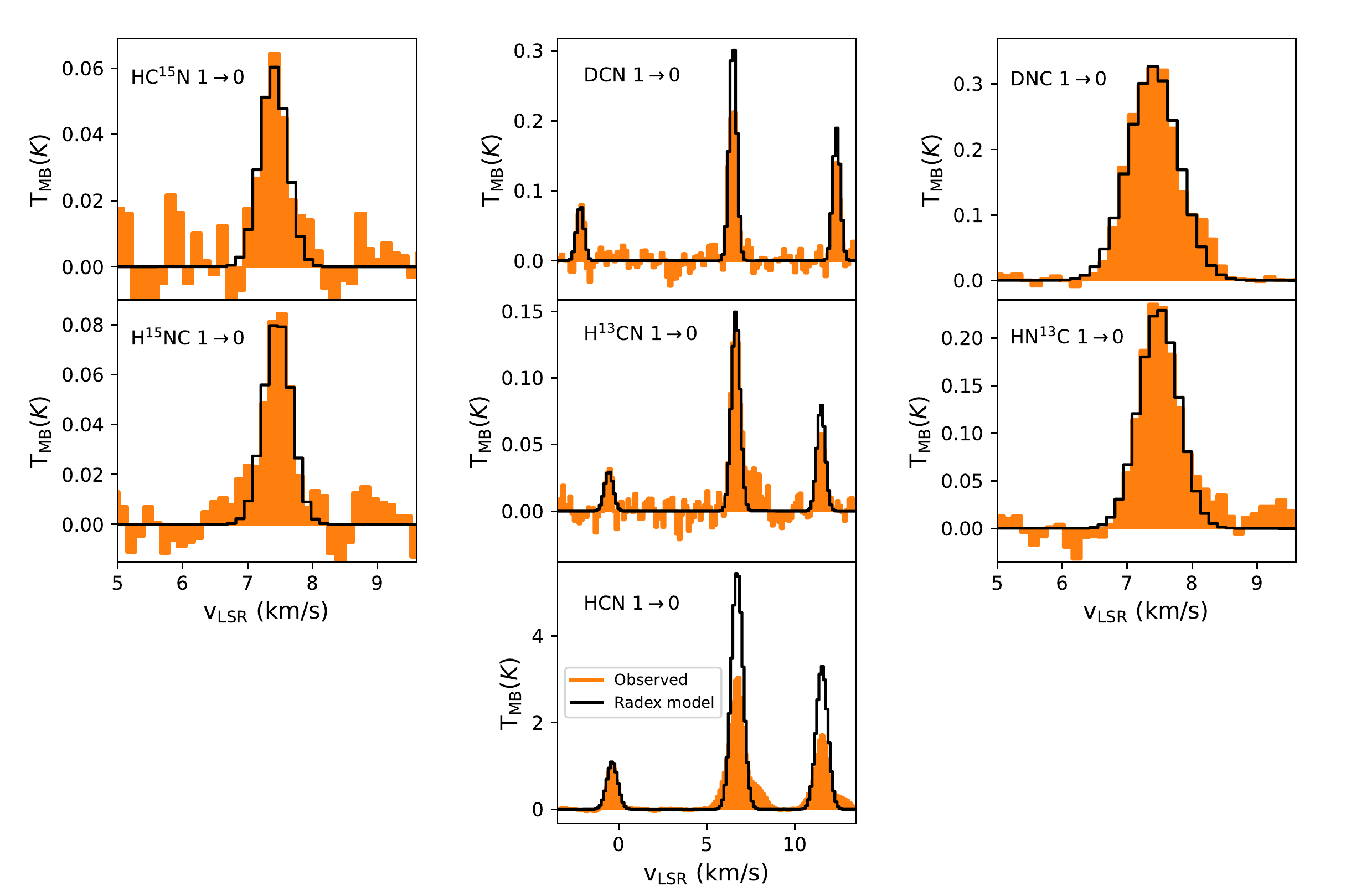}
    	    \caption{Line spectra of the rotational transitions considered in this paper observed towards the NGC1333-C7-2 position. The frequency taken as reference in each panel is given in Table \ref{tab:summarylines}. The solid black lines represent the RADEX fitting over each line to compute column densities.}
	        \label{fig:spectraNGC1333_C7_2}
        \end{figure}
        
        \begin{table}[h!]
		\centering
		\caption{Properties, main beam temperatures, and integrated intensities of the spectral lines in NGC1333-C7-2. We include the opacities of the resolved hyperfine structure components.}
		\resizebox{\textwidth}{!}{
			\begin{tabular}{lccccrrcccc}
				\toprule
				\multirow{2}{*}{{Species}} &  \multirow{2}{*}{{Transition}} &  {Frequency} & \multirow{2}{*}{{E$_{\rm up}$} {(K)}} & \multirow{2}{*}{{log(A$_{\rm{ij}}$)}} & \multirow{2}{*}{Peak position (km s$^{-1}$)} & \multirow{2}{*}{Width (km s$^{-1}$)} & & \multirow{2}{*}{{T}$_{\rm MB}$ {(K)}} & $\int${T}$_{\rm MB}\ dv$ & \multirow{2}{*}{${\tau}$}\\
					&  & {(MHz)} & & & & & & & {(K km s}$^{{-1}}${)} & \\
				 \midrule\midrule
					 D$^{13}$CN & $1\rightarrow 0$ & 71175.07 & 3.4 & -4.90 & $-$ & $-$ & & \multicolumn{2}{c}{$\rm rms = 2.223\times 10^{-2}$ K} & \\
					\midrule
					\multirow{3}{*}{DCN} & $1_{1}\rightarrow 0_{1}$   & 72413.50 & \multirow{3}{*}{3.5} & \multirow{3}{*}{-4.88} & $12.385\pm 0.018$ & $0.542\pm 0.047$ & & $0.133\pm 0.011$ & $0.077\pm 0.005$ & $0.060\pm 0.012$ \\
					 & $1_{2}\rightarrow 0_{1}$ & 72414.93 & & & $6.489\pm 0.011$ & $0.522\pm 0.026$ & & $0.216\pm 0.011$ & $0.120\pm 0.005$ & $0.100\pm 0.020$\\
					 & $1_{0}\rightarrow 0_{1}$ & 72417.03 & & & $-2.160\pm 0.030$ & $0.328\pm 0.060$ & & $0.063\pm 0.011$ & $0.022\pm 0.004$ & $0.020\pm 0.004$\\\midrule
					 DN$^{13}$C & $1\rightarrow 0$ & 73367.75 & 3.5 & -4.90 & $-$ & $-$ & & \multicolumn{2}{c}{$\rm rms = 8.828\times 10^{-3}$ K} &\\
					\midrule
					DNC & $1\rightarrow 0$ & 76305.70 & 3.7 & -4.80 & $7.425\pm 0.006$ & $0.980\pm 0.016$ & & $0.316\pm 0.007$ & $0.327\pm 0.016$ \\\midrule
					HC$^{15}$N & $1\rightarrow 0$ & 86054.97 & 4.1 & -4.66 & $7.505\pm 0.037$ & $0.583\pm 0.079$ & & $0.058\pm 0.011$ & $0.036\pm 0.004$ \\\midrule
					\multirow{3}{*}{H$^{13}$CN} &  $1_{1}\rightarrow 0_{1}$ & 86338.74 & \multirow{3}{*}{4.1} & \multirow{3}{*}{-4.65} & $11.601\pm 0.039$ & $0.722\pm 0.151$ & &  $0.064\pm 0.010$ & $0.049\pm 0.006$ & $0.060\pm 0.102$\\
					&  $1_{2}\rightarrow 0_{1}$ & 86340.17 & & & $6.646\pm 0.016$ & $0.668\pm 0.042$ & & $0.132\pm 0.010$ & $0.094\pm 0.005$ & $0.100\pm 0.170$\\
					&  $1_{0}\rightarrow 0_{1}$ & 86342.25 & & & $-0.583\pm 0.160$ & $0.589\pm 0.488$ & & $0.049\pm 0.010$ & $0.029\pm 0.010$ & $0.020\pm 0.034$\\\midrule
					HN$^{13}$C & $1\rightarrow 0$ & 87090.83 & 4.2 & -4.62 & $7.499\pm 0.009$ & $0.688\pm 0.023$ & & $0.203\pm 0.009$ & $0.149\pm 0.004$ \\\midrule
					\multirow{3}{*}{HCN} &  $1_{1}\rightarrow 0_{1}$ & 88630.42 & \multirow{3}{*}{4.2} & \multirow{3}{*}{-4.62} & $11.555\pm 0.049$ & $1.105\pm 0.156$ & &  $1.592\pm 0.188$ & $2.326\pm 0.146$ & $0.060\pm 0.047$\\
					&  $1_{2}\rightarrow 0_{1}$ & 88631.85 & & & $6.725\pm 0.029$ & $1.200\pm 0.091$ & & $2.819\pm 0.188$ & $4.241\pm 0.143$ & $0.100\pm 0.078$\\
					&  $1_{0}\rightarrow 0_{1}$ & 88633.94 & & & $-0.402\pm 0.057$ & $0.728\pm 0.161$ & & $1.101\pm 0.188$ & $1.196\pm 0.134$ & $0.020\pm 0.016$\\
					\midrule
					H$^{15}$NC & $1\rightarrow 0$ & 88865.69 & 4.2 & -4.70 & $7.469\pm 0.026$ & $0.462\pm 0.060$ & & $0.080\pm 0.012$ & $0.039\pm 0.004$ \\
    			\bottomrule
			\end{tabular}
			}
			\label{tab:lineResultsNGC1333_C7_2}
        \end{table}

\newpage
    \subsection{NGC1333-C7-3}
    
        \begin{figure}[h!]
	        \centering
	        \includegraphics[width=\textwidth,keepaspectratio]{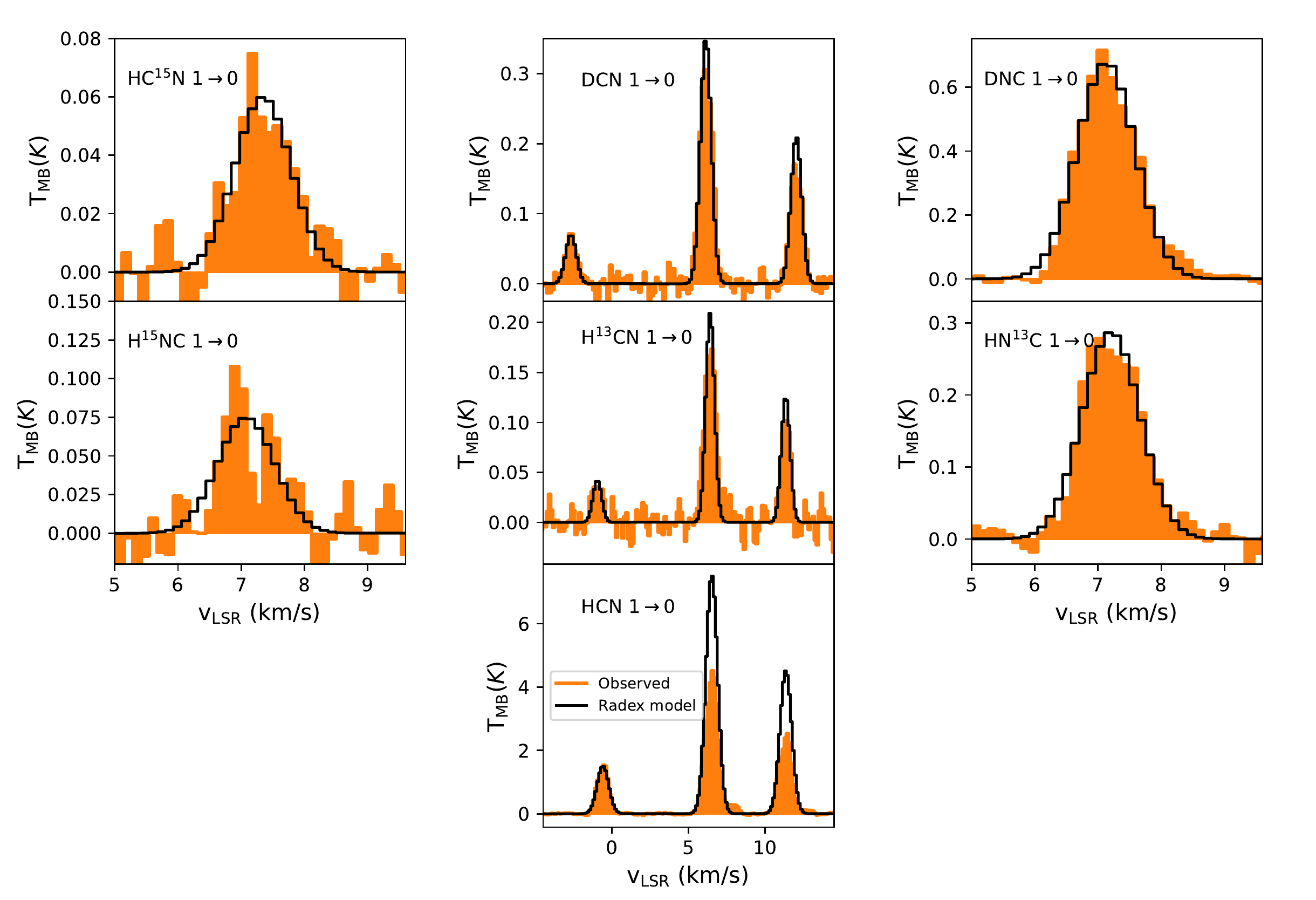}
	        \caption{Line spectra of the rotational transitions considered in this paper observed towards the NGC1333-C7-3 position. The frequency taken as reference in each panel is given in Table \ref{tab:summarylines}. The solid black lines represent the RADEX fitting over each line to compute column densities.}
	        \label{fig:spectraNGC1333_C7_3}
        \end{figure}
        
        \begin{table}[h!]
		\centering
		\caption{Properties, main beam temperatures, and integrated intensities of the spectral lines in NGC1333-C7-3. We include the opacities of the resolved hyperfine structure components.}
		\resizebox{\textwidth}{!}{
			\begin{tabular}{lccccrrcccc}
				\toprule
				\multirow{2}{*}{{Species}} &  \multirow{2}{*}{{Transition}} &  {Frequency} & \multirow{2}{*}{{E$_{\rm up}$} {(K)}} & \multirow{2}{*}{{log(A$_{\rm{ij}}$)}} & \multirow{2}{*}{Peak position (km s$^{-1}$)} & \multirow{2}{*}{Width (km s$^{-1}$)} & & \multirow{2}{*}{{T}$_{\rm MB}$ {(K)}} & $\int${T}$_{\rm MB}\ dv$ & \multirow{2}{*}{${\tau}$}\\
					&  & {(MHz)} & & & & & & & {(K km s}$^{{-1}}${)} & \\
				 \midrule\midrule
					 D$^{13}$CN & $1\rightarrow 0$ & 71175.07 & 3.4 & -4.90 & $-$ & $-$ & & \multicolumn{2}{c}{$\rm rms = 3.712\times 10^{-2}$ K} & \\
					\midrule
					\multirow{3}{*}{DCN} & $1_{1}\rightarrow 0_{1}$   & 72413.50 & \multirow{3}{*}{3.5} & \multirow{3}{*}{-4.88} & $12.034\pm 0.014$ & $0.931\pm 0.034$ & & $0.188\pm 0.010$ & $0.186\pm 0.006$ & $0.060\pm 0.056$ \\
					 & $1_{2}\rightarrow 0_{1}$ & 72414.93 & & & $6.132\pm 0.008$ & $0.915\pm 0.018$ & & $0.328\pm 0.010$ & $0.319\pm 0.006$ & $0.100\pm 0.009$\\
					 & $1_{0}\rightarrow 0_{1}$ & 72417.03 & & & $-2.684\pm 0.035$ & $0.832\pm 0.102$ & & $0.070\pm 0.010$ & $0.062\pm 0.006$ & $0.020\pm 0.018$\\\midrule
					 DN$^{13}$C & $1\rightarrow 0$ & 73367.75 & 3.5 & -4.90 & $-$ & $-$ & & \multicolumn{2}{c}{$\rm rms = 1.439\times 10^{-2}$ K} &\\
					\midrule
					DNC & $1\rightarrow 0$ & 76305.70 & 3.7 & -4.80 & $7.130\pm 0.005$ & $1.101\pm 0.012$ & & $0.709\pm 0.020$ & $0.814\pm 0.012$ \\\midrule
					HC$^{15}$N & $1\rightarrow 0$ & 86054.97 & 4.1 & -4.66 & $7.346\pm 0.048$ & $0.991\pm 0.091$ & & $0.058\pm 0.011$ & $0.062\pm 0.006$ \\\midrule
					\multirow{3}{*}{H$^{13}$CN} &  $1_{1}\rightarrow 0_{1}$ & 86338.74 & \multirow{3}{*}{4.1} & \multirow{3}{*}{-4.65} & $11.348\pm 0.022$ & $0.755\pm 0.048$ & &  $0.111\pm 0.011$ & $0.090\pm 0.005$ & $0.060\pm 0.001$\\
					&  $1_{2}\rightarrow 0_{1}$ & 86340.17 & & & $6.413\pm 0.015$ & $0.857\pm 0.036$ & & $0.172\pm 0.011$ & $0.156\pm 0.006$ & $0.100\pm 0.001$\\
					&  $1_{0}\rightarrow 0_{1}$ & 86342.25 & & & $-0.962\pm 0.058$ & $0.674\pm 0.098$ & & $0.042\pm 0.011$ & $0.030\pm 0.010$ & $0.020\pm 0.034$\\\midrule
					HN$^{13}$C & $1\rightarrow 0$ & 87090.83 & 4.2 & -4.62 & $7.193\pm 0.009$ & $0.984\pm 0.020$ & & $0.306\pm 0.011$ & $0.321\pm 0.006$ \\\midrule
					\multirow{3}{*}{HCN} &  $1_{1}\rightarrow 0_{1}$ & 88630.42 & \multirow{3}{*}{4.2} & \multirow{3}{*}{-4.62} & $11.354\pm 0.046$ & $0.987\pm 0.120$ & &  $2.438\pm 0.246$ & $2.056\pm 0.239$ & $0.128\pm 0.141$\\
					&  $1_{2}\rightarrow 0_{1}$ & 88631.85 & & & $6.514\pm 0.024$ & $1.021\pm 0.062$ & & $4.465\pm 0.246$ & $4.852\pm 0.242$ & $0.213\pm 0.235$\\
					&  $1_{0}\rightarrow 0_{1}$ & 88633.94 & & & $-0.592\pm 0.069$ & $0.896\pm 0.162$ & & $1.508\pm 0.246$ & $1.440\pm 0.219$ & $0.043\pm 0.047$\\
					\midrule
					H$^{15}$NC & $1\rightarrow 0$ & 88865.69 & 4.2 & -4.70 & $7.060\pm 0.034$ & $0.932\pm 0.083$ & & $0.084\pm 0.011$ & $0.083\pm 0.006$ \\
    			\bottomrule
			\end{tabular}
			}
			\label{tab:lineResultsNGC1333_C7_3}
        \end{table}

\newpage

    \subsection{NGC1333-C7-4}
    
        \begin{figure}[!h]
	        \centering
	        \includegraphics[width=\textwidth,keepaspectratio]{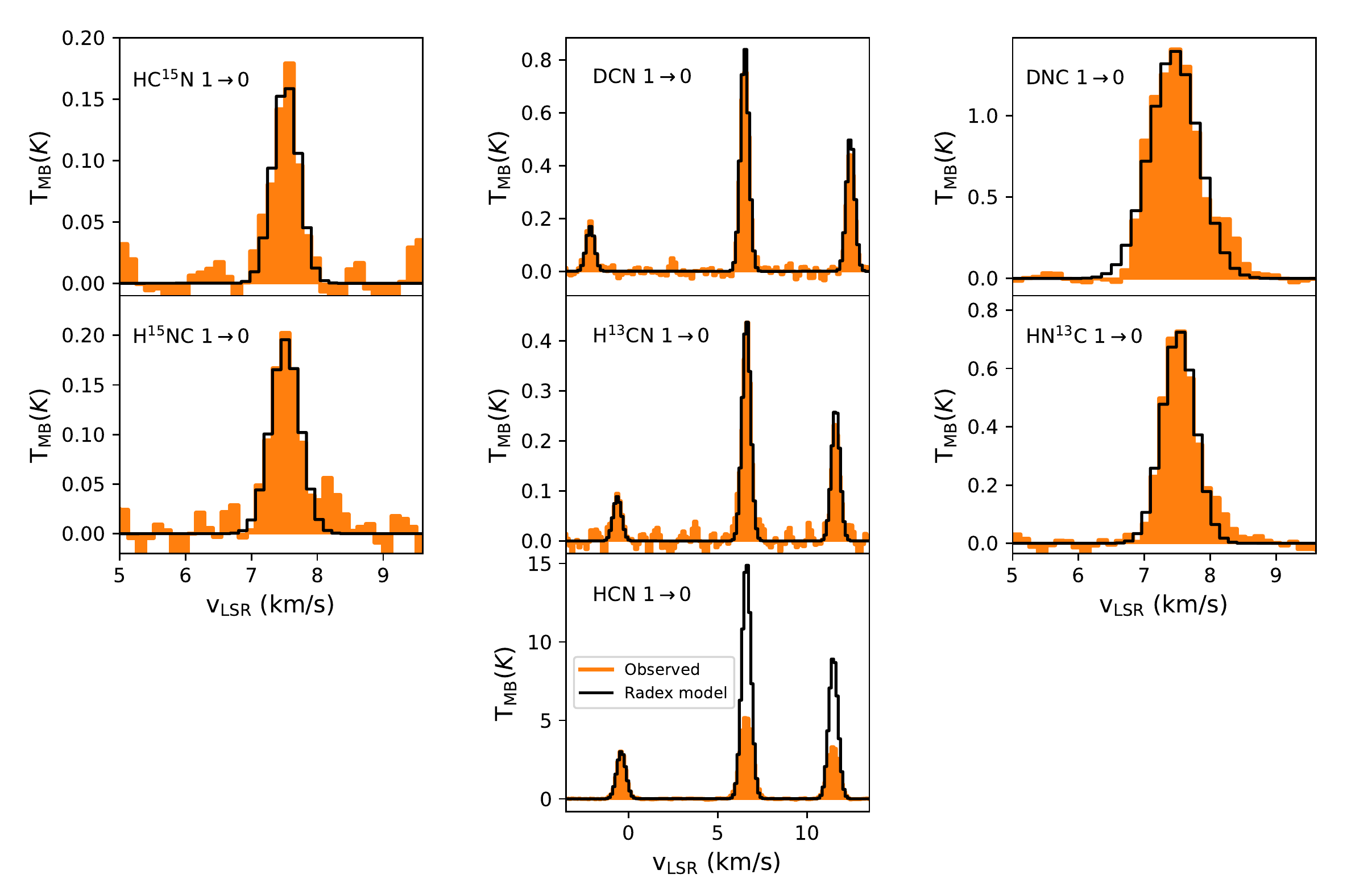}
	        \caption{Line spectra of the rotational transitions considered in this paper observed towards the NGC1333-C7-4 position. The frequency taken as reference in each panel is given in Table \ref{tab:summarylines}. The solid black lines represent the RADEX fitting over each line to compute column densities.}
            \label{fig:spectraNGC1333_C7_4}
        \end{figure}
        
        \begin{table}[h!]
		\centering
		\caption{Properties, main beam temperatures, and integrated intensities of the spectral lines in NGC1333-C7-4. We include the opacities of the resolved hyperfine structure components.}
		\resizebox{\textwidth}{!}{
			\begin{tabular}{lccccrrcccc}
				\toprule
				\multirow{2}{*}{{Species}} &  \multirow{2}{*}{{Transition}} &  {Frequency} & \multirow{2}{*}{{E$_{\rm up}$} {(K)}} & \multirow{2}{*}{{log(A$_{\rm{ij}}$)}} & \multirow{2}{*}{Peak position (km s$^{-1}$)} & \multirow{2}{*}{Width (km s$^{-1}$)} & & \multirow{2}{*}{{T}$_{\rm MB}$ {(K)}} & $\int${T}$_{\rm MB}\ dv$ & \multirow{2}{*}{${\tau}$}\\
					&  & {(MHz)} & & & & & & & {(K km s}$^{{-1}}${)} & \\
				 \midrule\midrule
					 D$^{13}$CN & $1\rightarrow 0$ & 71175.07 & 3.4 & -4.90 & $-$ & $-$ & & \multicolumn{2}{c}{$\rm rms = 3.606\times 10^{-2}$ K} & \\
					\midrule
					\multirow{3}{*}{DCN} & $1_{1}\rightarrow 0_{1}$   & 72413.50 & \multirow{3}{*}{3.5} & \multirow{3}{*}{-4.88} & $12.449\pm 0.007$ & $0.482\pm 0.018$ & & $0.449\pm 0.018$ & $0.253\pm 0.008$ & $0.060\pm 0.065$ \\
					 & $1_{2}\rightarrow 0_{1}$ & 72414.93 & & & $6.515\pm 0.004$ & $0.526\pm 0.011$ & & $0.755\pm 0.010$ & $0.436\pm 0.006$ & $0.100\pm 0.108$\\
					 & $1_{0}\rightarrow 0_{1}$ & 72417.03 & & & $-2.136\pm 0.019$ & $0.546\pm 0.055$ & & $0.170\pm 0.010$ & $0.084\pm 0.006$ & $0.020\pm 0.022$\\\midrule
					 DN$^{13}$C & $1\rightarrow 0$ & 73367.75 & 3.5 & -4.90 & $-$ & $-$ & & \multicolumn{2}{c}{$\rm rms = 1.532\times 10^{-2}$ K} &\\
					\midrule
					DNC & $1\rightarrow 0$ & 76305.70 & 3.7 & -4.80 & $7.452\pm 0.003$ & $0.893\pm 0.006$ & & $1.320\pm 0.017$ & $1.243\pm 0.009$ \\\midrule
					HC$^{15}$N & $1\rightarrow 0$ & 86054.97 & 4.1 & -4.66 & $7.531\pm 0.015$ & $0.477\pm 0.035$ & & $0.155\pm 0.013$ & $0.079\pm 0.005$ \\\midrule
					\multirow{3}{*}{H$^{13}$CN} &  $1_{1}\rightarrow 0_{1}$ & 86338.74 & \multirow{3}{*}{4.1} & \multirow{3}{*}{-4.65} & $11.617\pm 0.015$ & $0.503\pm 0.039$ & &  $0.264\pm 0.017$ & $0.131\pm 0.006$ & $0.060\pm 0.030$\\
					&  $1_{2}\rightarrow 0_{1}$ & 86340.17 & & & $6.611\pm 0.008$ & $0.556\pm 0.023$ & & $0.431\pm 0.017$ & $0.238\pm 0.007$ & $0.100\pm 0.049$\\
					&  $1_{0}\rightarrow 0_{1}$ & 86342.25 & & & $-0.625\pm 0.039$ & $0.556\pm 0.089$ & & $0.074\pm 0.017$ & $0.040\pm 0.006$ & $0.020\pm 0.010$\\\midrule
					HN$^{13}$C & $1\rightarrow 0$ & 87090.83 & 4.2 & -4.62 & $7.524\pm 0.004$ & $0.609\pm 0.009$ & & $0.761\pm 0.013$ & $0.493\pm 0.006$ \\\midrule
					\multirow{3}{*}{HCN} &  $1_{1}\rightarrow 0_{1}$ & 88630.42 & \multirow{3}{*}{4.2} & \multirow{3}{*}{-4.62} & $11.486\pm 0.004$ & $0.729\pm 0.008$ & &  $3.427\pm 0.310$ & $2.716\pm 0.132$ & $1.134\pm 0.220$\\
					&  $1_{2}\rightarrow 0_{1}$ & 88631.85 & & & $6.611\pm 0.002$ & $0.831\pm 0.006$ & & $5.442\pm 0.310$ & $4.823\pm 0.141$ & $1.890\pm 0.367$\\
					&  $1_{0}\rightarrow 0_{1}$ & 88633.94 & & & $-0.404\pm 0.004$ & $0.584\pm 0.009$ & & $3.032\pm 0.310$ & $1.911\pm 0.121$ & $0.378\pm 0.073$\\
					\midrule
					H$^{15}$NC & $1\rightarrow 0$ & 88865.69 & 4.2 & -4.70 & $7.511\pm 0.018$ & $0.513\pm 0.048$ & & $0.220\pm 0.014$ & $0.114\pm 0.006$ \\
    			\bottomrule
			\end{tabular}
			}
			\label{tab:lineResultsNGC1333_C7_4}
        \end{table}
        
\newpage    
    \subsection{NGC1333-C7-5}
    
\begin{figure}[!h]
	\centering
	\includegraphics[width=\textwidth,keepaspectratio]{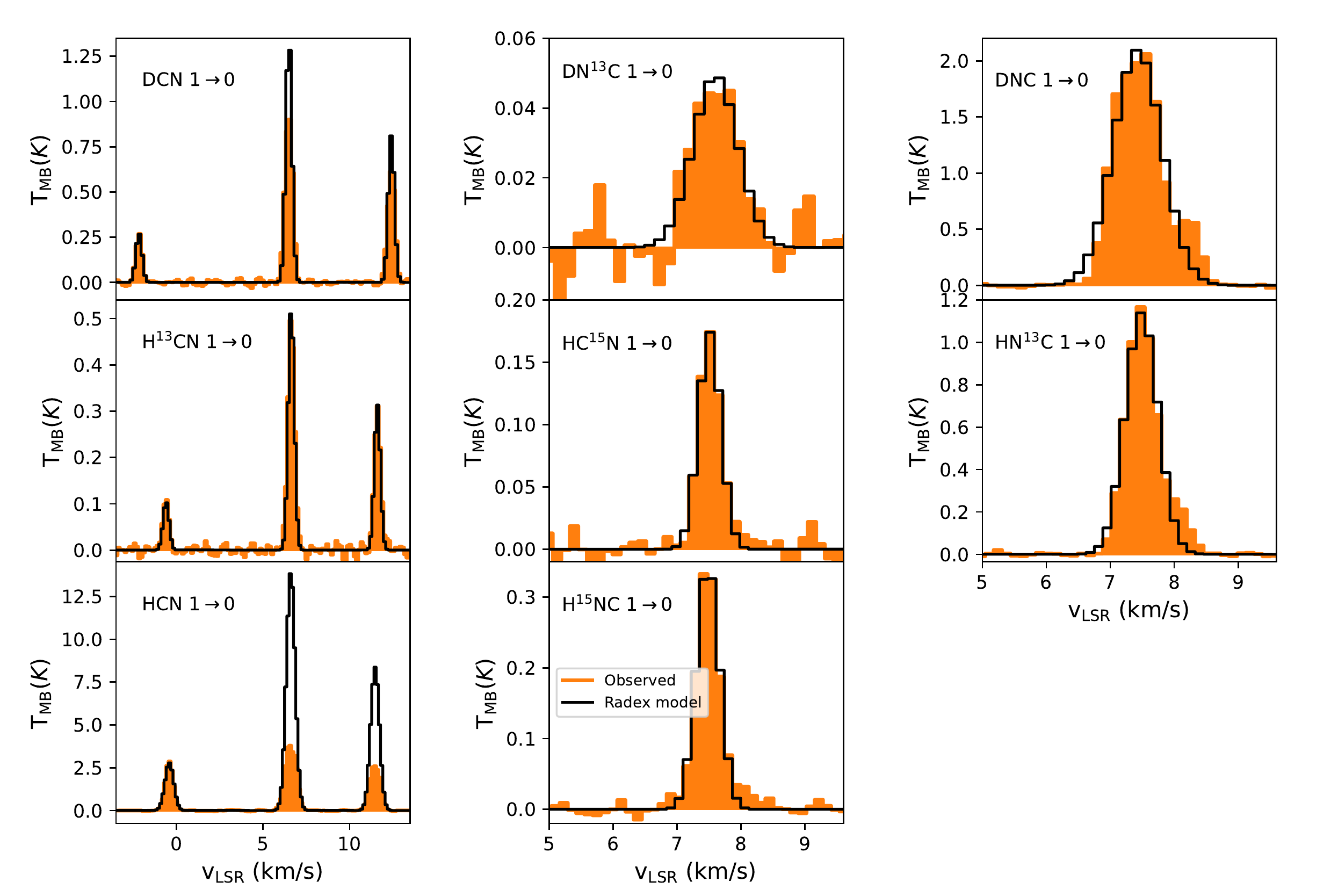}
	\caption{Line spectra of the rotational transitions considered in this paper observed towards the NGC1333-C7-5 position. The frequency taken as reference in each panel is given in Table \ref{tab:summarylines}. The solid black lines represent the RADEX fitting over each line to compute column densities.}
	\label{fig:spectraNGC1333_C7_5}
\end{figure}

\begin{table}[!h]
		\centering
		\caption{Properties, main beam temperatures, and integrated intensities of the spectral lines in NGC1333-C7-5. We include the opacities of the resolved hyperfine structure components.}
		\resizebox{\textwidth}{!}{
			\begin{tabular}{lccccrrcccc}
				\toprule
				\multirow{2}{*}{{Species}} &  \multirow{2}{*}{{Transition}} &  {Frequency} & \multirow{2}{*}{{E$_{\rm up}$} {(K)}} & \multirow{2}{*}{{log(A$_{\rm{ij}}$)}} & \multirow{2}{*}{Peak position (km s$^{-1}$)} & \multirow{2}{*}{Width (km s$^{-1}$)} & & \multirow{2}{*}{{T}$_{\rm MB}$ {(K)}} & $\int${T}$_{\rm MB}\ dv$ & \multirow{2}{*}{${\tau}$}\\
					&  & {(MHz)} & & & & & & & {(K km s}$^{{-1}}${)} & \\
				 \midrule\midrule
					 D$^{13}$CN & $1\rightarrow 0$ & 71175.07 & 3.4 & -4.90 & $-$ & $-$ & & \multicolumn{2}{c}{$\rm rms = 3.606\times 10^{-2}$ K} & \\
					\midrule
					\multirow{3}{*}{DCN} & $1_{1}\rightarrow 0_{1}$   & 72413.50 & \multirow{3}{*}{3.5} & \multirow{3}{*}{-4.88} & $12.412\pm 0.001$ & $0.495\pm 0.008$ & & $0.623\pm 0.021$ & $0.337\pm 0.006$ & $0.356\pm 0.077$ \\
					 & $1_{2}\rightarrow 0_{1}$ & 72414.93 & & & $6.487\pm 0.000$ & $0.524\pm 0.005$ & & $0.961\pm 0.021$ & $0.544\pm 0.008$ & $0.593\pm 0.129$\\
					 & $1_{0}\rightarrow 0_{1}$ & 72417.03 & & & $-2.175\pm 0.000$ & $0.421\pm 0.018$ & & $0.272\pm 0.021$ & $0.126\pm 0.007$ & $0.119\pm 0.026$\\\midrule
					 DN$^{13}$C & $1\rightarrow 0$ & 73367.75 & 3.5 & -4.90 & $7.464\pm 0.037$ & $0.829\pm 0.074$ & & $0.051\pm 0.007$ & $0.045\pm 0.004$ \\
					\midrule
					DNC & $1\rightarrow 0$ & 76305.70 & 3.7 & -4.80 & $7.427\pm 0.001$ & $0.897\pm 0.002$ & & $2.082\pm 0.010$ & $2.022\pm 0.006$ \\\midrule
					HC$^{15}$N & $1\rightarrow 0$ & 86054.97 & 4.1 & -4.66 & $7.509\pm 0.009$ & $0.416\pm 0.022$ & & $0.179\pm 0.008$ & $0.082\pm 0.003$ \\\midrule
					\multirow{3}{*}{H$^{13}$CN} &  $1_{1}\rightarrow 0_{1}$ & 86338.74 & \multirow{3}{*}{4.1} & \multirow{3}{*}{-4.65} & $11.603\pm 0.008$ & $0.469\pm 0.018$ & &  $0.314\pm 0.013$ & $0.157\pm 0.005$ & $0.097\pm 0.074$\\
					&  $1_{2}\rightarrow 0_{1}$ & 86340.17 & & & $6.630\pm 0.005$ & $0.474\pm 0.011$ & & $0.494\pm 0.013$ & $0.250\pm 0.005$ & $0.161\pm 0.124$\\
					&  $1_{0}\rightarrow 0_{1}$ & 86342.25 & & & $-0.605\pm 0.021$ & $0.406\pm 0.047$ & & $0.105\pm 0.013$ & $0.045\pm 0.005$ & $0.032\pm 0.025$\\\midrule
					HN$^{13}$C & $1\rightarrow 0$ & 87090.83 & 4.2 & -4.62 & $7.495\pm 0.002$ & $0.606\pm 0.005$ & & $1.124\pm 0.011$ & $0.724\pm 0.005$ \\\midrule
					\multirow{3}{*}{HCN} &  $1_{1}\rightarrow 0_{1}$ & 88630.42 & \multirow{3}{*}{4.2} & \multirow{3}{*}{-4.62} & $11.468\pm 0.023$ & $0.767\pm 0.059$ & & $2.731\pm 0.28$ & $2.230\pm 0.135$ & $2.616\pm 0.504$\\
					&  $1_{2}\rightarrow 0_{1}$ & 88631.85 & & & $6.600\pm 0.015$ & $0.855\pm 0.039$ & & $4.019\pm 0.280$ & $3.657\pm 0.138$ & $4.360\pm 0.840$\\
					&  $1_{0}\rightarrow 0_{1}$ & 88633.94 & & & $-0.420\pm 0.019$ & $0.578\pm 0.048$ & & $2.791\pm 0.280$ & $1.716\pm 0.113$ & $0.872\pm 0.168$\\
					\midrule
					H$^{15}$NC & $1\rightarrow 0$ & 88865.69 & 4.2 & -4.70 & $7.480\pm 0.006$ & $0.425\pm 0.014$ & & $0.336\pm 0.008$ & $0.149\pm 0.003$ \\
    			\bottomrule
			\end{tabular}
			}
			\label{tab:lineResultsNGC1333_C7_5}
        \end{table}
        
\end{appendix}
\end{document}